# When Energy and Information Revolutions Meet 2D Janus

*Long Zhang,*[1,2] *Ziqi Ren,*[1,3] *Li Sun,*[1,3] *Yihua Gao,*[1,3] *Deli Wang,*[4] *Junjie He,*[5] *and Guoying Gao*[1,2,a)]

[1]School of Physics, Huazhong University of Science and Technology, Wuhan 430074, China
[2]Wuhan National High Magnetic Field Center, Huazhong University of Science and Technology, Wuhan 430074, China
[3]Wuhan National Laboratory for Optoelectronics, Huazhong University of Science and Technology, Wuhan 430074, China
[4]Key Laboratory of Material Chemistry for Energy Conversion and Storage (Ministry of Education), Hubei Key Laboratory of Material Chemistry and Service Failure, School of Chemistry and Chemical Engineering, Huazhong University of Science and Technology, Wuhan 430074, China
[5]Faculty of Science, Charles University, Prague 12843, Czech Republic

[a)]Author to whom all correspondence should be addressed: guoying_gao@mail.hust.edu.cn

**ABSTRACT:** The depletion of energy sources, worsening environmental issues, and the quantum limitations of integrated circuits for information storage in the post-Moore era, are pressing global concerns. Fortunately, two-dimensional (2D) Janus materials, possessing broken spatial symmetry, with emerging non-linear optical response, piezoelectricity, valley polarization, Rashba spin splitting and more, have established a substantial platform for exploring and applying modifiable physical, chemical, and biological properties in material science and offered a promising solution for these energy and information issues. To provide researchers with a comprehensive repository of 2D Janus family, this review systematically summarizes their theoretical predictions, experimental preparations, and modulation strategies. It also reviews the recent advances in tunable properties, applications, and inherent mechanisms in optics, catalysis, piezoelectricity, electrochemistry, thermoelectricity, magnetism, and electronics, with a focus on experimentally realized hexagonal and trigonal Janus structures. Additionally, their current research state is summarized, and potential opportunities and challenges that may arise are highlighted. Overall, this review aims to serve as a valuable resource for designing, fabricating, regulating, and applying 2D Janus systems, both theoretically and experimentally. This review will strongly promote the advanced academic investigations and industrial applications of 2D Janus materials in energy and information fields.

**TABLE OF CONTENTS**

## I. INTRODUCTION

Two-dimensional (2D) materials at the quantum-confinement limit hold significant advantages in energy and information transport and storage compared with three-dimensional (3D) materials.[1,2] Since the mechanical exfoliation of graphene from highly oriented pyrolytic graphite in 2004,[3] 2D materials have been extensively explored. These include 2D transition metal carbides, nitrides and carbonitrides (MXenes),[4] transition metal dichalcogenides (TMDs),[5-7] transition metal halides,[8] metal-organic frameworks (MOFs),[9] covalent organic frameworks (COFs)[10] and their derivatives,[11-14] etc. The term “Janus” stemming from the Roman mythology, is an ancient deity with auspicious metaphor. It possesses two faces, looking at the past and the future, which also symbolizes duality and diversity of objects.[15,16] Research into Janus began in 1989 with amphiphilic glass beads, they are hydrophilic on one side while hydrophobic on the other side, and are also known as Janus beads.[17] Inspired by such asymmetric structures and composite functions, Janus materials have entered into the research spotlight for their exhilarating physical, chemical and biological properties.[18-20] Janus was considered both as a specialized material type[21] and a strategy for structural engineering.[22] The 2D Janus family combines the merits from 2D materials and Janus structures, a wealth of properties may emerge due to its broken spatial symmetry, encompassing optics, catalysis, electrochemistry, thermoelectricity, mechanics, magnetism and valleytronics.[21,23]

As two core pillars of modern society, energy and information provided power support and intelligent decision-making, respectively. Together, they promoted technological development, economic growth, and profoundly influencing human lifestyles. However, amidst the backdrop of dwindling traditional energy sources and escalating environmental issues, energy concerns have become a global focal point. This has spurred the utilization of  diverse energy forms, such as renewable solar and thermal energy, as well as mechanical, hydrogen and chemical energy, while striving to minimize energy losses during transport and storage.[24,25] The search for materials and the design of devices with high energy utilization efficiency have become imperative. In addition, information encoding, transport and storage, essentially require high security, low consumption, strong non-volatility and large specific surface.[26] Moore's law[27] for integrated circuits is hindered by the quantum tunneling effect, promoting the exploration of inaugurating avenues to overcome this limitation. This raises a crucial question – how to address the urgent demands in energy and information fields?

To answer this, 2D Janus materials emerge as a promising solution. Since the experimental fabrications of Janus hexagonal- (H-) and distorted trigonal- (T'-) MoSSe, H-WSSe, trigonal- (T-) CrSeTe, T-VSeTe and T-PtSSe,[28-36] extensive experimental and theoretical studies have been devoted to their design, modulation and employment.[37-39] It's evident from the rising annual number of reports on 2D Janus materials [Fig. 1(a)], with data sourced from Web of Science. To understand the current research landscape, we performed the VOSviewer package[40] to organize all the relevant articles extracted from the Web of Science database using "Janus" and "2D" as keywords. Our co-occurrence network with overlay visualization [Fig. 1(b)] reveals research focuses on properties and applications, like optical (including optoelectronic and photocatalytic) performance, electronic, piezoelectric, and magnetic properties, adsorption, and energy utilization. The unique interactions of 2D Janus materials with matter facilitate microscopic particles transport, water treatment, and gas adsorption/separation.[23,41,42] Their distinctive Janus structure gives rise to an unusual electronic structure, optimizing light absorption coefficients and carrier mobility, for efficient energy conversion in photoelectric, thermoelectric and others.[42-44] By breaking mirror and inversion symmetries, 2D Janus systems generate a built-in electric field owing to different electronegativity, potentially enabling the Rashba effect, valley polarization and out-of-plane piezoelectricity.[21,45,46] While 2D Janus materials possess many attractive properties, they may not be sufficient for practical applications, requiring further modulation. This is confirmed by the co-occurrence network [Fig. 1(b)], with frequent mentions of modulation strategies such as strain, heterostructure and interface. Other approaches involving electric and magnetic fields, force field/strain, doping, and intercalation are also useful.[22,47-50] With their diverse and tunable properties, 2D Janus materials hold great promise for energy and information revolution (See Supporting Information in detail). Building a comprehensive resource library for them is of substantial significance for researchers.

To date, reviews exclusively focused on 2D Janus systems remain scarce. In 2018 and 2025, Li et al. and Ahmad et al. reported on the progress of 2D Janus TMDs, respectively.[51,52] In 2020, Zhang et al. reviewed Janus graphene and other materials' prediction and preparation,[23] while Yagmurcukardes et al. summarized their quantum properties.[19] However, these reviews[19,23,51] did not encompass the numerous and primary reports [Figs. 1(a) and 1(b)] published after their respective online dates. In 2023, Jiang et al. focused on predicted magnetic Janus materials founded on density functional theory (DFT) calculations, yet no experimentally prepared or non-magnetic ones were included.[21] Given these gaps, a comprehensive review is urgently needed to cover the latest developments in 2D Janus family, involving modulation, adequate quantitative results and underlying mechanisms, and to reveal current opportunities and challenges.

In this review, we systematically summarize the comprehensive and latest progress and perspectives of 2D Janus materials. Section 2 addresses theoretical prediction, experimental preparation, and manipulation. Sections 3 and 4 present tunable properties and applications, in energy and information, respectively, with a detailed review of inherent mechanisms and recent advances. Section 5 summarizes the current research status of 2D Janus H- and T-systems and envisions future opportunities and challenges. We hope this review provides an exhaustive knowledge resource and a comprehensive perspective on the theoretical and experimental design, regulation, and application, thereby promoting their advanced academic exploration and industrial deployment in energy conversion and storage as well as information encoding and memory.

## II. PREDICTION, FABRICATION AND MANIPULATION OF 2D JANUS MATERIALS

We classified the properties of 2D Janus materials based on their applications [Fig. 2(a)]. The first category is energy utilization, which includes optics and catalysis, thermoelectricity, piezoelectricity, and electrochemistry. The second category is information memory, which involves magnetic anisotropy, magnetic state and critical temperature, Dzyaloshinskii-Moriya interaction (DMI), spin polarization, valley splitting, and Rashba, Dresselhaus and Zeeman splitting. Explorations on 2D Janus family began with those experimentally available ones like MoSSe, WSSe, CrSeTe, VSeTe and PtSSe[28,30,32-36] (H- or T-$MX_2$-type derivatives). Thus, we focus on these phases in theoretical and experimental reports with manipulations. In the diagrammatic keyboard of the 2D Janus H- and T-MXY family [Fig. 2(b)], M stands for metals (primarily transition metals, some lanthanide elements, and carbon-group metals), while X and Y are oxygen-group (O, S, Se and Te) and halogen (F, Cl, Br, and I) elements. The left M part is paired with the right X and Y parts. Orange, green, and grey colors represent 2D Janus materials reported by both experiment and theory, only theory, and neither, respectively.

To facilitate relevant and profound design and modulation of 2D Janus family, this section provides a systematic summary (Fig. 3) of theoretical prediction methods, experimental preparation techniques, and prevalent manipulations in both theory and experiment.

### A. Theoretical Prediction

Theorists have predicted many materials including 2D Janus ones using various approaches,[22,23] with the hope of analyzing mechanisms and guiding experiments. The methods of theoretical calculations can be streamlined and categorized into several types: quantum mechanical calculations,[53] classical mechanical simulations,[54] statistical methods,[55] numerical analysis,[56] and machine learning (ML).[57]

(1) Quantum mechanical calculations explore physical and chemical properties at the atomic and molecular levels using Hartree-Fock (HF)[58] and DFT[59] built on quantum mechanics. These involve first-principles calculations[60] and quantum chemical calculations.[61] (2) Classical mechanical simulations model the dynamic behavior of macroscopic matter, such as molecular dynamics (MD) simulations.[62] (3) Statistical methods investigate the statistical properties of large particle systems, including Monte Carlo (MC) simulations[63] and statistical physics methods.[64] (4) Numerical analysis is applied to solve continuous problems in engineering and physics, like finite element analysis[65] and computational fluid dynamics (CFD).[66] (5) ML leverages algorithms and statistical models to process data, make predictions, and optimize problems, emerging as a powerful tool for large-scale data processing.[67]

Numerous computational packages and programs have been widely recognized to support these theoretical approaches. (1) Gaussian is a popular program for quantum chemical calculations, including electronic structure and reaction kinetics.[68] Vienna Ab initio Simulation Package (VASP) is widely implemented for quantum mechanics and ab initio molecular dynamics (AIMD) simulations, utilizing plane wave basis sets and pseudopotentials.[69] Other computational programs such as Quantum Wise Atomic Tool Kit (Quantum ATK),[70] Quantum ESPRESSO (QE),[71] Materials Studio (MS),[72] CP2K,[73] Siesta,[74] and WIEN2k,[75] are also extensively used. (2) In the field of molecular dynamics simulation, Large-scale Atomic/Molecular Massively Parallel Simulator (LAMMPS) offers atomistic modeling;[76] GROningen MAchine for Chemistry Simulations (GROMACS) integrates high-performance MD simulation and analysis for biochemical molecules such as proteins, lipids, and nucleic acids;[77] Assisted Model Building with Energy Refinement (AMBER) is designed for biomolecules MD simulations;[78] Simulation Package tOward Next GEneration molecular modeling (SPONGE), stands out with Graphics Processing Unit (GPU) acceleration and efficient enhanced sampling methods, integrated with MindSpore's deep learning framework.[79] (3) Python,[80] Julia,[81] Fortran[82] and C/C++[83,84] are commonly used programming languages, while MATLAB,[85] Wolfram

Mathematica[86] and Maple[87] bestow powerful numerical calculation and processing functions. (4) ML enables computer systems to learn and improve from data and algorithms, is employed into image recognition, financial market prediction, weather analysis, and healthcare.[88] Scikit-learn,[89] Keras[90] and PyTorch[91] are widely used ML libraries.

The rapid development of artificial intelligence (AI) technology has recently furnished new assistance. Notably and amazingly, John J. Hopfield and Geoffrey E. Hinton were awarded with the 2024 Nobel Prize in Physics for their foundational discoveries and inventions in ML with artificial neural networks, and the 2024 Nobel Prize in Chemistry recognized David Baker, Demis Hassabis and John M. Jumper for computational protein design and prediction in AI.[92] The black-box AI is performed to high-throughput screening for molecules in materials science.[93] Language models like Chat Generative Pre-trained Transformer (ChatGPT)[94] of OpenAI organization, and emerging DeepSeek[95] are influential AI assistants for reading, writing, and research, though they still require human oversight to ensure accuracy and appropriateness. The AI and high-throughput computing are concretely accelerating the development of Janus materials. (1) Accelerated discovery. They can rapidly screen vast chemical spaces to identify stable materials with target properties. (2) High-throughput screening. They can automate lots of calculations to build databases and identify promising candidates. (3) Advanced modeling. They can create accurate and transferable models that bypass the high cost of direct quantum mechanics calculations. (4) Property Prediction. They can use material composition and structure to directly predict key performance metrics. Integrating AI with theoretical and experimental approaches enhances large-scale efficiency[96] and promises transformative evolution.

Utilizing the aforementioned methods of theoretical calculations and computational tools, various properties of 2D Janus materials can be investigated. Below are some specific theories and principles applied in the investigation of designated properties. One perspective is energy utilization. (1) Optical properties can be bestowed by predicting the dielectric function, electronic structure, and light-matter interaction. Exploiting Langmuir adsorption theory,[97] transition-state theory[98] and Marcus electron-transfer theory,[99] combined with the Brønsted-Evans-Polanyi relation,[100] provides deep insights into catalytic mechanisms, helping design and optimize more efficient catalysts. (2) Piezoelectric properties can be predicted using DFT, density functional perturbation theory (DFPT)[101] and modern theory of polarization.[102] (3) For thermoelectric properties, through DFT combined with Boltzmann transport equation[103] and deformation potential theory,[104] the nature of atomic bonding and phonon vibrations can be analyzed. These approaches also help study phonon and electron transport, as well as thermoelectric conversion efficiency. Importantly, for thermoelectric, optical, and catalytic applications, the electronic structure especially the band gap, is valuable. While the Perdew-Burke-Ernzerhof (PBE) of generalized gradient approximation (GGA)[105] and local density approximation (LDA)[106] functionals usually underestimate band gaps, the DFT+U[107] and Heyd-Scuseria-Ernzerhof (HSE)[108] methods yield more accurate values. (4) For batteries and capacitors, electrochemical models such as the Newman, Tiedemann, Gu and Kim (NTGK) model,[109] the equivalent circuit model (ECM)[110] and Newman's Pseudo-Two-Dimensional Model (P2D) model[111] help capture ion migration within cells. The Gouy-Chapman-Stern (GCS) layer,[112] Tafel dynamics[113] and Marcus electron-transfer[99] theories are used to explore the solid-liquid interface. Coupling the Nernst-Planck equation with Poisson equation[114] allows the investigation of substances diffusive behavior under electric fields. DFT calculations can be employed for structural stability, reaction free energy, electronic structure, ion diffusion and adsorption kinetic simulations.

The other significant perspective is compilation, transport, and storage for information. (1) Magnetic anisotropy originates from spin-orbital coupling (SOC) and anisotropic dipole-dipole (D-D) interaction,[115,116] which is vital for the stable existence of 2D long-range magnetic ordering. DFT calculations can directly explore magnetocrystalline anisotropy (MCA) energy with SOC, while numerical methods extract the magnetic shape anisotropy (MSA) energy from D-D interactions.[117] (2) The magnetic ground state is primarily determined by the strengths of exchange interaction and DMI. In calculations, trivial magnetic states can be artificially selected, but limited cell size restricts the number of magnetic states considered. The magnetic critical temperature can be simulated using MC methods with exchange interaction and magnetic anisotropy energies,[118] or roughly evaluated by mean field theory (MFT),[119] though it often overestimates the value. (3) DMI derives from SOC, and is a key driver of non-trivial topological spin structures. Moriya rule offers a concise criterion for determining the DMI vector direction,[120] and Fert-Levy theory explains the DMI at ferromagnetic-heavy metal interfaces.[121] Reciprocal-spin spiral calculations fix the SOC operator in the direction of spin-helix rotational axis, allowing the first-order perturbation of the generalized Bloch theory to be handled by self-consistent calculations.[122] Real-space spin spiral method in supercell is utilized to calculate the DMI.[123] (4) Spin splitting is reflected by spin-polarized DFT calculations of electronic structures like band structure and density of states (DOS). High spin polarization is useful for low-depletion information transport, and combining DFT calculations with non-equilibrium Green's function[124] can simulate magnetoresistance and spin-resolved currents. (5) Valley polarization arises from broken spatial inversion symmetry, a condition satisfied by Janus systems. DFT calculations with Wannier functions in tight-binding (TB) models are used to analyze Berry curvature and anomalous valley Hall effect (AVHE).[125] (6) Rashba, Zeeman and Dresselhaus spin splitting manifests as a spin and momentum coupling in the Hamiltonian and can be employed in information transport and encoding.[126,127] Subsequent sections will provide a detailed overview of theoretical investigations into diverse properties of 2D Janus materials.

The integration of theoretical and experimental approaches with AI tools is driving transformative evolution in the study of 2D Janus materials. By leveraging quantum mechanical calculations, classical mechanical simulations, statistical methods, numerical analysis, and ML, researchers can efficiently investigate the diverse properties of these materials for energy utilization and information-related applications. The advancements in computational tools and programming languages, coupled with the emergence of AI technologies such as ML libraries and language models, are paving the avenue for breakthroughs in materials science. These developments hold significant promise for enhancing efficiency, enabling the design of more effective catalysts, and optimizing materials for applications in batteries, capacitors, and electronic devices. Furthermore, the exploration of magnetic properties and spin-related phenomena could offer pathways for information storage and transport, potentially revolutionizing the fields of electronics, spintronics, and valleytronics.

**B. Experimental Fabrication**

Based on theoretical predictions of 2D Janus materials, various experimental preparations have been carried out. The experimental synthesis of 2D materials involves diverse techniques. Here, we provide a brief overview of some common method, involving mechanical exfoliation (ME), molecular beam epitaxy (MBE), chemical vapor deposition (CVD), liquid phase exfoliation (LPE) and pulsed laser deposition (PLD).[128-132]

(1) ME method uses adhesive tapes to overcome the van der Waals (vdW) force between neighboring films, stripping the film from the bulk.[133,134] While ME produces 2D materials with flat surfaces and few defects, its low efficiency and limited controllability make it unsuitable for large-scale industrial production.[128,133,134] (2) MBE involves spraying specific atoms or molecules as a beam stream onto the substrate surface in an ultra-vacuum environment.[129,135] This method allows precise control over the thickness, composition, and concentration of 2D materials. However, MBE is prone to defects, technically complex, and not conducive to massive production.[129,135,136] (3) CVD generates thin films via chemical reactions or decomposition of gaseous compounds or monomers on the substrate surface.[130] It enables controlled preparation of large-area films but has drawbacks such as high costs, toxic gases, and slow deposition speeds.[130,137] (4) LPE prepares 2D materials by stripping them in the liquid medium using methods like sonication, ion-insertion, and selective-etching. While widely applicable, LPE suffers from low yield and uneven thickness.[131,138-140] (5) PLD employs high power pulsed lasers to evaporate a target surface and form plasma, which is then transported to the substrate to generate a film.[132] This method offers directed, high rate, and controllable deposition, but the film quality is often unsatisfactory, and large-area preparation is challenging.[141,142]

These fabrication approaches can be categorized into two groups. ME, LPE and PLD are classified as up-down strategies, typically preferring to relatively simple 2D materials with few elements. MBE and CVD belong to bottom-up strategies, which are better suitable for relatively complex and multi-element 2D materials.[143] Each method has its own advantages and limitations. The choice of method depends on the specific requirements of the material and its application.

The experimentally synthesized 2D Janus monolayers mainly include Janus H- and T'-MoSSe, H-WSSe, T-CrSeTe, T-VSeTe and T-PtSSe,[28-36] which were obtained by varying fabrication methods. As presented in Fig. 4(a), the earliest breakthrough began in 2017, when Lu et al. proposed three principal steps for the preparation of Janus H-MoSSe monolayer: using CVD to prepare $MoS_2$ on the sapphire, then adopting remote hydrogen plasma to replace S on one side with H to obtain MoSH, maintaining vacuum and finally replacing H in MoSH with Se utilizing thermal selenization to yield Janus H-MoSSe [Fig. 5(a)].[28] Additionally, synthesis of Janus H-MoSSe with optimized surface quality and electrical back contacts can be performed on highly oriented pyrolytic graphite (HOPG).[28] Zhang et al. attained Janus H-MoSSe monolayer through well-controlled sulfurization using gaseous sulfur molecules for $MoSe_2$ on $SiO_2$/Si substrate at atmosphere pressure [Figs. 4(b) and 4(c)], characterized as presented in Figs. 5(b)-5(g).[29] In 2023, Shi et al.[30] derived Janus T'- and H-MoSSe by room-temperature (RT) atomic-layer substitution method similar to the strategies about highly reactive hydrogen radicals and selenium vapor.[28] Janus H-WSSe [Fig. 5(h)] was also fabricated utilizing similar atomic substitution methodology[28,29] by Tongay's group in 2020.[31] Trivedi et al. performed RT selective epitaxy atomic replacement (SEAR) process driven by kinetics rather than heat to obtain Janus H-WSSe and MoSSe and their vertical heterostructures (VHSs) and lateral heterostructures (LHSs) [Figs. 4(d)-4(f) and 5(i)-5(n)].[32] Lin et al. exploited controllable implantation of hyperthermal species from PLD plasma to realize the selective replacement of sulfur with selenium to obtain Janus H-WSSe at low temperatures (300 °C).[33]

Sant et al. obtained Janus T-PtSSe on Pt (111) substrate by sulfurization of $PtSe_2$ under $H_2S$ atmosphere [Fig. 4(g)].[34] Very recently, Nie et al. fabricated Janus T-CrSeTe by MBE selective selenization of $CrTe_2$ on graphene-covered SiC (0001) substrate with critical temperature of ~350 K.[35] Xu et al. proposed to selenize the surface of $VTe_2$ monolayer on the bilayer graphene/SiC (0001) substrate and replace an atomic layer of Te with Se to obtain Janus T-VSeTe at ~523 K.[36] It is noteworthy that CrSSe was prepared by chemical deintercalation of iodine-oxidized NaCrSSe and applied as a high-rate cathode material in lithium-ion batteries (101.7 mAh $g^{-1}$ at 200 C), but the authors are not sure whether or not the synthesized CrSSe is a Janus material as they claimed that it's difficult to differentiate S and Se atoms by scanning transmission electron microscopy (STEM) images.[144] The uncertainty regarding the Janus structure of the synthesized CrSSe deserves more scrutiny from both experimentalists and theorists.

The experimental fabrication of 2D Janus systems has witnessed remarkable progress, with various methods such as ME, MBE, CVD, LPE, and PLD being employed. Each technique comes with its own advantages and limitations, and the selection of methods hinges on the specific demands of the material and its intended application. Experimental breakthroughs have successfully led to the synthesis of several 2D Janus monolayers, multilayers, and heterostructures. These fabrication strategies and fabrication reports inform the direction of experimental preparation of 2D Janus materials. The primary experimental difficulties can be categorized as follows. (1) Thermodynamic instability. The different chemical compositions on the two surfaces create a surface energy imbalance, driving the structure towards spontaneous symmetrization at elevated temperatures. (2) Demanding synthesis Precision. Achieving atomically precise, selective deposition on each side of a monolayer is extremely difficult. Conventional methods often cause atomic mixing, and managing precursor compatibility and process control remains a major hurdle. (3) Characterization difficulties. Conventional techniques struggle to provide non-destructive, precise verification of the 3D atomic structure and out-of-plane chemical asymmetry. There are promising avenues for overcoming these challenges. (1) Advanced synthesis techniques. Developing methods like plasma-assisted MBE for finer atomic-level control. (2) Alternative structures. Designing more stable structures (e.g., 3R-stacked bilayers) that mimic key Janus properties. (3) Improved characterization. Utilizing advanced techniques like high-resolution STEM with energy-dispersive X-ray spectroscopy (EDS) mapping to unambiguously confirm the asymmetric structure. The experimental advancements not only validate theoretical predictions but also provide valuable insights for future research and potential applications in fields such as energy storage and catalysis. As researchers continue to explore and refine these fabrication techniques, there is significant promise for the large-scale production and practical implementation of 2D Janus materials across various industries.

### C. Manipulation Trail

The 2D Janus family has garnered extensively attention from both theoretical and experimental perspectives. However, the properties of pristine 2D Janus materials may not suffice for practical applications. Fortunately, various manipulation methods can enhance their properties and guide subsequent preparations and applications in both theory and experiment.[22,145] These modulation strategies can be broadly categorized into internal and external regulations.

Internal regulation focuses on the composition and structure of the material itself, typically involving the material synthesis and changes in internal microstructure. This includes elemental doping (substitution and embedding), alloying, defects, structural phase changes, grain size modification, interface engineering, and stacking.[146-150] Internal modulation directly and permanently alters the chemical composition or structure to control properties. However, the modulation process during preparation process can be complex, with precise control being challenging. It's also poorly reversible, costly, and difficult to scale up for massive production.

External regulation involves changing the external environment to affect material properties. Common external modulation include magnetic field, electric field, force field/strain, temperature, humidity, light, chemical environment, electrostatic doping, and ion intercalation.[151-156] External modulation is more flexible, non-invasive and reversible, but less stable, has a limited scope and consumes more energy.

Janus can be recognized as an internal strategy, and it can also be regulated by other modulation trails as a material type.[21,22] Concrete instances of manipulation and underlying mechanisms of action on 2D Janus materials will be depicted in later sections. The internal and external regulation methods are not mutually exclusive. Each has its own advantages and disadvantages, and the choice depends on the specific property requirements of the material and its application. Combining the two regulation types can achieve better performance and supplies more possibilities for modulating various properties in 2D Janus materials.

The exploration of 2D Janus materials has made significant strides, with a variety of property-enhancement methods emerging to maximize their utility in practical applications. This unified strategy promises to unlock the full spectrum of their multifunctional potential, catalyzing meaningful scientific and technological advancement.

## III. ENERGY IN 2D JANUS MATERIALS

Approximately ninety percent of global electrical power relies on the incineration of fossil fuels. The finite nature of these resources constrains the burgeoning demands for energy. Regarding the universally concerned issue of energy and environmental protection, utilizing diverse forms of energy and reducing consumption are highly desired. Energy from optical, thermal, mechanical and chemical sources in 2D Janus materials is enticing,[28,30,157,158] and catalysts can significantly reduce energy consumption.[159] Additionally, sensors convert these energy forms into electrical signals and detect changes in non-electrical values. This benefits monitoring processes of energy conversion and pollutant emission or leakage.[160] In this section, we present properties and applications of optics, catalysis, thermoelectricity, piezoelectricity, and electrochemistry in 2D Janus family, stimulating multiple energy utilization.

### A. Optics and Catalysis

Optics, a discipline with a lengthy and illustrious history, has captivated human since ancient times. Early applications include using concave mirrors to focus sunlight for fire, imaging through small holes, and observing natural phenomena like mirages. Einstein proposed the concept of light quanta as energy packets and theoretically analyzed the photoelectric effect in 1905,[161] for which he was awarded the Nobel Prize in Physics in 1921. The interaction of light with matter is invaluable for both fundamental research and practical applications. One substantial application of optics is photocatalysis. Other catalytic processes, such as electrocatalysis, sonocatalysis, mechanocatalysis, rely on different energy sources and activation methods, and can be combined in various ways.[162-165] This part summarizes the research progress in optics and catalysis of 2D Janus materials. We review investigations of optics excluding photocatalysis, and then provide a summary of catalysis including photocatalysis.

#### 1. Optics

The interaction of light (a form of electromagnetic energy) with matter offers multidimensional value in scientific research, technological innovation, industrial applications, and daily life.[166] By employing the generation, modulation, propagation, absorption and conversion of light, we can design various devices and systems, such as emitters, detectors, super-resolution optical microscopy, optical sensors, logic and memory devices, and catalysts.[167-176] Traits like an appropriate band gap, which enables optical response across a broad spectrum from ultraviolet (UV) to radio frequencies, and strong light-matter interaction, are particularly favorable for optical applications.[145] 2D Janus materials, with their anisotropic nature and different chemical compositions and surfaces on their two sides, exhibit distinct absorption, scattering, and emission properties for light.[177,178] This asymmetry can enhance local electric and magnetic fields, thereby strengthening Raman scattering and surface plasmon resonance.[179-181] Additionally, photoexcited electron-hole pairs can separate to produce photocurrents.[182,183] By precisely controlling the design of 2D Janus materials, their optical response can be accurately tailored.[145,184,185]

The optical response of materials depends on their physical and chemical properties, as well as specific light-matter interactions. Research on the optical properties of 2D Janus materials, such as MoSSe and WSSe monolayers [Figs. 6(a)-6(j)], has explored various aspects including harmonic generation, emission, reflection, absorption, detection, and conversion. Here are some key findings. Attributed to augmented structural asymmetry and topological band-mixing, Janus T'-MoSSe exhibits giant RT non-linear optical properties.[30] It demonstrates over fifty (twenty) times higher than H-$MoS_2$ for eighteenth-order harmonic generation (terahertz emission).[30] Janus T'-MoSSe presents a topologically non-trivial semiconductor with molybdenum-*d* orbitals bearing band inversion with chalcogen-*p* orbitals [Figs. 6(g) and 6(h)].[30] Its Janus H-phase with a prismatic T-structure is topologically trivial, leading to a substantial increase in high harmonic generation (HHG) for 1T' phase, second-order THz and infrared (IR) second harmonic generation (SHG) emission frequencies.[30] Janus T'-MoSSe shows a band inversion between the conduction and valence bands accompanied by a non-zero Berry curvature. This nontrivial topological character enables electrons to accumulate an additional geometric phase in their momentum-space wave functions under intense optical fields, thereby markedly enhancing the nonlinear current response. This intensified nonlinear optical process ultimately leads to the dramatic enhancement of eighteenth-order harmonic generation and terahertz emission efficiency. The photoluminescence (PL) energy of Janus H-MoSSe is predominantly dictated by the $\mathrm{d}_{x^2-y^2}/d_{xy}$ orbitals of valence band maximum (VBM) at the K point, the optical SHG [Figs. 6(a)-6(c)] uncovers the out-of-plane optical diploe transition due to the unbalanced electronic wavefunction of S and Se atoms.[28] The complex optical dielectric functions of Janus H-MoSSe and WSSe monolayers with excitonic resonances and tunable SOC directly linked to their broken mirror symmetry was also experimentally established.[186]

Janus H-MoSSe and WSSe show second and third harmonic generation with exciton resonances.[187] They possess in-plane second harmonic non-linear susceptibility up to 200 pm $V^{-1}$, and the out-of-plane non-linear susceptibility of $\chi^{(2)}_{\mathrm{xxz}} = \chi^{(2)}_{\mathrm{zxx}}$ = 6 pm $V^{-1}$ and $\chi^{(2)}_{\mathrm{zzz}}$=1.5 pm $V^{-1}$ for MoSSe.[187] Both materials also exhibit blue shifts in PL under high pressures, with competing effects from ab-plane (blue shift) and c-axis (red shift) compression influencing the optical response [Figs. 6(d)-6(f)].[31] Both T- and H-MoSeTe exhibit high solar energy utilization efficiencies with absorption coefficients of $1 \times 10^6$ $cm^{-1}$, average light absorbance of 2 % in the visible region, and corresponding average transmittance of approximately 80 %.[188] Janus H-MoSTe and WSTe exhibit negative differential resistance (NDR) effects, enabling low threshold voltages (< 1.5 V) and peak-to-valley ratios (PVRs) of 161 and 247, respectively.[189] This suggests their promising applications as photovoltaic devices.

The above pristine 2D Janus MoSSe and WSSe, and their derivative homologues, exhibit favorable optical responses, which can be further modulated by regulatory instruments to tune their electronic structures. Biaxial strain can change Janus H-MoSSe monolayer from a direct semiconductor to an indirect one and even a metal.[48] Based on quantum confinement and interlayer interactions, Janus H-MoSSe bilayers with different stacks are all indirect semiconductors, and the indirect band gap of MoSSe decreases with increasing layers.[48] Janus H-WSSe monolayer exhibits an out-of-plane second-order non-linear photocurrent response, which can be modulated by biaxial strain or an external electric field.[190] Experimentally prepared nanoscrolls of

Janus H-MoSSe and WSSe display strong interlayer interactions and anisotropic optical responses.[191]

The Janus H-MoSSe/multiwalled carbon nanotubes (MWCNTs) heterostructure exhibits charge transfer from C-2*p* to Mo-4*d* orbitals compared to Janus H-MoSSe monolayer.[184] These modulated electronic properties enhance conductivity, reduce work function (WF) and significantly increase carrier concentration.[184] This heterostructure enables the construction of an emitter that offers enhanced field electron emission (FEE), lower turn-on and threshold field values, and large emission current density.[184]

The light absorption of 2D Janus molybdenum and tungsten dichalcogenides can be enhanced through multiple approaches. External electric field and tensile strain can transform the band gap from indirect to direct.[44] In the visible, IR and UV regions, the absorption efficiencies of Janus H-MoSTe, WSeTe and WSTe monolayers can reach 80-90 %.[44] Mn-doped WSSe achieves an absorption coefficient of 64 % in the IR region.[192] Pristine, and Mn-, P- and As-doped Janus H-WSSe are suitable for photovoltaic applications in the visible and IR regions, with a maximum quantum efficiency of 0.62 % at 307.5 nm.[192] Additionally, interface-, stacking- and layer-dependent electronic structure can modulate optical absorption. In Janus H-MoSSe trilayers, excitons are spatially separated by the center layer.[193] The reduction of exciton binding energy prolongs their lifetimes, and the intrinsic electric field produces a large interlayer band offset, which may dissociate excitons into free carriers.[193] This suggests potential applications of vdW structural MoSSe in light harvesting, Bose-Einstein condensation, and superfluidity.[193]

Beyond homogeneous structures, heterogeneous structures have also been extensively explored. Janus H-MoSSe and WSSe display many-body effects (e-e and e-h interactions).[194] LHS composed of them shows optical activity in photo-response and absorption coefficients across the visible range.[194] Both LHS and VHS enable exciton separation and dominate the optical response, making them suitable for energy conversion.[194] The absorption coefficients in the visible region for VHS and LHS of Janus H-MoSTe/WSTe reach 6.91 × $10^5$ and 4.18 × $10^5$ $cm^{-1}$, respectively.[189] The Shockley-Queisser (SQ) efficiencies of WSTe monolayer and MoSTe/WSTe LHS are 33.88 % and 33.75 %, respectively.[189] Optomechanically tunable second-harmonic generation anisotropy in Janus H-MoSSe/$MoS_2$ heterostructures is achieved through wavelength-dependent optical resonance and optically induced strain.[195] Furthermore, the carrier mobility of Janus H-MoSSe/GaN and MoSSe/AlN heterostructures can reach up to 3651.83 $cm^2$ $V^{-1}$ $s^{-1}$ for holes along the zigzag direction.[196] The recombination of photogenerated electron-hole pairs is prevented by the built-in electric field.[196] The energy conversion efficiencies of BP/Janus H-MoSSe and BAs/Janus H-MoSSe heterostructures are calculated to be 22.97 % and 20.86 %, respectively.[197] These efficiencies benefit from the sharp band dispersion near the VBM in the BP and BAs, resulting in carrier mobility of approximately $10^3$ $cm^2$ $V^{-1}$ $s^{-1}$, which is advantageous for advanced excitonic solar cells.[197]

For regulation of detection in optics, constructing heterojunctions is a primary approach to modify electronic structures through charge redistribution. By leveraging the inhomogeneous distribution of Janus H-MoSSe sheets on graphene between the source and drain, the Janus/graphene photodetector presents efficient photovoltaic behavior.[198] It achieves a responsivity of 10.78 A $W^{-1}$ and a power density of 830 fW $\mu m^{-2}$ under zero source-drain voltage and 800 nm excitation.[198] It can also operate low power with a responsivity of 0.6 × $10^2$ A $W^{-1}$ at a small power of 17 fW $\mu m^{-2}$ and a low bias of 10 mV.[198] Abid et al. proposed that Janus H-MoSSe/black-phosphorene and Janus H-WSeTe/black-phosphorene heterojunctions can serve as IR detectors and optical absorbers, benefiting from low effective mass, high carrier mobility, and favorable absorption spectra.[199] Founded on Janus H-WSSe/$MoSe_2$ heterojunctions, Cui et al. developed armchair self-powered photodetectors, the photocurrent varies sensitively with the polarization angle and photon energy of linearly and circularly polarized light.[185] For light conversion, uniaxial tensile (compressive) strain decreases (increases) the band gap of Janus H-WSSe, causing red (blue) shift.[200] Tensile strain can effectively reduce exciton binding energy and enhance energy conversion efficiency.[200] Janus H-WSSe/$MoSe_2$ heterojunctions exhibit an electron mobility of 1131 $cm^2$ $V^{-1}$ $s^{-1}$ and a power conversion efficiency (PCE) of 17.0 % for pristine samples and an improved 22.7 % under 2 % strain.[185]

In addition to the optical response mentioned above, 2D Janus materials also have many cross-disciplinary applications. Janus molybdenum and tungsten dichalcogenides are picked as representatives. Janus H-MoSSe nanotubes present potential hysteresis-free steep-slope transistors and multi-valued logic devices.[201] Wang et al. also found that the interplay of flexoelectricity and deformation potential in these nanotubes results in a diameter-dependent behavior of band gap.[201] The decoupling effect caused by the flexoelectric field, rather than the quantum confinement effect, creates red or blue shifts in exciton transition energy.[201] Negative compressibility of electrons and holes without electron correlation effect realizes negative quantum capacitance.[201] Furthermore, Meng et al. proposed a flexible optoelectronic artificial retinal perception device.[202] This device utilizes $Al_2O_3/Li^+/Al_2O_3$ as the dielectric and Janus H-MoSSe as the channel.[202] It has functions such as pre-processing, light adaptation, and pattern recognition due to the photosensitivity of Janus H-MoSSe.[202] The device also exhibits stable flexibility, scalability, and energy efficiency, tendering an approach for optoelectronic integrated sensing-memory-processing components with artificial perception [Figs. 6(i) and 6(j)].[202] Additionally, introducing chalcogen defects in Janus H-WSSe produces a notable potential difference, enhancing sensitivity to gas molecules.[41] It features a NDR effect and high PVR, and shows optical response in the visible region, exhibiting optical gas sensing properties for $CH_4$, $C_3H_8$ and $C_4H_{10}$.[41] These cross-disciplinary applications demonstrate the broad potential of 2D Janus materials in various fields.

Other 2D Janus materials beyond the initial molybdenum and tungsten dichalcogenides[28,31] also show significant potential for light-matter interactions and related applications. Some details for optical properties and applications in 2D Janus materials are presented in Supporting Information.

In summary, several approaches can enhance the optical properties of 2D Janus family. These mainly include: (1) Strain engineering. Applied strain modifies the electronic structure, particularly the band gap, to adjust light absorption and PL. (2) Constructing heterostructure. Leveraging interactions between different materials influences WF and carrier concentration. (3) Elemental doping. Introducing impurity atoms alters the electronic structure. (4) Multilayer cascading. Increasing the number of material layers exploits interlayer interactions for modulating exciton dissociation. (5) External stimulation. Electric and other fields may change the band gap. These strategies enhance optical responses of 2D Janus systems, by regulating band structures, modifying exciton behavior, and improving light absorption and emission.

These findings underscore the diverse optical responses of 2D Janus materials, making them promising for advanced optoelectronic devices like emitters, detectors, and energy conversion devices. Their anisotropic nature and asymmetric chemical compositions further highlight their potential in cross-disciplinary applications, such as flexible optoelectronic devices and gas sensors. Ongoing experimental and theoretical research into light-matter interactions in these materials is driving the development of next-generation miniaturized optical and energy conversion technologies.

## 2. Catalysis

Catalysts play a crucial role in enhancing energy efficiency, accelerating reaction rates, and improving reaction selectivity, making them vital components in strategies for sustainable development and for addressing global challenges in energy, environment, and health.[203-205] Catalysis involves microscopic mechanisms such as chemical kinetics and surface science, primarily accomplished through processes like adsorption, activation, reaction, desorption, and diffusion.[206,207] On account of their spatial symmetry breaking, 2D Janus structures exhibit differing electronegativity, creating a built-in electric field that facilitates charge transfer and elevates electronic interactions. Their diverse interfaces and surfaces offer varied active sites, while multiple elements provide synergistic effects. These characteristics unlock the potential of 2D Janus materials to enhance catalytic efficiency.[145,208] This segment reviews photocatalysis, electrocatalysis, and others, as well as their multiple combinations, to supply a description of catalysis in 2D Janus materials.

In photocatalytic water-splitting, catalysts are mostly solid semiconductors with photosensitivity. Key requirements include sufficient light absorption, efficient separation and transfer of photogenerated carriers, and suitable overpotentials for the water aqueous redox reactions.[209] Janus H-MoSSe, with a band gap around 2 eV, serves as an efficient photocatalyst for water-splitting with a wide solar-spectrum.[42] Its VBM (CBM) is lower (higher) than that of the oxidation level of oxygen (reduction level of hydrogen) by 0.83 (0.74) eV, outperforming $MoS_2$ and favoring water redox reductions.[42] Owing to its intrinsic defects and strain synergy, Janus H-MoSSe exhibits high basal plane activity for the hydrogen evolution reaction (HER) as presented in Figs. 7(a) and 7(b).[29]

Janus H-TMD structures, with broken mirror symmetry and in-built electric fields, display altered *d*-orbital energies near the Fermi level compared with the pristine ones [Figs. 7(c) and 7(d)].[159] The asymmetry in Janus H-WSSe introduces mid-gap states and shifts the Fermi level, stimulating HER activity.[159] Its modulated catalytic performance with Se and S vacancies versus other HER catalysts are displayed in Figs. 7(e) and 7(f).[159] The charge density difference for $H_2O$ adsorption on Janus H-WSSe [Fig. 7(g)], with purple and orange regions indicating charge accumulation and depletion, respectively, and the isosurface value is $5 \times 10^{-4}$ e Å$^{-3}$.[200] Figs. 7(h) and 7(i) outline the photocatalytic pathways and free energy steps for HER and oxidation evolution reaction (OER) on Janus H-WSSe monolayer, with insets illustrating the proposed photocatalytic HER and OER pathways.[200] $H^*$, $OH^*$, $O^*$, and $OOH^*$ represent the most probable intermediates on the WSSe monolayer, and the gray and red balls are the H and O atoms, respectively.[200]

In electrocatalysis, high electrical conductivity, efficient and stable selective reactions, and suitable overpotential are essential. Catalytic properties can be regulated through artificial strategies, such as introducing other metal atoms to modify the electronic structure and provide active sites.[210] Pd- and Pt-anchored Janus H-MoSSe monolayers [Fig. 7(j)] possess excellent bifunctional electrocatalytic properties, with Pd-MoSSe displaying oxygen reduction reaction (ORR) and OER overpotentials of 0.43 and 0.50 V, respectively.[211] This performance is attributed to the large built-in electric field from electronegativity differences, enhancing the electrical conductivity and carrier separation.[211] Janus H-MoSMg is a topologically trivial obstructed atomic insulator, exhibits exceptional HER activity with near-zero Gibbs free energies at its edge sites, offering a design strategy for high-performance quantum electrocatalysts beyond traditional topological materials.[212] Janus H-WSSe with adsorbed Fe acts as a single-atom catalyst (SAC) for CO catalytic oxidation, with an oxidation reaction energy barrier of 0.47 eV.[213] Figs. 7(k) and 7(l) illustrate the corresponding CO oxidation process via the Eley-Rideal (ER) mechanism ($CO + O_2 \rightarrow OOCO \rightarrow O_{ads} + CO_2$ and $CO + O_{ads} \rightarrow CO_2$).[213]

In catalysis, heterojunction construction is broadly explored to realize specific charge transfer. This synergistic interface effect promotes the separation of photogenerated electrons and holes, reduces recombination, and provides more active sites.[214] Janus WSSe/$MoSe_2$ heterojunctions show high efficiency in electrocatalytic water-splitting, with a free energy change of just 0.066 eV when adsorbing two H atoms.[185] These heterojunctions enhance electron-hole pair separation through materials with different band gaps.[185] Differences in Fermi levels enable spontaneous carrier transfer at the interface via potential gradients.[185] Ju et al. proposed Janus bilayer tungsten chalcogenides (Janus H-WSSe/WSeTe heterojunctions), unveiling a solar-to-hydrogen (STH) conversion efficiency of 10.71 %.[215] Monolayer and multilayer Janus H-MoSSe are photocatalysts for solar water-splitting, with strain- and stack-tunable, and layer-dependent electronic and optical properties.[48] Considering the different contact interfaces in Janus H-MoSSe/GeC and Janus H-WSSe/GeC heterostructures, the Se-near-GeC structure favors water decomposition into $H^+/H_2$ and $O_2/H_2O$, while the S-near-GeC structure inhibits photocatalytic water-splitting.[216] Qian et al. suggested introducing C = C bonds on h-BN to regulate the band gap, achieving high carrier mobility, large redox over-potentials ($\chi_{H_2}$ = 2.56 eV and $\chi_{O_2}$ = 0.83 eV) in acidic environment, high quantum efficiency, and good solar hydrogen production efficiency (33.31 %). Both OER and HER are exothermic reactions in light conditions.[217]

The exploration of catalytic properties extends to experimentally prepared chromium, vanadium, and platinum dichalcogenides.[218-220] Janus H-CrSSe demonstrates a STH efficiency of 30.5 %, surpassing the 18 % theoretical limit for commercial semiconductors, driven by light absorption, high overpotential, and electric field effects.[221] Its band edge is near the redox potential, significantly impacting photocatalytic water-splitting.[222] Janus H-CrSSe/$Ti_2CO_2$ heterostructure, with redistributed charges, emerges as a promising photocatalyst with high light absorption from 300 to 1300 nm.[222] Janus H-CrSSe and CrSeTe heterostructure with Se-Te interface, leveraging the synergistic effects of intrinsic dipoles and interfacial electric fields, produces a moderate 0.49 eV vacuum level difference and enhances redox abilities with increased over-potentials (0.35 eV for $H_2$ and 0.73 eV for $O_2$), resulting in a 44 % STH efficiency.[218] Janus H-CrSSe/$MoS_2$ and Janus H-CrSSe/$WS_2$ heterostructures, exhibit proper band edges for water-splitting redox reactions under solar irradiation.[223]

Doping in Janus H-VSSe (As and Si atoms at S or Se sites and C and Ge atoms at Se site) can strengthen orbital interaction and improve bond strength, exhibiting high HER activity and yielding a near-zero hydrogen adsorption free energy such as -0.022 eV.[219] Janus T-PtSTe achieves 24.7% STH efficiency, with photo-triggered OER and HER in different acidities.[224] B-doping causes reduced band gap and orbital hybridization, broadening light absorption range, which enhances photocatalytic water-splitting in Janus T-PtSSe/g-$C_3N_4$ heterostructure.[225,226] Some other 2D Janus materials for catalysis are summarized in Supporting Information.

In conclusion, the primary methods to enhance the catalytic performance of 2D Janus systems are as follows: (1) Heterojunction construction promotes charge transfer, modulates band structure, and alters carrier separation and complexation. (2) Elemental doping optimizes electronic structure by introducing other atoms into 2D Janus materials, and provides more active sites, enhances orbital interactions, improves bond strength, enhances electrical conductivity, and facilitates carrier separation. (3) Strain engineering can modify the crystal lattice to adjust the band gap and carrier mobility. (4) Defect engineering may offer active sites and shifts the Fermi level.

The built-in electric fields and diverse active sites in 2D Janus materials facilitate charge transfer and elevate catalytic efficiency. These outstanding catalytic properties and potential applications of 2D Janus family are anticipated to significantly influence future catalytic technologies.

## B. Thermoelectricity

Most electrical power generation mechanisms have conversion efficiencies rarely exceeding forty percent, leading to considerable energy loss and unwanted thermal emissions. Consequently, there is an urgent requirement for avant-garde technologies to reclaim and utilize thermal energy. Thermoelectric properties involve the bidirectional conversion between heat and electricity. Specifically, a material can generate an electric potential difference when subjected to a temperature gradient, or produce a temperature gradient in an electric field. Thermoelectricity contributes to targets for achieving carbon-neutrality and green-clean energy goals. It finds employment in power generation, refrigeration, and precise temperature control across medicine, aerospace, and defense.[227,228] To quantitatively determine the thermoelectric performance, figure of merit (ZT) is concerned as:[229]

$$\mathrm{ZT} = \frac{\sigma S^2}{\kappa} T \tag{1}$$

in which $T$ stands for temperature in Kelvin, thermal conductivity $\kappa$ includes lattice and electrical components, which is $\kappa = \kappa_e + \kappa_l$ . $\sigma S^2$ represents power factor (PF) and $S$ is the Seebeck coefficient. $\sigma$ is electrical conductivity and can be described as $\sigma = \mathrm{ne}\mu$, where n, e and $\mu$ are carrier concentration, elemental charge and carrier mobility, respectively. Considering the Wiedemann-Franz law,[230] electronic thermal conductivity is obtained as $\kappa_e = L\sigma T$, where $L$ represents dependent Lorentz number. Therefore, ZT can be attained as:

$$\mathrm{ZT} = \frac{ne\mu S^2}{(Lne\mu T + \kappa_l)}T \tag{2}$$

Furthermore, the Seebeck coefficient $S$ can be interpreted by parabolic band and energy-independent scattering approximation in a degenerate semiconductor or metal as:[231]

$$S = \frac{8\pi^2 k_\mathrm{B}^2}{3eh^2} m^* T \left(\frac{\pi}{3n}\right)^{2/3} \tag{3}$$

in which $k_\mathrm{B}$ and $h$ stand for Boltzmann and Planck constants, respectively. m* represents the effective mass. Based on this form, many efforts have been devoted to enhancing PF by increasing carrier concentration and mobility, while reducing $\kappa_l$ to uncover inter-coupling and improve ZT. To achieve these goals, the search and design for structures with excellent thermoelectric transport properties has gained widespread attention.[232] Band engineering through doping, interface engineering, electric field, and strain can effectively modify thermoelectricity and enhance ZT in conventional thermoelectric systems.[233,234] It's noted that the role of four and higher orders of phonon scattering is paramount in some systems. High intrinsic anharmonicity exhibits significant scattering events involving four phonons.[235,236] These processes become dominant, especially at high temperatures, and can reduce the predicted lattice thermal conductivity compared to three-phonon scattering. Four-phonon scattering is particularly effective in scattering out-of-plane acoustic phonons. In 2D materials, these modes often carry a large portion of the heat, and their strong suppression by four-phonon processes is a primary mechanism for the dramatic reduction in lattice thermal conductivity. Four-phonon interactions can bypass the momentum-conservation selection rules that restrict three-phonon scattering for certain phonon modes, leading to additional resistive channels that were previously unaccounted for. The asymmetric structure of 2D Janus materials, with broken spatial symmetry, may lead to enhanced phonon scattering and reduced thermal conductivity $\kappa$. Additionally, adjusting the chemical composition compared with pristine one, can optimize their electron structure, realizing high conductivity and mobility.[237-239] These characteristics suggest that 2D Janus systems are promising candidates for thermoelectric applications.

The experimental and theoretical explorations of 2D Janus structures have confirmed their favorable thermoelectric transport properties. Figs. 8(a)-8(l) illustrate the thermoelectric performance in 2D Janus materials, and Table I lists representatives of ZT values under different conditions, which are monolayers unless otherwise specified. The detailed descriptions are as follows: Janus H-MoSSe possesses a short phonon lifetime, with group velocities between those of $MoS_2$ and $MoSe_2$ [Figs. 8(a) and 8(b)], which mainly leads to its thermal conductivity $\kappa$ being between those of $MoS_2$ and $MoSe_2$.[240] The RT thermal sheet conductance is 342.50 W $K^{-1}$, and isotope scattering causes a decrease in lattice thermal conductivity $\kappa_l$ of 5.8 %, reduced to half when the characteristic length is approximately 110 nm.[240] The infinitely long sample $\kappa_\infty$ and phonon mean free path $l_{MFP}$ of Janus H-MoSSe/WSSe and $MoS_2$/$WSe_2$ are compared in Fig. 8f, revealing their thermal transport properties.[241] The $\kappa_\infty$ of MoSSe (WSSe) is higher than $MoS_2$ (similar to $WSe_2$), while the $l_{MFP}$ is similar to $MoS_2$ (larger than $WSe_2$).[241] An increase in vacancy density enhances phonon scattering, thereby reducing their $\kappa$ values of thermal conductivity.[241] For MoSSe and WSSe, 2 % vacancies cause decreases of 16.03 % and 14.04 %, respectively.[241] An increase in temperature from 100 K to 300 K leads to a substantial decrease in thermal conductivity of MoSSe (by 28.4 %), while WSSe exhibits temperature-insensitive behavior (by 12.75 %), attributed to the weak temperature dependence of low-frequency phonons that significantly contribute to $\kappa$.[241]

Tensile strain can enhance scattering and reduce lattice thermal conductivity $\kappa_l$ to 9.90 from 25.37 W $m^{-1}$ $K^{-1}$ for Janus H-WSSe.[157] By modulating the flatness near the band edges to realize higher effective mass, 6 % tension (compression) increases the ZT values to 0.32 and 0.50 at 600 and 900 K in Janus H-WSSe, respectively.[157] However, the ZT value is still relatively low. Fortunately, after Te replacement, increases in atomic mass and strengthened electron-phonon interactions lead to improved ZT values. The band structures and DOS of Janus H-WSTe are presented in Figs. 8(c) and 8(d) to attain the electronic structure.[242] The lattice thermal component of Janus H-WSTe is lower than those of $WS_2$ and WSSe, resulting in high ZT values of approximately 2.25 and 0.74 at 600 and 300 K [Fig. 8(e)], respectively.[242]

A -3 % strain tailors the bands, resulting in an optimal ZT value of 1.62 for p-type doped Janus H-WSeTe/MoSSe heterojunction at 300 K.[243] Constructing heterostructures can suppress phonon propagation, reduce lattice thermal conductivity, and modulate electrical transport performance through thickness and interface engineering.[244] Heterostructures also provide enhanced structural stability and flexibility, valuable for flexible and wearable devices. The thermal vibrational properties of Janus tungsten chalcogenides change when interlayer hybrid phonon modes are introduced during stacking.[245] There is large longitudinal-transverse optical (LO-TO) splitting at the Γ point.[245] Janus H-WSeTe/WSTe, WSSe/WSeTe, and WSSe/WSTe heterojunctions all exhibit ultra-low lattice thermal components of 0.01, 0.02 and 0.004 W $m^{-1}$ $K^{-1}$ at 300 K [Fig. 8(i)], respectively.[245] Graphene/Janus H-MoSSe nanoribbons and symmetric armchair MoSSe nanoribbons display thermoelectric ZT values of 2.01 and 1.64 at 300 K [Figs. 8(g) and 8(h)], respectively.[43] VdW interactions restrict electron and phonon transport and thermal conductivity in 2D normal direction.[43] Lattice thermal conductivity and temperature are the pivotal determinants for ZT values of MoSSe nanoribbons and their heterojunctions.[43]

Ongoing exploration into thermoelectric applications has identified additional 2D Janus materials with potential (Table I). Janus T-platinum and palladium dichalcogenides are promising candidates.[246,247] Electron and hole doping enhances their ZT values through adjusting relaxation time and chemical potential.[246] Among them, thanks to relatively small thermal conductivity from low group velocity and converged phonon scattering, the ZT of Janus T-PtSeTe is 0.91 and 2.54 at 300 and 900 K, respectively.[247] Interestingly, Carrete et al. found a marked higher ZT in T-PtSTe than its parents $PtS_2$ and $PtTe_2$, attributed to a drastic decrease in thermal conductivity.[248] The reduction in thermal conductivity can be interpreted by the lower symmetry, and a relaxation of selection rules for more intense three-phonon scattering in Janus T-PtSTe.[248] Janus T-ZrSSe outperforms T-$ZrS_2$ due to its smaller group velocity, shorter phonon lifetime, and lower thermal conductivity.[249] It displays a ZT value of 4.88 at 900 K under 6 % strain.[250] Strain engineering can optimize electronic structure and phonon scattering, potentially boosting thermal and electrical conductivity, thus may enhance thermoelectric conversion efficiency.

Beyond 2D Janus molybdenum, tungsten, platinum, palladium, and zirconium dichalcogenides with thermoelectric performance, titanium, hafnium, nickel, stannum, and plumbum dichalcogenides and zirconium and hafnium dihalides are also being investigated (Table I). For instance, as presented in Figs. 8(j)-8(l), Janus T-SnSSe bilayer and 4 % strained T-PbSSe monolayer display ZTs of 2.55 and 3.77 at 900 K, respectively.[238,251] Others are detailed in Supporting Information.

To summarize, the key approaches to enhancing the thermoelectric properties of 2D Janus materials are: (1) Strain engineering modifies band flatness to improve effective mass, enhances phonon scattering, and reduces the lattice thermal conductivity. (2) Elemental doping adjusts carrier concentration and mobility to boost PF. (3) Heterostructure building inhibits phonon propagation, lowers thermal conductivity, and improves structural stability and flexibility. (4) Element substitution alters atomic mass and electron-phonon interactions.

2D Janus materials present great promise in thermoelectric applications. Their asymmetric structures enhance phonon scattering, reducing thermal conductivity, while their adjustable chemical compositions optimize electron structures for high conductivity and mobility. These findings highlight the significant thermoelectric potential of 2D Janus materials, paving the avenue for miniaturized thermoelectric devices. This is crucial for achieving carbon-neutrality and green energy goals, with applications in power generation, refrigeration, and temperature control across various industries.

## C. Piezoelectricity

The piezoelectric effect refers to the accumulation of electric charge in a material when it is subjected to mechanical stress. This phenomenon is reversible, like the thermoelectricity mentioned earlier, meaning the material undergoes mechanical deformation when an external voltage is applied. The underlying mechanism arises from the non-centrosymmetric atomic structure, where the positive and negative charge centers do not coincide. This imbalance creates an internal polarization, resulting in piezoelectric behavior.[252] Wang et al. proposed utilizing the piezoelectric effect to change interfacial barriers between metals and semiconductors and to regulate transport properties in p-n junctions.[253] They also invented self-powered piezoelectric nanogenerators with 17-30 % efficiency from mechanical energy.[253] Piezoelectricity can be applied in sensors, imaging technology, energy harvesting, and electronic musical instruments.[254,255] 2D Janus family, with its natural broken spatial symmetry, meets the pre-conditioning requirements for piezoelectricity and is deserving of investigation and application. Inversion and mirror symmetries breaking and modified elemental composition cause electronegativities of the two sides to differ, generating a built-in electric field that triggers a polarized structure and produces piezoelectricity.[256] Generally, the piezoelectric effect is stronger with a larger difference in electronegativities between elements at the two sides.

Piezoelectricity is an intrinsic electromechanical coupling between the strain or stress of a system lacking spatial inversion symmetry and the electric polarization or electric field,[257] which can be described by piezoelectric stress and strain tensors from ions and electrons, their formulas are expressed as:

$$e_{ijk} = \frac{\partial P_i}{\partial \varepsilon_{jk}} = \frac{\partial \sigma_{jk}}{\partial E_i} = e_{ijk}^{electron} + e_{ijk}^{ion} \tag{4}$$

$$d_{ijk} = \frac{\partial P_i}{\partial \sigma_{jk}} = \frac{\partial \varepsilon_{jk}}{\partial E_i} = d_{ijk}^{electron} + d_{ijk}^{ion} \tag{5}$$

in which $\varepsilon_{jk}$ and $\sigma_{jk}$ stand for strain and stress tensors, $P_i$ and $E_i$ are vectors of piezoelectric polarization and electric field, respectively. The subscripts *i*, *j*, and *k* can all take values of 1, 2, and 3, and represent the spatial orientations of three representative axes, which are armchair (*x*), zigzag (*y*) and vertical (*z*) directions, respectively. Employing Voigt notation, the third-rank tensors of $e_{ijk}$ and $d_{ijk}$ can be reduced to two-rank $e_{il}$ and $d_{il}$ based on crystal symmetry, respectively.[258] With symmetry of at least 3*m* point-group of 2D Janus H- and T-phases (*P*3*m*1),[259,260] the tensors $e_{il}$ and $d_{il}$ can be defined as:

$$e_{il} = \begin{pmatrix} e_{11} & -e_{11} & 0 & 0 & e_{15} & 0 \\ 0 & 0 & 0 & e_{15} & 0 & -0.5e_{11} \\ e_{31} & e_{31} & e_{33} & 0 & 0 & 0 \end{pmatrix} \tag{6}$$

$$d_{il} = \begin{pmatrix} d_{11} & -d_{11} & 0 & 0 & d_{15} & 0 \\ 0 & 0 & 0 & d_{15} & 0 & -2d_{11} \\ d_{31} & d_{31} & d_{33} & 0 & 0 & 0 \end{pmatrix} \tag{7}$$

where *l* is an integer from 1 to 6. The relation between $e_{il}$ and $d_{il}$ can be written as $e_{il} = d_{ik}C_{kl}$, in which $C_{kl}$ represents the elastic stiffness tensor and can be described as:

$$C_{kl} = \begin{pmatrix} C_{11} & C_{12} & C_{13} & C_{14} & 0 & 0 \\ C_{12} & C_{11} & C_{13} & -C_{14} & 0 & 0 \\ C_{13} & C_{13} & C_{33} & 0 & 0 & 0 \\ C_{14} & -C_{14} & 0 & C_{44} & 0 & 0 \\ 0 & 0 & 0 & 0 & C_{44} & C_{14} \\ 0 & 0 & 0 & 0 & C_{14} & 0.5(C_{11} - C_{12}) \end{pmatrix} \tag{8}$$

the vertical direction is strain and stress free for monolayers, thus $\varepsilon_3 = \varepsilon_4 = \varepsilon_5 = 0$ and $\sigma_3 = \sigma_4 = \sigma_5$ is presented with negligible $e_{15}$, $e_{33}$, $d_{15}$ and $d_{33}$, the tensors $e_{il}$ and $d_{il}$ can be abbreviated, herein $d_{11}$, $d_{22}$ and $d_{31}$ can be attained by:

$$d_{11} = \frac{e_{11}}{(C_{11} - C_{12})} \tag{9}$$

$$d_{22} = \frac{e_{22}}{(C_{11} - C_{12})} \tag{10}$$

$$d_{31} = \frac{e_{31}}{(C_{11} + C_{12})} \tag{11}$$

when the number of layers exceeds one, the stress and strain in the vertical direction are worth considering for elastic stiffness and anisotropy, and $e_{11}$, $e_{31}$, $e_{33}$ and $e_{15}$ are not neglected, the $d_{11}$ is written consistently while the corresponding $d_{31}$, $d_{33}$ and $d_{15}$ can be obtained by:

$$d_{31} = \frac{C_{33}e_{31} - C_{13}e_{33}}{(C_{11} + C_{12})C_{33} - 2C_{13}^2} \tag{12}$$

$$d_{33} = \frac{(C_{11} + C_{12})e_{33} - 2C_{13}e_{31}}{(C_{11} + C_{12})C_{33} - 2C_{13}^2} \tag{13}$$

$$d_{15} = \frac{e_{15}}{C_{44}} \tag{14}$$

but Janus T'-structure possesses space group of $Pmn2_1$ differing from those of Janus H-and T-phases,[261,262] $e_{14}$, $e_{21}$ and $e_{22}$ are distinct, and $d_{14}$, $d_{21}$ and $d_{22}$ can be given as:

$$d_{14} = \frac{e_{14}}{C_{44}} \tag{15}$$

$$d_{21} = \frac{C_{22}e_{21} - C_{12}e_{22}}{C_{11}C_{22} - 2C_{12}^2} \tag{16}$$

$$d_{22} = \frac{C_{11}e_{22} - C_{21}e_{21}}{C_{11}C_{22} - 2C_{12}^2} \tag{17}$$

$d_{il}$ and $e_{il}$ can be solved numerically to obtain quantitative in-plane and out-of-plane piezoelectricity.

The relationship between the direction of strain (stress) and that of electric polarization (electric field) is indeed critical in piezoelectric materials. The direction of polarization is mainly determined by crystal structure. When the polarization direction is parallel to the applied stress, it can be effectively utilized in piezoelectric sensors, piezoceramics, and motors.[263,264] These applications benefit from the direct piezoelectric effect, where mechanical stress generates electricity, or the converse piezoelectric effect, where an electric field induces mechanical deformation. When the polarization direction is perpendicular to the applied stress, it can be employed in acoustic wave devices,[265] where the piezoelectric material converts electrical signals into acoustic waves or vice versa. This property is essential for devices such as filters, resonators, and sensors that operate at high frequencies. The distinct directional dependencies of piezoelectric materials enable their versatile use in a wide range of technological applications.

To evaluate the piezoelectric conversion efficiency and explore the potential piezoelectric applications of 2D Janus materials, some representative instances are presented in Figs. 9(a)-9(m) and Table II, illustrating cases where the direction of strain (stress) is parallel or perpendicular to the direction of electric polarization (electric field). The earliest intrinsic vertical piezoelectric response in a single-molecular-layer crystal is demonstrated in Janus H-MoSSe.[28] Piezo-response force microscopy combined with resonance enhancement yields a qualitative result of approximately 0.1 pm $V^{-1}$, indicating that piezoelectricity is sensitive to electrical properties and spatial variations [Figs. 9(a)-9(c)].[28] Breaking spatial symmetry enables six species of 2D Janus molybdenum and tungsten chalcogenides to exhibit out-of-plane piezoelectric polarization in addition to conventional in-plane piezoelectric polarization.[259] Uniaxial strain enhances in-plane piezoelectricity while weakening out-of-plane one.[259] Taking H-MoSTe as an example [Figs. 9(d) and 9(e)], red and blue colors indicate electron accumulation and depletion, respectively, revealing the electronegativity difference and induced piezoelectricity.[259] The S-Te exchange in Janus H-MoSTe generates out-of-plane piezoelectricity with $d_{13}$ of 0.40 pm $V^{-1}$, and out-of-plane symmetry breaking results in a large piezoelectric coefficient $d_{14}$ of -17.80 pm $V^{-1}$ for 1T'-MoSTe.[262] Multilayer Janus H-MoSTe exhibits strong out-of-plane piezoelectric polarization regardless of stacking orders, with the out-of-plane piezoelectric coefficient $d_{33}$ reaching 5.7-13.5 pm $V^{-1}$ [Figs. 9(f) and 9(g)].[259]

In heterojunction construction, a built-in electric field induces charge transfer and redistribution at the asymmetric interface, altering the position of piezoelectric response. The out-of-plane piezoelectricity $d_{33}$ values in Janus H-MoSSe/BP and Janus H-MoSSe/BAs vdW heterostructures are 14.91 and 7.63 pm $V^{-1}$, respectively.[197] The piezoelectric coefficient $d_{33}$ of Janus H-MoSeTe/WSTe heterojunction reaches 13.91 pm $V^{-1}$.[266] Additionally, strain can be employed to alter the structural changes and dielectric constant, thereby controlling the piezoelectric effect.[261] As presented in Figs. 9(h) and 9(i), Janus H-CrSSe, CrSTe, and CrSeTe exhibit intrinsic piezoelectric responses with out-of-plane piezoelectric coefficients $d_{31}$ of 0.40, 0.83, and 0.44 pm $V^{-1}$, which can be increased to 0.61, 1.58, and 0.72 pm $V^{-1}$ under 6 %, 4 %, and 6 % tension, respectively.[261]

Janus H-VSSe possesses $e_{11}$ of 3.303 × $10^{-10}$ C $m^{-1}$ and $e_{13}$ of 0.948 × $10^{-10}$ C $m^{-1}$ with out-of-plane piezoelectric polarization [Fig. 9(j)].[46] For 3D Janus H-VSSe multilayers with different stacking orders, $e_{33}$ can reach 0.49 C $m^{-1}$ with $d_{33}$ of 4.92 pm $V^{-1}$.[267] Other studies have also investigated that the Janus H-VSeTe and VSTe monolayers exhibit in-plane $e_{11}$ values of 2.9 × $10^{-10}$ and 2.0 × $10^{-10}$ C $m^{-1}$ and out-of-plane $e_{31}$ values of 0.11 × $10^{-10}$ and 0.36 × $10^{-10}$ C $m^{-1}$, respectively.[268]

Platinum dichalcogenides, including dioxides in Janus T-phase, display both in-plane and out-of-plane piezoelectricity, proportional to charge and electronegativity differences, and can be modulated by compressive or tensile strain.[269] As shown in Figs. 9(k) and 9(l), Raman activity of $A_1^{\mathrm{O}}$ corresponds to changes in in-plane piezoelectricity. The Bader charge difference $\Delta Q_{X-Y} = |Q_X - Q_Y|$ and the electronegativity difference ratio $\Delta\chi_{\max} / \Delta\chi_{\min} = \frac{[|\chi(X)-\chi(M)|,|\chi(Y)-\chi(M)|]_{\max}}{[|\chi(X)-\chi(M)|,|\chi(Y)-\chi(M)|]_{\min}}$ between the adjacent layers are proportional to out-of-plane piezoelectricity [Fig. 9(m)], where M is the middle Pt atom.[269] Taking Janus T-PtOSe as an instance, 6 % tension strengthens $d_{11}$ to approximately 24 from 8.80 pm $V^{-1}$, while -4% compression enhances $d_{31}$ to 2.32 from 1.54 pm $V^{-1}$.[269] These findings highlight the potential of 2D Janus molybdenum, tungsten, chromium, vanadium and platinum dichalcogenides for piezoelectric applications.

The exploration of piezoelectricity in 2D Janus materials has identified several promising candidates that could expand the piezoelectric library. Some instances are presented in Supporting Information. These include zirconium, hafnium, germanium, and tin dichalcogenides including dioxides,[260,270,271] as well as vanadium, scandium, yttrium, titanium, nickel, zinc, and cerium dihalides.[272-278] Most of them possess relatively large in-plane piezoelectric coefficients but small out-of-plane values (Table II).

To briefly summarize, the primary strategies to enhance the piezoelectric properties of 2D Janus materials are as follows: (1) Strain engineering can effectively change the crystal symmetry and enhance the polarization effect. (2) Heterojunction modifies the crystal field and electron distribution through interface effects and interlayer interactions. (3) Elemental substitution adjusts electronegativity differences within the material. (4) Multilayer stacking modulates the vibrational modes of phonons, thereby influencing piezoelectric behavior. (5) External stimuli, such as electric fields, can tune the polarization state and charge distribution. These results illustrate the substantial promise of 2D Janus materials in piezoelectric applications. They are likely to play a crucial role in the development of next-generation piezoelectric devices, including sensor technology, energy harvesting systems, and flexible electronics.

Akin to piezoelectricity through shared requirements for non-centrosymmetric crystal structures, ferroelasticity is a reversible mechanical phenomenon,[279] determined by crystal structure. It involves a non-linear internal structure response,[280] enabling materials to revert to their original shape upon strain removal. While piezoelectric materials often exhibit ferroelasticity, the converse is not always true. For instance, Janus H-VSSe monolayer demonstrates reversible ferroelastic strain of up to 73 %, with a ferroelastic switching energy barrier of roughly 0.23 eV under uniaxial strain.[46] Modern technological applications of ferroelastic materials include non-volatile memory devices through domain variation, mechanical sensors and actuators, waveguides, and smart or flexible textiles.[280,281] This suggests that the same structural asymmetry that grants strong piezoelectricity also facilitates robust ferroelastic domain switching. The co-existence of these properties opens exciting possibilities for multi-functional devices, where a single material can simultaneously act as a mechanical sensor (via piezoelectricity), a non-volatile memory element (via ferroelastic domain variation), and a component in flexible textiles and actuators.

Therefore, the interplay between piezoelectric and ferroelastic properties in 2D Janus materials represents a promising frontier for developing next-generation intelligent and responsive electronic systems.

**D. Electrochemistry**

Electrochemistry, a multidisciplinary field encompassing chemistry, physics, and materials science, explores charge transfer processes and their interrelation with chemical reactions. It serves as a cornerstone of modern energy technology and sustainable development. Implementations of electrochemistry are extensive, including batteries, capacitors, electroplating, and sensors.[282-285] At its core lies redox reactions, which typically involve the exchange of ions and electrons at interfaces. A basic electrochemical system comprises anode and cathode materials along with electrolytes.

Electrochemical performance is characterized by parameters such as charge transfer rate, ionic diffusion coefficient, electrochemical window, energy and power density, cycling stability, and capacity.[286-289] These parameters can be assessed using techniques like cyclic voltammetry (CV), linear sweep voltammetry (LSV), electrochemical impedance spectroscopy (EIS), potentiostatic chronoamperometry, and galvanostatic chronoamperometry.[290-294]

Theoretically, aspects such as adsorption energy, charge transfer, open circuit voltage (OCV), storage capacity, ion diffusion, reaction kinetics, and stability can also be examined.[13,295,296] The charge/discharge processes of metal atom/ion pairs like $Li/Li^{+}$, $Na/Na^{+}$, $K/K^{+}$, $Zn/Zn^{2+}$, $Mg/Mg^{2+}$, and $Al/Al^{3+}$ for electrode materials can generally be described as: $L + nM^{i+} + ine^{-} \leftrightarrow LM_n$, in which L represents the pristine electrode material, M is the involved metal, n and i stand for the numbers of adsorbed ions and the charge lost by per metal atom, respectively. In accordance with the reaction, from a state of $LM_n$ to $LM_m$, the OCV can be defined by:

$$OCV = -\frac{E_{LM_n} - E_{LM_m} - (n-m)E_M}{i(n-m)e} \tag{18}$$

where the $E_{\mathrm{LM_n}}/E_{\mathrm{LM_m}}$ represents the energy of state $LM_n/LM_m$, and $E_M$ is the energy of one metal atom. It is closely related to the average value of adsorption energy, and the average value of OCV is usually focused on, disclosing the electric potential of the electrode with the designated metal. In addition, the storage capacity is another essential prerequisite for electrochemical performance, which is attained by:

$$C = \mathrm{ni}F/W \tag{19}$$

in which $F$ and $W$ stand for the Faraday constant of 26801 mAh $mol^{-1}$ and the total atomic weights of the substrate, respectively.

Designing efficient and ecology-friendly electrochemical systems is of paramount importance and paves avenues in electrochemical technology. The diverse and tunable surfaces of 2D Janus materials provide different active sites, favoring reaction kinetics. Their presence of various elements helps achieve wide electrochemical windows and integrates multiple functions, such as catalysis, thermoelectricity, piezoelectricity, and ferroelasticity. Their asymmetric structure allows the construction of self-driven nano-motors, where electrochemical reactions can be driven by ionic concentration gradients. Overall, 2D Janus family furnishes vast potential for the development of miniaturized electrochemical devices.

Sensing and energy storage are closely related, as both involve the adsorption, and migration of ions, atoms, or molecules. Thus, we put sensing and energy storage in one part. Seeing as optical, thermoelectric, and piezoelectric properties have been described in detail above, in this part, we focus on batteries, capacitors, and sensors, excluding those related to solar, thermoelectric, and piezoelectric applications.

Structural asymmetry generates internal dipole moments, Chaney et al. employed the ML method combined with first-principles calculations to explore the potential of six species of Janus H-molybdenum and tungsten chalcogenides as anode materials for lithium-ion batteries [Figs. 10(a)-10(d)].[158] Their chalcogen sides with higher electronegativity favor Li adsorption.[158] After Li adsorption, their conductivity improves from semiconductor to metal.[158] Multilayer Janus structures undergo less volumetric expansion/contraction during charge/discharge, enhancing storage capacity.[158] Janus H-MoSSe, MoSTe, WSSe, and WSTe can be utilized as cathode materials for sodium-sulfur batteries.[297] MoSTe binds strongly with $Na_2S_n$ than electrolytes, reducing shuttling, and improving conductivity and electrochemical processes.[297] Janus H-WSSe serves as a high-performance anode material, with theoretical capacities and small potential barriers of 477.8 mAh $g^{-1}$ and 0.18 eV, 371.5 mAh $g^{-1}$ and 0.04 eV, and 156.0 mAh $g^{-1}$ and 0.038 eV for lithium-, sodium-, and potassium-ion batteries, respectively.[295] Ion diffusion is faster on the Se layer than the S layer, providing higher charge/discharge rates [Figs. 10(e)-10(h)], and all of them exhibit OCVs of less than 1 V.[295] Wu et al. explored that Ni-doped Janus H-MoSSe electrodes in a non-aqueous rechargeable sodium-oxygen battery demonstrate high selectivity and Faradaic efficiency near 100 % for ORR, with ORR and OER over-potentials of 0.49 and 0.59 V, respectively.[208] The trifunctional electrocatalytic activity of Janus H-MoSSe for ORR, OER and sodium-oxygen batteries arises from the synergy of built-in electric fields and transition-metal doping.[208] Zhang et al. used Janus H-MoSSe/$Ti_3C_2T_x$ MXene heterostructures as anode materials for sodium-ion batteries, achieving an OCV of 1.54 V and a storage capacity of 593.3 mAh $g^{-1}$, exceeding $MoSe_2$/MXene anode electrodes.[298] It exhibits a low diffusion barrier of 0.45 eV, and the MXene component provides good conductivity.[298] This superior electrochemical performance of Janus H-MoSSe heterostructure is attributed to strong charge transfer, intensive interactions, and structural stability from broken symmetry.[298] A scalable Janus H-WSSe/$WS_2$ heterostructure-based surface-enhanced Raman spectroscopy (SERS) platform is presented for ultrasensitive, rapid, and reagent-free detection of electrochemical ammonia.[299]

The electrochemical performance of 2D Janus materials, beyond molybdenum and tungsten dichalcogenides, has also been explored for various battery applications. Janus T-CrSSe, CrSTe, and CrSeTe can serve as anode materials for lithium- and sodium-ion batteries,[300] with adsorption energies up to -2.15 eV, and diffusion barriers as low as 0.10-0.15 eV, enabling fast charge/discharge rates. They offer average OCVs and storage capacities of 1.00 V and 348.0 mAh $g^{-1}$, and 0.51 V and 260.0 mAh $g^{-1}$ for lithium, and sodium, respectively.[300] Janus H-VSSe can stably adsorb alkali metals with maximum adsorption energy around -2 eV,[301] achieving an OCV of 0.57 V and specific capacity around 331.0 mAh $g^{-1}$.[301] Janus T-stannum dichalcogenides, including dioxides, can be employed as anode materials in sodium-ion batteries.[302] Janus T-SnSSe has a diffusion barrier of 0.12 eV and sodium-ion storage capacity of 1380.0 mAh $g^{-1}$,[302] while its heterojunction with graphene can be utilized as an anode material for lithium-ion batteries with a capacity of 472.7 mAh $g^{-1}$.[303] These findings highlight the potential of 2D Janus systems in enhancing the energy efficiency of batteries and capacitors through their unique physicochemical properties.

In addition to batteries and capacitors in energy storage and conversion, sensing, detection, and removal of atoms and molecules are critical for environmental monitoring, medical diagnostics, industrial process control, and identification of explosive materials and drugs.[304-306] 2D Janus family shows potential in this area, with interfacial conductive signaling and different surfaces facilitating environmental adaptability. Babar et al. utilized Janus H-MoSSe for Hachimoji deoxyribonucleic acid (DNA) detection [Figs. 10(i)-10(k)], with its vertical diploe moment enhancing the adsorption of base Guanine, affecting electron flow and showing superior sensing performance. This is beneficial for medical applications like DNA sequencing.[307] Janus H-MoSSe and MoSTe can be used as conductive and resistive sensors for volatile organic compounds (VOCs), especially acetone, due to their changing band gaps upon adsorption.[308]

Beyond the sensing of these molecules above, attention has been devoted to the detection and adsorption of gases. Ir adsorbed Janus H-MoSSe can detect gases like NO and CO through chemical bonding compared with the physical adsorption of pristine one.[211] Its gas sensitivity, surface, and strain selectivity make it capable of forming ultra-high sensitivity sensors.[211] Atomic modifications introduce additional electrons and create more active sites on surfaces, enhancing gas-material interactions. Ag- and Au-doped Janus H-MoSeTe monolayers can detect gases related to lithium-ion thermal runaway.[309] Its corresponding recovery times for $C_2H_4$, $CH_4$ and CO are $8.125 \times 10^{-7}$, $1.297 \times 10^{-12}$ and $5.009 \times 10^{-7}$ s, respectively, and strain can enhance adsorption properties and facilitate charge transfer, favoring gas sensing.[309] Janus H-WSSe is highly sensitive to $H_2S$, $NH_3$, $NO_2$, and NO, with a fast recovery time (a microsecond order at RT).[310] The S/Se defect and substitution cause charge redistribution, further enhancing adsorption performance and sensitivity.[310]

The energy storage and sensing capabilities of 2D Janus materials have been demonstrated for various energy storage and sensing applications. For example, the CuO and $(CuO)_2$ doped Janus T-ZrSSe can detect dissolved CO, $CO_2$, and $C_2H_4$ gases in oil-immersed transformers [Figs. 10(l)-10(n)].[311] Some representatives are displayed in Supporting Information.

The primary avenues to enhance the electrochemical performance of 2D Janus materials mainly include: (1) Heterojunction construction enhances charge transfer, and the multilayer structure reduces volume change during charge/discharge cycles. (2) Elemental doping improves conductivity and catalytic activity. (3) Strain engineering optimizes ion diffusion and charge transfer. The mechanism of electrochemical performance enhancement primarily works through modulating charge transfer and interfacial interactions, as well as optimizing ion diffusion paths and charge transfer efficiency.

In summary, 2D Janus structures, with their structural asymmetry and diverse surfaces, offer advantages such as improved reaction kinetics and wide electrochemical windows. Due to their unique physicochemical properties, 2D Janus structures show notable potential in energy storage, sensing, and scavenging applications. They are paving the way for advanced sensing and environmental technologies.

## IV. INFORMATION IN 2D JANUS MATERIALS

Beyond addressing energy and environmental challenges, 2D Janus materials offer wide-ranging properties and applications in information technology. As Moore's law[27] faces limitations due to quantum tunneling and hardware size constraints, additional degrees of freedom, like spin and valley, are crucial for low-depletion transport and strong non-volatility memory.[312,313] Magnetic anisotropy is key for stable low-dimensional long-range magnetic order and valley polarization.[22,314] Spintronics and valleytronics provide emerging avenues for information transport,[315,316] while topologically protected spin structures facilitate resistance to interference.[317,318]

In this section, we demonstrate magnetic anisotropy, trivial/non-trivial magnetic states (exchange interactions and DMI), critical temperature, spin polarization, valley polarization, and Rashba/Dresselhaus/Zeeman splitting in 2D Janus family, highlighting prospects for information coding, transport, and storage.

## A. Magnetic Anisotropy

According to the Mermin-Wagner theory,[319] 2D magnetism cannot be sustained, as thermal perturbations disrupt long-range magnetic ordering at finite temperatures, based on the isotropic 2D Heisenberg model. Fortunately, magnetic anisotropy can induce a spin excitation gap, effectively resist thermal perturbations, enabling the stabilization of 2D long-range magnetic ordering. This concept is supported by the earliest 2D antiferromagnetic (AFM) $FePS_3$,[320,321] and ferromagnetic (FM) $CrI_3$ monolayer and $CrGeTe_3$ bilayer.[133,322] Magnetic anisotropy energy (MAE) normally includes two components: MSA energy caused by anisotropic D-D interactions, and MCA energies caused by SOC.[115] The energy of D-D interaction is attained as:[116,117]

$$E_{\mathrm{D-D}} = \frac{1}{2}\frac{\mu_0}{4\pi}\sum_{i\neq j}\frac{1}{\boldsymbol{r}_{ij}^3}\left[\boldsymbol{M}_i\boldsymbol{M}_j - \frac{3}{r_{ij}^2}\left(\boldsymbol{M}_i\boldsymbol{r}_{ij}\right)\left(\boldsymbol{M}_j\boldsymbol{r}_{ij}\right)\right] \tag{20}$$

in which $\boldsymbol{M}_i$ and $\boldsymbol{r}_{ij}$ stand for the magnetic moment and the vector connecting the $i$ and $j$ sites. In a collinear system with parallel magnetic moments, the form can be simplified as:

$$E_{\mathrm{D-D}}^{\parallel} = \frac{1}{2}\frac{\mu_0\boldsymbol{M}^2}{4\pi}\sum_{i\neq j}\frac{1}{\boldsymbol{r}_{ij}^3}\left[1-3\cos^2\theta_{ij}\right] \tag{21}$$

where the magnetic moments are parallel to the atomic plane. If those are perpendicular to the plane, the equation is obtained as:

$$E_{\mathrm{D-D}}^{\perp} = \frac{1}{2}\frac{\mu_0 \boldsymbol{M}^2}{4\pi}\sum_{i\neq j}\frac{1}{\boldsymbol{r}_{ij}^3} \tag{22}$$

accordingly, the MSA energy is attained as:[323]

$$E_{\mathrm{MSA}} = E_{\mathrm{D-D}}^{\parallel} - E_{\mathrm{D-D}}^{\perp} = \frac{3}{2}\frac{\mu_0 \boldsymbol{M}^2}{4\pi}\sum_{i\neq j}\frac{1}{\boldsymbol{r}_{ij}^3}\cos^2\theta_{ij} \tag{23}$$

where $\theta_{ij}$ is the angle between $\boldsymbol{M}$ and $\boldsymbol{r}_{ij}$. The MSA energy is ignored unless the SOC is weak or the magnetic atoms are not in the same plane, and the MAE is usually generalized to the MCA energy. Importantly, the elements with relatively large atomic masses usually possess strong SOC strength, of which intraorbital hybridizations dominate over MCA.

To compare these values without misunderstanding, the MAE herein is uniformly simplified as MAE = $E_{x/y}$ - $E_z$ unless otherwise specifically stated, where the $E_{x/y}$ and $E_z$ represent the energies of magnetic states with in-plane and out-of-plane easy axes, respectively. In accordance with the second-order perturbation theory,[324,325] the MAE is expressed as:

$$\mathrm{MAE} = \xi^2 \sum_{\mathrm{o,u}}\sum_{\alpha,\beta}\left(1-2\delta_{\alpha\beta}\right)\frac{\left|\left\langle \mathrm{o}^{\alpha}\right|L_x\left|\mathrm{u}^{\beta}\right\rangle\right|^2 - \left|\left\langle \mathrm{o}^{\alpha}\right|L_z\left|\mathrm{u}^{\beta}\right\rangle\right|^2}{E_{\mathrm{u}}^{\beta} - E_{\mathrm{o}}^{\alpha}} \tag{24}$$

where $\xi$ is the SOC constant, $\delta_{\alpha\beta}$ represents the Kronecker delta, which is 1 when $\alpha = \beta$ and 0 elsewhere. $E$ is the energy level. $\alpha$ and $\beta$ stand for the spin components, o and u are the occupied and unoccupied states, respectively.

For the hybridization of *d* and *p* orbitals, the effect of the unoccupied spin-up and spin-down states are neglected, respectively. Consequently, the contribution is sourced from two terms, which cancel each other. The differences of spin-orbital angular momentum matrix elements $\left|\left\langle o^{+}\right|L_x\left|u^{-}\right\rangle\right|^2 - \left|\left\langle o^{+}\right|L_z\left|u^{-}\right\rangle\right|^2$ and $\left|\left\langle o^{-}\right|L_x\left|u^{-}\right\rangle\right|^2 - \left|\left\langle o^{-}\right|L_z\left|u^{-}\right\rangle\right|^2$ of *d* orbitals ( $\left|\left\langle o^{+}\right|L_x\left|u^{+}\right\rangle\right|^2 - \left|\left\langle o^{+}\right|L_z\left|u^{+}\right\rangle\right|^2$ and $\left|\left\langle o^{-}\right|L_x\left|u^{+}\right\rangle\right|^2 - \left|\left\langle o^{-}\right|L_z\left|u^{+}\right\rangle\right|^2$ of *p* orbitals) are equal in absolute value, a negative sign can be extracted for simplicity and the values

are listed in Tables III and IV. The difference of energy levels $E_{u}^{\beta} - E_{o}^{\alpha}$ can be obtained from the orbital-resolved DOS. Give the form, the underlying mechanism of the changes in MAE under regulation can be probed through intraorbital hybridization.

Furthermore, perpendicular magnetic anisotropy (PMA) is critical for stabilizing valley polarization and magnetic skyrmions,[326-328] and it offers advantages such as high density, and strong storage stability with reduced energy consumption compared to in-plane magnetic anisotropy (IMA).[329-332] While a significant challenge for PMA-based devices has been the relatively high current required for magnetization switching, this also motivates intense research into more efficient switching mechanisms (e.g., leveraging spin-orbit torques). However, most current 2D systems either have in-plane easy axes or minimal PMA,[22,333] hindering pragmatic applications of 2D magnets. There is an urgent need to design structures or modulate existing systems to attain sufficiently strong PMA. Utilizing Janus engineering/structure, the orbital interaction, electronic structure, and SOC strength can be regulated on account of the asymmetric structure and varying compositions. This approach shows great potential for enhancing magnetic anisotropy.

Some representative MAEs of 2D Janus materials are listed in Table V to visualize the elaborate values. The earliest 2D Janus molybdenum and tungsten dichalcogenides are non-magnetic,[28,32] and studies have not concentrated the modulation of magnetic anisotropy in them, so they are not discussed here. Fu' group found that Janus T-CrSeTe on SiC (0001) substrate exhibits IMA, with the magnetic moment tilting 47° off the *z*-axis after a charge density wave (CDW) phase transition.[35] Theoretically, originating from the dedication of the Te-$p_y$ and $p_z$ orbitals, the MAE of Janus T-CrSeTe monolayer is -0.176 meV f.u.$^{-1}$, increasing to 1.110 and 0.438 meV f.u.$^{-1}$ at 6 % strain and 0.1 h doping, respectively.[334] The marked difference between experiment and theory may stem from the interface and strained lattice effects from experimental substrate, and the choice of computational parameters. Electric field effect on Janus T-CrSeTe is minimal, with MAE decreasing by merely 0.040 meV f.u.$^{-1}$ at electric field of 0.2 V Å$^{-1}$.[334] Janus T-CrSTe bilayer maintains a robust in-plane easy axis under different stackings and strains [Figs. 11(a) and 11(b)].[50] However, Janus T-CrSSe provides a PMA of ~1.400 meV f.u.$^{-1}$, robust within ±4 % strain owing to the absence of Te-orbital contributions.[335]

The magnetic anisotropy of Janus T-VSeTe monolayer on graphene/SiC (0001) substrate remains unmeasured experimentally.[36] Theoretically, Janus H-VSSe monolayer shows IMA with a MAE of -0.576 mJ m$^{-2}$ (~-0.330 meV f.u.$^{-1}$), stable from -8 % to 8 % strain. A 3 % tension induces an out-of-plane easy axis in Janus H-VSeTe, with a MAE of 2.97 erg cm$^{-2}$ (~1.950 meV f.u.$^{-1}$) [Fig. 11(c)].[336] 2D Janus materials with metal elements in same group behave similarly, Janus T-NbSeTe displays IMA with MAE value of -1.140 meV f.u.$^{-1}$.[337] The MAE primarily stems from the SOC of heavy-atom orbitals like V(Nb)-*d*, Se-*p* and Te-*p*.[338,339]

Heterostructures enable interfacial orbital reconfiguration and symmetry breaking, which in turn modify electron occupation states and MAE. Different stackings of Janus T-VSTe (2 × 2 × 1 supercell) and Janus $Cr_2I_3Br_3$ heterostructure results in Ising and XY ferromagnetism, with MAEs of 0.040 (0.013 f.u.$^{-1}$) and -0.490 (-0.133 f.u.$^{-1}$) meV, respectively.[340] Twist angle may alter the chemical potential, leading to modified SOC for MCA, and changing magnetic D-D interactions for MSA. The IMA of graphene/Janus H-VSeTe heterostructure varies with twist angle, the largest (smallest) MAE is -2.143 (-1.562) mJ m$^{-2}$ at twist angle of 0° (7.6°).[341] Electric fields modulate carrier concentration, and electronic states near the Fermi level, tuning magnetic anisotropy. The IMA of graphene/Janus H-VSeTe heterostructure linearly varies with electric field, realizing -2.198 (-1.541) mJ m$^{-2}$ at electric field of -0.3 (0.3) V Å$^{-1}$.[342] Interlayer distance of graphene/Janus H-VSeTe heterostructure can affect MAE, decreasing to -1.381 mJ m$^{-2}$ when reduced by 0.6 Å.[342] Strain can directly change the lattice and interatomic distances, modulating electronic energy levels, and manipulating magnetic anisotropy. But biaxial strain from -4 % to 4 % does not alter the IMA of graphene/Janus H-VSeTe heterostructure.[342] Additionally, compression strain and electron doping can switch the easy axes in Janus T-MnSeTe and MnSTe from out-of-plane to in-plane, with Te-*p* orbitals dominating MAE changes [Figs. 11(d)-11(k)].[343]

In addition to metal dichalcogenides, the MAEs of 2D Janus dihalides were also extensively explored (Table V). Janus T-FeBrI transitions from IMA to PMA under compression or tension strain exceeding 2 %, while the MAE of FeClI is significantly enhanced by compressive strain.[344] These changes are mainly attributed to the modified intraorbital hybridization of iodine atoms. The angular dependence of total and atom-resolved MAEs for Janus T-FeBrI are presented in Figs. 11(l)-11(o) as an instance.[344] The MAE of Janus T-FeBrI/$In_2S_3$ multiferroic heterostructures increases by 167 % through ferroelectric (FE) polarization compared to the monolayer, but all different stacking orders retain IMA.[345] The easy axis of Janus H-FeClF monolayer shifts from out-of-plane to in-plane as the Hubbard U value increases, and the MAE is 0.132 meV f.u.$^{-1}$ without considering Fe-3*d* electronic correlation.[346] Under the reasonable electronic correlation strength of U = 2.5 eV, whose aligns with HSE06 method results, the IMA of Janus H-FeClF monolayer (-0.760 meV f.u.$^{-1}$) can be altered to PMA (0.062 meV f.u.$^{-1}$) under -8 % strain, and reaches 0.172 meV f.u.$^{-1}$. under -10 % strain.[347] The hybridizations of $d_{x^2-y^2}$ and $d_{xy}$ orbitals, $d_{yz}$ and $d_{z^2}$ orbitals of Fe plays a primary role in Janus ferrum dihalides.[347] For bilayer systems, the stacking effect significantly influences magnetic anisotropy. Janus H-FeClF bilayers possess PMA (IMA) with interlayer Cl-F (F-F and Cl-Cl) couplings.[348] This difference can be interpreted by the reduced PMA contribution from intraorbital hybridization of $d_{xy}$ and $d_{x^2-y^2}$, $d_{xy}$ and $d_{xz}$ orbitals, as its stacking order turns from interlayer Cl-F to F-F and Cl-Cl.[348] Electric field can induce a PMA-IMA-PMA transition in Janus H-FeClF bilayer, attributed to sensitive changes in intraorbital hybridization of positive (negative) contributions from $d_{xy}$ and $d_{x^2-y^2}$ ($d_{yz}$ and $d_{xz}$), $d_{xy}$ and $d_{xz}$ ($d_{yz}$ and $d_{z^2}$).[348] Some other representatives are presented in Supporting Information.

The main methods to improve the magnetic anisotropy of 2D Janus materials are: (1) Strain engineering modifies the lattice structure, thereby regulating orbital hybridization and SOC. (2) Elemental doping possibly enhances SOC and orbital hybridization by introducing additional interactions. (3) Heterostructure building causes interfacial orbital reconfiguration and symmetry breaking, can tune the electron occupancy state and magnetic interactions. (4) Applied electric field modulates carrier concentration and electronic states. (5) Torsion changes the chemical potential and SOC. In conclusion, the extensively tunable magnetic anisotropy in 2D Janus materials offers vast potential for miniaturized and high-density magnetic storage, magneto-optical devices, and quantum bit designs.[22,349]

### B. Magnetic State

Among topologically trivial magnetic systems, major exchange interactions include direct exchange,[350] super-exchange,[351] double exchange,[352] and interlayer exchange interactions like super-super exchange[353,354] and multi-intermediate double exchange.[355,356] The topological magnetism and non-collinear DMI will be discussed in later part. Direct exchange arises from short-range interactions due to the direct overlap of metal-electron wavefunctions.[350] Super-exchange is an indirect interaction mediated by non-magnetic ions.[351] According to the Goodenough-Kanamori-Anderson (GKA) rule,[357] the AFM state is favored when the bonding angle between the orbitals of two magnetic and non-magnetic atoms is near 180°, while near 90°, the FM state is preferred. Double exchange occurs between magnetic ions of different valence states.[352] Interlayer super-super exchange and multi-intermediate double exchange involve longer hoping distances with multiple non-magnetic ions as mediators, based on super-exchange and double exchange, respectively.[353-356]

Common magnetic arrangements include para-magnetism, ferromagnetism, antiferromagnetism, ferrimagnetism, and emerging altermagnetism,[358-362] which have broad applications. Fig. 12 illustrates the conventional intralayer FM, zigzag- and stripy-AFM, interlayer FM and A-type AFM, altermagnetic, and ferrimagnetic configurations. (1) Para-magnets can achieve ultra-low temperatures via adiabatic demagnetization, and are employed in noisy microwave quantum amplifiers and electron paramagnetic resonance imaging.[359] (2) Ferromagnets, with high permeability, high saturation magnetic induction, and low coercivity, are widely implemented in spintronics.[362] (3) Antiferromagnets, resistant to external perturbations with zero stray fields, offer stable data storage for spintronic devices.[358] (4) Ferrimagnets induce small eddy current in variable magnetic fields, useful in isolators, circulators, and other microwave devices.[363] (5) Altermagnets combine both properties of the macroscopic antiferromagnet's compensated anti-parallel magnetic orders, and the microscopic ferromagnet's non-relativistic spin splitting, resisting external magnetic field interference and featuring high resonance frequency, promising a revolution in spin-correlated quantum information technology.[361,364]

For practical spintronic applications, achieving a high magnetic critical temperature, preferably above RT, is essential. The magnetic phase transition temperature can be measured through various experimental techniques, including magnetization, resistance, Raman scattering, and magneto-optical effect. However, these measurements can be sensitive to factors such as sample impurity composition, substrate type, and environmental air conditions.

Theoretically, the magnetic critical temperature can be simulated using methods like MC simulations and MFT. These approaches help in understanding and predicting the magnetic behavior of 2D Janus materials, providing valuable insights for their potential applications in spintronics. The spin Hamiltonian of the Heisenberg model can be written as:

$$H = -\sum_{i,j} J_{ij} \boldsymbol{S}_i \boldsymbol{S}_j - \sum_i A_i (S_i^z)^2 \quad (25)$$

where $i$ and $j$ label magnetic atoms, $J_{ij}$ and $A_i$ stand for the exchange and anisotropy parameters, respectively. $\boldsymbol{S}_{i/j}$ and $S_i^z$ are the spin vector and spin component parallel to the $z$ direction, respectively. The spin vector can be normalized as $S_N = 1$ to simplify the parameterization without affecting the energy determination in MC simulations. The magnetic critical temperature can be simulated through the crystal structure, energies of different magnetic states, and magnetic anisotropy. MC approach generally presents agreement with experimental measurements, but MFT strategy[365] usually overestimate the magnetic phase transition temperature, which can be calculated by $T = -2\Delta E/(3Nk_B)$, where $\Delta E$ is the energy difference between different magnetic configurations, $N$ is the number of magnetic atoms, and $k_B$ is Boltzmann constant.

Most established 2D magnetic materials exhibit low magnetic phase transition temperatures. For example, the earliest 2D AFM $FePS_3$ monolayer exhibits Néel temperature ($T_N$) of ~118/104 K,[320,321] while the earliest 2D $CrI_3$ monolayer and $CrGeTe_3$ bilayer possess Curie temperatures ($T_C$s) of only 45 and 30 K, respectively.[133,322] Although a few 2D magnetic materials present RT ferromagnetism, such as $Fe_3GaTe_2$ few-layers,[366] $MnSe_2$ monolayer,[367] and $VSe_2$ monolayer,[368] most 2D magnetic materials have transition temperatures substantially below RT,[22,362] posing challenges for practical implementation. Therefore, it is highly demanded to design materials or modulate the existing systems for RT equipment. The 2D Janus family, with its broken spatial symmetry and diverse components, offers promise for enhancing magnetic phase transition temperatures by shifting atomic orbital energy levels.[21] In this part, we review the trivial magnetic states and critical temperatures of 2D Janus materials to provide feasible design and regulation strategies for RT-compatible devices.

In the realm of 2D Janus materials incorporating magnetic constituents, the bonding angles of magnetic cations mediated by non-magnetic anions are predominantly close to 90°, leading to a preference for FM super-exchange interactions and generally manifesting a FM ground state. This phenomenon is similar in 2D Janus H and T phases, which shall not be subjected to redundant analysis unless exceptional circumstances herein. However, their $T_C$s are often below RT, necessitating strategies to enhance $T_C$ by changing FM and AFM exchange interactions through modulatory techniques. The magnetic critical temperatures of 2D

Janus materials are summarized in Fig. 13, with most values representing $T_C$s unless otherwise specified.

Key representative findings are presented as follows. Spin-polarized scanning tunneling microscopy (SP-STM) has experimentally verified that the Janus T-CrSeTe monolayer on graphene/SiC (0001) substrate has a zigzag-AFM as the ground state.[35] But the detached monolayer of Janus T-CrSeTe is simulated with $T_C$ of 167 K and 188 K with or without considering the zigzag-AFM state, respectively.[334] Its $T_C$ is assessed to 306 K at combined 4 % strain and 0.1 e doping [Figs. 14(a) and 14(b)].[334] Strain modifies interatomic distances and changes bonding strengths, further manipulating exchange interactions, while the electric field has a minimal effect on thin monolayers like Janus T-CrSeTe.[334] The magnetic states of 2D bilayers and multilayers may be obviously modified by electric field.[369-371] Janus T-CrSTe exhibits a $T_C$ of 295 K without strain, increasing to 410 K with 5 % strain.[372] For its bilayers, different stackings result in intralayer FM and interlayer AFM configurations, with tensile strain transforming intralayer ferromagnetism to antiferromagnetism and achieving a $T_C$ of 310 K at 2 % tensile strain.[50] The 5 % tensile strain and 0.3 h doping can raise the $T_C$ of Janus T-CrSSe to 496 and 370 K from 272 K of the pristine one, respectively, while compression strain exceeding -4 % induces an AFM state.[335] Unlike their Janus T-phase counterparts, H-phase chromium dichalcogenides are non-magnetic. Janus chromium dichalcogenides of H-phase hold strong covalency, resulting in the absence of spin splitting or magnetic moments. However, strain-induced bond length changes can induce charge transfer, and different bonds respond differently to strain, so the magnetism can be uncovered.[373,374] For instance, the non-magnetic ground state of Janus H-CrSTe can turn to FM with tensile strain over 2 %, with a $T_C$ of approximately 275 K under 6 % strain.[374] Hydrogen doping at the top of the Se atoms in Janus H-CrSSe can also induce ferromagnetism with a $T_C$ (~554 K).[373] Although Janus T-VSeTe was fabricated, its magnetic properties were not experimentally investigated.[36] Structural transitions from T- to H-phases can redistribute the charges and amend the direct exchange and super-exchange interactions, enhancing ferromagnetism in Janus chromium dichalcogenides.

The $T_C$ of Janus H-VSeTe is estimated to 350 K,[339] increasing to 360 K under 4 % strain.[336] However, excessive tensile strain can reduce magnetic moments, the $T_C$ is decreased to 263 K under 10 % strain, while compressive strain enhances AFM coupling, with $T_N$ of 135 K under -13 % strain [Figs. 14(c) and 14(d)].[336] Due to the IMA in Janus H-VSeTe, VSTe, and VSSe,[375] their magnetic transitions are described by the classical XY model, which exhibits quasi-long-range order at low temperatures.[376] The critical temperatures of the Berezinskii-Kosterlitz-Thouless (BKT) transition ($T_{BKT}$) are 82, 46 and 106 K for Janus H-VSeTe, VSTe and VSSe, respectively.[375] Combining 5% strain and 0.1 h doping, the $T_C$s are augmented to 345 K for MnSeTe and 290 K for MnSTe [Fig. 14(e)].[343]

Numerous 2D Janus dihalides, including those containing ferrum, ruthenium, osmium, cobalt, scandium, lanthanum, cerium, and gadolinium groups,[278,346,377-382] have been simulated to study their magnetic states and critical temperatures. Janus T-FeBrI/$In_2S_3$ multiferroic heterostructure realizes a $T_C$ of 41 K owing to interfacial charge redistribution modulated by FE polarization.[345] Janus H-FeClBr monolayer possesses a high $T_C$ of 651 K,[383] while that of Janus H-FeClF monolayer is 311 and 63 K with Hubbard U of 1.5 and 2.5 eV, respectively.[346] Under reasonable electronic correlation, Zhang et al. reported the $T_C$ of Janus H-FeClF monolayer is 56 K, which can be considerably enhanced to 286 K under collective -10 % strain and 0.02 e doping.[347] Notably, these huge differences in $T_C$s for 2D pristine Janus ferrum dihalides are attributed to the different strengths of electronic correlation for Fe-3*d* orbitals considered in these studies. The electronic correlation significantly affects their exchange interactions, dominating the AFM and FM states' stabilities and further causing the substantial difference in $T_C$s.[346,347] As displayed in Figs. 14(f)-14(h), in addition to changes of magnetic moments, the negative-integrated crystal orbital Hamilton population (-ICOHP) values of Janus T-FeClBr, FeClI and FeBrI,[344] and H-FeClF monolayers[347] suggest the FM super-exchange interactions mediated by halogen atoms are relatively enhanced under compression compared with the weak AFM direct Fe-Fe interactions, resulting in more stable ferromagnetism and higher $T_C$. Furthermore, Janus H-FeClF bilayer with different stacking orders all present robust interlayer AFM and intralayer FM couplings against electric field.[348] Ultrafast laser pulses induce a transient net magnetization in a 2D fully compensated ferrimagnetic Janus T-NiICl bilayer within 50 femtoseconds, due to asymmetric interlayer spin transfer caused by its Janus structure.[384] As presented in Figs. 14(i)-14(k), the electrostatic potential and electron localization function (ELF) of Janus H-CeClBr uncover the electronegativity difference and bond character, and the magnetic moment and capacity varying with temperature are displayed through MC methods to verify its $T_C$ of 540 K.[278] For simplicity, other examples for trivial magnetic states and critical temperatures in Fig. 13 are discussed in Supporting Information.

The strategies to enhance the magnetic stability of 2D Janus materials are as follows: (1) Strain engineering can modulate the lattice and exchange interactions, potentially inducing a magnetic phase transition. (2) Elemental doping can modulate the electronic structure; foreign atoms may enhance the magnetic moment. (3) Constructing heterojunctions can generate interactions at interfaces and modulate magnetic coupling. (4) Electric field modulates carrier concentrations and electronic states near the Fermi level. (5) Torsion can affect magnetic interactions through modifying chemical potentials and interface effects. In summary, these results broaden potential horizons in 2D Janus materials for regulating magnetism and spin transport, offering opportunities for spintronic applications.

### C. Dzyaloshinskii-Moriya Interaction

In addition to the exchange interactions and collinear magnetism mentioned above, the DMI is also a significant magnetic mechanism. Non-trivial and non-collinear magnetism furnishes broad prospects. Spin topological structures, formed spontaneously by atomic magnetic moments, are stable without external disturbances.

Skyrmions, stabilized by the interplay of DMIs, exchange interactions, and magnetic anisotropy, are local and stable spin structures formed by non-collinear magnetic moments.[317] They exist in systems with broken center-reversal symmetry and strong spin-orbit coupling.[385] Néel (or hedgehog) skyrmions feature magnetic moments primarily rotating within the plane perpendicular to the skyrmion center, while Bloch(or spiral) skyrmions have magnetic moments mainly rotating around the skyrmion center.[318] Phase transitions between them can be induced through external approaches like magnetic field and temperature.[317,318] Other spin topological structures include merons, vortices, and anti-skyrmions, which can also be manipulated by current, temperature, and magnetic and electric fields.[386-388]

High-resolution magnetic imaging techniques such as Lorentz transmission electron microscopy (Lorentz TEM) can observe these structures.[389] Theoretical investigations can determine the strength of DMIs and exchange interactions, and further simulate the visual patterns of spin topological structures.[317,326] Topological magnetism shows potential in quantum computing, non-volatile information storage, skyrmion Hall effect, and logic devices.[317,390]

To describe the spin topological structures, the spin Hamiltonian is presented as:

$$H = -\sum_{i,j} \boldsymbol{D}_{ij}(\boldsymbol{S}_i \times \boldsymbol{S}_j) - \sum_{i,j} J_{ij}\boldsymbol{S}_i\boldsymbol{S}_j - \lambda\sum_i (S_i^z S_j^z) - K\sum_i (S_i^z)^2 - \mu B\sum_i S_i^z \qquad (26)$$

where $\boldsymbol{D}_{ij}$, $J_{ij}$, $\lambda$ and $K$ are the interatomic DMI, Heisenberg isotropic exchange, anisotropic symmetric exchange and single ion anisotropy, respectively. The magnetic moment and external magnetic field are represented by $\mu$ and $B$, respectively. The $\boldsymbol{D}_{ij}$ can be attained as $\boldsymbol{D}_{ij} = \mathrm{d}_{//}(\hat{\boldsymbol{u}}_{ij} \times \hat{\boldsymbol{z}}) + \mathrm{d}_{ij,z}\hat{\boldsymbol{z}}$, where $\mathrm{d}_{ij,z} \approx \mathrm{d}_{//} / \tan\tilde{\theta}_{ij}$, and $\tilde{\theta}_{ij}$ is the bonding angle, $\hat{\boldsymbol{u}}_{ij}$ is the unit vector pointing from *i* to *j*. The approach via chirality-dependent total energy difference, can be utilized to obtain in-plane component $\mathrm{d}_{//}$ along with the associated SOC energy $\Delta E_{SOC}$.[123] For skyrmion formation, the DMI magnitude is typically about 10 % of the exchange interaction, and can reach 20 % and 30 % based on analytical and ab initio calculations, respectively.[391] Thus, the DM/exchange $|d_{//}/J|$ ratio is usually considered to be from 0.1 to 0.2 (or 0.3) for skyrmion existence.

The topological charge $Q$, quantitatively reflecting the trivial and non-trivial chiral states, and the skyrmion stability, is expressed as $Q = (1/4\pi)\int \boldsymbol{S}(\partial_x \boldsymbol{S} \times \partial_y \boldsymbol{S})dxdy$, in which $\boldsymbol{S}$ stands for the normalized magnetization vector, and $x$ and $y$ are the coordinates. The Fert-Levy mechanism explains the DMI generation through indirect interactions between magnetic and heavy atoms at interfaces, involving the coupling of electron clouds in atomic layers.[121] Lacking of spatial inversion symmetry, 2D Janus materials satisfy one of the prerequisites for stable spin topological structures, promise non-collinear magnetism if they have the strong SOC.[326,372] This makes them highly attractive for topological magnetism research.

As presented in Figs. 15(a)-15(k), some 2D Janus materials, including experimental Janus chromium and vanadium dichalcogenides, were predicted to possess prominent DMI.[342,372] Janus T-CrSeTe and CrSTe exhibit DMI/exchange coupling ratios $|d_{//}/J|$ of approximately 0.14 and 0.02, respectively.[372] Consequently, Janus T-CrSTe lacks chiral spin textures, while CrSeTe features wormlike domains separated by chiral Néel domain walls (DWs) due to in-plane magnetic anisotropy.[372] External magnetic field can induce skyrmion states, and tensile strain enhances the PMA, reducing the required magnetic field and the density and diameters of skyrmions [Fig. 15(a)].[372] The spin-spiral configurations of left- and right-hands are illustrated in Fig. 15(b).[326] Moreover, the atom-resolved localization of the associated SOC energy are displayed in Fig. 15(c), and Janus T-VSeTe possesses a DMI parameter of 1.25 meV, with a strong IMA contributed by Te atoms, preventing skyrmion formation and resulting only in magnetic domains [Fig. 15(e)].[326] The $d_{//}$ values of Janus H-VSeTe and VSSe are -0.33 and 0.49 meV, the 8 % and -8 % strains bring them to the maximums of 1.42 and 0.89 meV, respectively, indicating the emergence of spin-chirality configurations, including local DWs.[338]

Heterojunctions enhance spatial asymmetry and provide interfacial interatomic interactions, which can strengthen the DMI. The h-BN/Janus H-VSeTe heterostructure exhibits a $d_{//}$ of 0.54 meV.[49] With twist angles, there is a significant effect on symmetry breaking, electron orbital hybridization and interface quality, modulating the DMI. Graphene/Janus H-VSeTe heterostructures exhibit $d_{//}$ values of -0.58 and -0.40 meV at twist angles of 0/19.1° and 8.9°, respectively, which can stabilize chiral magnetic structures like DWs.[341] Electric fields can induce electron movement and modulate interatomic interactions, regulating the DMI. An external electric field of -0.30 (0.30) V $Å^{-1}$ adjusts the $d_{//}$ value of graphene/Janus H-VSeTe to -0.71 (-0.43) meV.[342] The $d_{//}$ value increases with decreasing interlayer spacing, reaching a maximum of -0.72 meV when the interlayer distance is reduced by 0.60 Å.[342] A -4 % strain leads to a $d_{//}$ of -0.85 meV, indicating interactions can be modified by structural changes.[342] The SOC of Se and Te atoms is relatively strong and dominates the DMI of Janus H-VSeTe and its heterostructure.

Other 2D Janus dichalcogenides were also predicted to have stable spin topological structures.[326,337] For Janus manganese dichalcogenides, merely the T-phase has been investigated. Yuan et al. found that the large intrinsic DMI in MnSTe and MnSeTe are dominated by Te atoms, stabilizing sub-50-nm skyrmions, while the DMI of MnSSe is negligible due to the weak SOC [Fig. 15(d)], which is interpreted by the atom-resolved SOC energy [Fig. 15(c)].[326] The $d_{//}$ values were calculated as -0.04, 5.58, and 4.34 meV for MnSSe, MnSTe, and MnSeTe, respectively.[326] As the topological charge $Q$ changes from -1 to 0, MnSTe transitions from a non-trivial skyrmion to a trivial FM state under a high magnetic field.[326] Meanwhile, Liang et al. also explored skyrmion states induced from FM states with worm-like magnetic domains under a magnetic field, with DMI parameter $d_{//}$ of 2.14 meV for MnSeTe and 2.63 meV for MnSTe, while MnSSe exhibits a small value of -0.39 meV [Fig. 15(f)].[327] The DM/exchange $|d_{//}/J|$ ratios of MnSSe, MnSTe and MnSeTe are 0.03, 0.16 and 0.25, respectively.[327] Below 125 K and above 1 T magnetic field, the topological charge $Q$ becomes negative, indicating skyrmion formation in MnSTe [Fig. 15(g)].[327] These two studies display consistency, while their differences in calculated

DMI may arise from varying electron correlation strengths for Mn-3*d* orbitals and other calculational parameters.[326,327]

Beyond pristine monolayers, Wang et al. found strain-induced diverse chiral spin textures in MnSeTe, involving FM and AFM spiral, skyrmion, skyrmionium, and bimeron.[392] Ga et al. demonstrated that multilayer MnSTe with strong interlayer exchange coupling and Bethe-Slater curve-like behaviors can host layer-dependent DMI and field-free magnetic skyrmion.[393] Dou et al. proposed a MnSTe/$In_2Se_3$ hetero-bilayer with ferroelectrically controllable skyrmions, enhancing DMI parameters and DMI/exchange ratios, from 1.88 meV and 0.18 for pristine MnSTe, to 2.18 meV and 0.39, 2.50 meV and 0.31 for two phases with opposite FE polarizations, respectively.[394] With adjustable magnetoelectric coupling, the MnSeTe/$In_2Se_3$ and $In_2Se_3$/MnSeTe/$In_2Se_3$ heterostructures display DMI parameters of 4.38 and 3.94 meV, respectively.[395] The similar FE polarization modified-DMI can be observed in the MnSeTe/$Hf_2Ge_2Te_6$ and MnSeTe/bilayer-$In_2Se_3$ heterostructures, while the DMI parameters of $MnSe_2$/monolayer-$In_2Se_3$ appear quite close to zero considering different FE directions.[395]

In addition to the aforementioned dichalcogenides, certain 2D Janus halides have been investigated for their non-linear magnetism.[344,396-398] Janus T-NiClBr exhibits DMI competing with exchange frustration, featuring a skyrmion Hall effect with a helicity of 139°, which can be controlled by strain and magnetic field [Figs. 15(h) and 15(i)], varying from ~0° to 50°, and from 120° to 180°.[396] This competition between DMI and exchange frustration results in both Néel and Bloch types of skyrmions [Fig. 15(j)]. The skyrmion Hall effect devices are proposed in Fig. 15(k), where the non-centro-symmetric frustrated magnetic materials like Janus T-NiClBr are put on a piezoelectric material to control strain and current, realizing modulation of skyrmion helicity.[396] DMIs in some other 2D Janus materials are discussed in Supporting Information.

The DMI arises in systems with broken inversion symmetry and strong SOC. To enhance it in in 2D Janus materials: (1) Strain engineering changes the lattice structure and SOC, and modulates the PMA. (2) Heterostructure building alters interlayer symmetry and electron orbital hybridization. (3) Applied electric field modulates electron distribution and magnetic interaction strength. (4) Multilayer cascading may improve magnetic coupling. Overall, these results expose that 2D Janus dichalcogenides and halides in T- and H-phases manifest topological magnetic structures. This presents cutting-edge pathways for high-density and rapid data storage and transport.

### D. Spin and Valley

Spin and valley serve as additional degrees of freedom for carriers beyond charge, providing principles for designing integrated circuit components that reduce energy consumption and overcome quantum limits.[399,400]

The 2D Janus family, with diverse components and asymmetric structures, may exhibit a wide range of electronic configurations. These span from insulators with large band gaps to semiconductors with narrow band gaps, topological insulators, semimetals, metals, etc., presenting rich electronic structures. As described earlier, the suitable and highly tunable band gaps provide potential for optical response, catalytic and thermoelectric properties. What's more, considering spin and valley splitting in electronic structures benefits spin dynamics, spintronics, and valleytronics equipment.[401,402] Energy and momentum degeneracy changes in different spin channels enable spin-polarized currents and spin control, crucial for magnetic random-access memory and low-dimensional device development in information processing and data storage.[316,403] The broken spatial inversion symmetry in 2D Janus materials necessitates valley polarization and the Rashba effect, promising superior electronic structures.

Although valley is a distinct degree of freedom from spin, valley splitting shares similarities with spin splitting in altering energy and momentum degeneracies. Thus, they are discussed together here for better comparative analysis. This part explores spin polarization, valley polarization, and Rashba, Zeeman and Dresselhaus effects in 2D Janus materials to advance information encoding, transport, and storage.

## 1. Spin Polarization

Spin polarization is essential for enhancing efficiency and accuracy in spintronics. It is defined by $\mathrm{SP} = \left| N_{\alpha}(E_{\mathrm{F}}) - N_{\beta}(E_{\mathrm{F}}) \right| / [N_{\alpha}(E_{\mathrm{F}}) + N_{\beta}(E_{\mathrm{F}})]$, in which $N_{\alpha}(E_{\mathrm{F}})$ and $N_{\beta}(E_{\mathrm{F}})$ stand for the DOS of spin-up and spin-down channels at the Fermi level, respectively. Yielding a purely spin-polarized current requires tailored electronic structures.

Beyond traditional metal, semiconductor, and insulator, several specialized electronic structures have been proposed (Fig. 16). (1) Half-metal (HM) possesses a metallic channel in one spin direction and a semiconducting or insulating channel in the other, enabling selective carrier transport.[404] (2) Spin gapless semiconductor (SGS)[405] features four electronic configurations. The first item is similar to HM but differs in that one of spin channels is not metallic but semiconducting without a band gap, which also produces a completely spin-polarized current. The second item behaves as a zero-band gap semiconductor when the spin orientations are not considered, but there are band gaps in both spin channels, which can acquire 100 % spin-polarized electrons in one spin direction and holes in the other. The third (fourth) item resembles the first item, but the differences are that the top (bottom) of the valence (conduction) band of the spin channel with a band gap touches the Fermi level, which only gets fully spin-polarized holes (electrons). All these SGSs facilitate carrier excitation without extra energy input.[405,406] (3) For the second SGS, if the band gap is not zero when the spin directions are neglected, a bipolar magnetic semiconductor (BMS) emerges. BMS can achieve pure spin polarization and allows spin orientation control via a gate voltage. (4) Related concepts include unipolar magnetic semiconductor (UMS) and half-semiconductor (HSC),[406,407] which both convey similar implications, and derive from comparison with BMS and HM, respectively. The band gap in one spin direction is not zero and significantly smaller than that in the other direction as HSC/UMS differentiated from HM.[406] Unlike BMSs, a UMS can generate a spin-polarized current of merely one spin direction, irrespective of whether the Fermi level is raised or lowered by electrical gating.[407] (5) The asymmetric antiferromagnetic semiconductor (AAFMS) is proposed as the magnetic moments stem from different species of magnetic ions, which act in opposite directions and are completely cancelled out by each other.[406,408] The mismatched energy levels of magnetic orbitals from different magnetic ions motivates high spin polarization near the Fermi level.[408] (6) Similarly, HM antiferromagnet[409] exhibits completely spin-polarized with compensated magnetic moments from different ions. These complex electronic structures highlight the potential for advanced spin transport applications. Note that, altermagnets exhibit nonrelativistic and momentum-dependent spin splitting due to specific crystal and spin symmetries, but the overall spin polarization is not high, and the high spin polarization needs special paths in momentum space.[410,411] So, the spin polarization in altermagnets is not within the scope of our present review.

While pristine 2D materials have limited electronic properties, 2D Janus materials introduce diverse chemical components and spatial distortions, enriching the electronic structure landscape. This segment reviews spin polarization in 2D Janus family to promote the development of miniaturized spintronic devices.

The experimentally existing 2D Janus systems are worth being concerned with electronic structures, but MoSSe, WSSe and PtSSe are non-magnetic without spin polarization.[28,30,32,34] Fe- and Cr-doped Janus H-MoSSe monolayers are HM and BMS, exhibiting perfect thermal spin-filtering and spin Seebeck effects, respectively, making them promising candidates for energy-efficient spin-caloritronic devices.[412] Liu et al. studied that spin polarization of Janus T-CrSeTe monolayer is only 33 %, but improvable to around 50 % via strain or holes doping.[334] Janus H-CrSeTe, CrSTe, and CrSSe monolayers are semiconductors [Figs. 17(a)-17(c)], but become metals under tensile strain or electric field.[261] Janus H-CrSTe transforms to a HM at a small strain of 1.6 % and remains so up to 10 % strain.[374] Janus H-VSeTe monolayer is a BMS with an indirect band gap of 0.254 eV, where V-*d* orbitals chiefly contribute to the VBM and CBM [Figs. 17(d)-17(h)].[339] Strain alters lattice and affects bonding, thus adapting band structure and electron occupation. As strain varies from -13 % to 10 %, Janus H-VSeTe monolayer can transform to spin-unpolarized metal, HM, SGS and HSC/UMS, unfolding the strain-tuned plentiful electronic structures.[336]

Heterojunctions introduce compositions and occupations, causes charge redistribution, allowing modulation of electronic structures. Semiconductor and metal phases can be switched by different stacking configurations in Janus H-VSSe/$VSe_2$ and VSeTe/$VSe_2$ heterostructures.[413] Twist angles modulate interlayer coupling strength and electronic interaction, possessing manipulation on band structure. The graphene/Janus H-VSeTe heterostructures produce nearly 100 % spin polarization at twist angles of 0°, 7.6°, 10.9° and 19.1°, while the spin polarization is zero with twist angles of 5.2°, 5.8°, 6.6° and 13.9°.[341] It's worth noting that the heterostructures can generate flat bands with twist angles of 5.2° and 10.9°, which makes tremendous sense for delving unconventional superconductivity and Mott-like insulating behavior.[341]

Electric field can transfer charges, affect chemical bonds and electron energy levels, and regulate electronic structures. Due to the combined effects of magnetic proximity coupling and intrinsic diploe moment, an external electric field exceeding 0.35 V $Å^{-1}$ can turn the light p-type doping in spin-down channel of graphene/Janus H-VSeTe heterostructure into n-type doping, while the spin-up bands around the Fermi level are mainly from graphene rather than VSeTe, and initial p-type doping is preserved against the electric field.[414] The positive (negative) electric field can shift the Fermi level upward (downward) for graphene/Janus H-VSeTe heterostructure.[342]

The spin field effect transistor (FET) is proposed to yield a purely spin-polarized current [Figs. 17(i) and (j)].[415] Chen et al. reported Janus T-MnSSe is a HM, becoming metal with doping concentration larger than 0.5 holes per atom ($1.4 \times 10^{14}$ $cm^{-2}$) [Fig. 17(k)].[415] Introduction of exotic elements has potential to bring magnetism and spin polarization in the non-magnetic Janus materials. In F- and Cl-doped Janus T-ZrOS and ZrOSe, HMs and magnetic semiconductors are induced at high and low concentrations, respectively.[416]

In addition to the aforementioned dichalcogenides, 2D Janus ferrum, cobalt, nickel, and gadolinium halides have been examined for their spin behavior.[347,382,417] Zhang et al. demonstrated the intrinsic HSC/UMS, -6 % compressive strain-induced SGS, and 0.02 e/h doping-induced HM characteristics in Janus H-FeClF monolayer [Fig. 17(l)].[347] Janus H-FeClF bilayers display electric field- and stacking-tunable BMS and AFM semiconductor features.[348] Among them, Fe-*d* orbitals make the main contribution near the Fermi level.[344,347,348,383] Additionally, Janus H-GdClBr, GdClI, and GdBrI monolayers are BMSs that can transform into semiconductors or HSCs/UMSs at 5% tension.[382] The strain dependence of total and spin-resolved band gaps in Janus H-GdClBr [Fig. 17(m)] are displayed as an instance.[382] Discussion on electronic structures of other 2D Janus materials are presented in Supporting Information.

These 2D Janus materials, with their diverse components and asymmetric structures, can exhibit a wide range of electronic configurations, making them suitable for various electronic structures with high spin polarization. Methods to enhance spin polarization include: (1) Strain engineering changes the lattice and band structure. (2) Heterostructure building realizes charge redistribution, and alters interlayer interactions due to interface effects. (3) Electric field modulates the Fermi level and electron distribution. (4) Elemental doping introduces other elements to change the electronic structure. In conclusion, these findings underscore the diverse and tunable electronic structures of 2D Janus materials, making them promising candidates for miniaturized applications in spin transport and storage.

## 2. Valley Splitting

Valley, as an emerging degree of freedom beyond charge and spin, facilitates information encoding, processing, and transmitting.[315] Valley polarization occurs with the incomplete equivalence of energy valleys, which are local extreme points in the electronic energy and momentum dispersion relation.[315,418]

In non-magnetic systems, valley polarization manifests as degenerate energy levels at valleys that are not completely isologous. This can be detected using circularly polarized light, in accordance with optical and spin selection rules.[419] Specifically, valley-dependent optical selection rules suggest that circularly polarized light can selectively excite carriers at different valleys. Linearly polarized light excites carriers regardless of K and K' points, without preferential selection.[420] Spin selection rules mean that a circularly polarized light with a specific frequency (e.g., left circularly polarized light) can excite spin-up carriers at K point, while the opposite circularly polarized light with the same frequency (right circularly polarized light) can excite spin-down carriers at K' point. This interrelation of valley, spin, and circularly polarized light, provides a practical avenue for controlling valley degree of freedom.[420]

Nevertheless, valley polarization in non-magnetic materials is typically unstable and temporary, depolarizing once external excitation ceases. For valleytronic applications favoring non-volatile properties, the transient state induced by circularly polarized light is insufficient. The degenerate energy levels of valleys pose a challenge for practical valleytronic utilization. To achieve spontaneous and stable valley polarization, it is necessary to break time reversal symmetry in addition to spatial inversion symmetry breaking. This creates an energy offset between valleys, lifting degeneracy, and enabling stable valley polarization. The spatial inversion symmetry is written as:

$$\boldsymbol{\Omega}_n(-\boldsymbol{k}) = \boldsymbol{\Omega}_n(\boldsymbol{k}),\ E_{n\uparrow}(\boldsymbol{k}) = E_{n\uparrow}(-\boldsymbol{k}),\ E_{n\downarrow}(\boldsymbol{k}) = E_{n\downarrow}(-\boldsymbol{k}) \tag{27}$$

and the time reversal symmetry requires:

$$\boldsymbol{\Omega}_n(-\boldsymbol{k}) = -\boldsymbol{\Omega}_n(\boldsymbol{k}),\ E_{n\uparrow}(\boldsymbol{k}) = E_{n\downarrow}(-\boldsymbol{k}) \tag{28}$$

Introducing an external magnetic field, doping magnetic atoms, and constructing heterostructures with magnetic materials are effective strategies to break time-reversal symmetry and achieve non-degenerate energies for non-magnetic systems.[420] For magnetic materials, time reversal symmetry is inherently broken without cumbersome external intervention. Thus, it is highly significant to search for materials that combine both magnetism and valley polarization, a concept known as ferrovalley (FV).[314] Fortunately, Janus magnets possess the nature of broken spatial inversion symmetry, can exhibit spontaneous valley polarization when there is adequate SOC strength.[421]

The quantitative value of valley splitting can be determined by $\Delta E^{v/c} = E_{K'}^{v/c} - E_{K}^{v/c}$, in which $E_{K'}^{v/c}$ and $E_{K}^{v/c}$ stand for the energies of the VBMs or CBMs at K' and K valleys, respectively. For RT practical valley physics applications, a valley splitting of up to 100 meV is necessary to effectively counteract thermal noise.[422,423] The SOC Hamiltonian for a valley-polarized system can be expressed as $H_{\mathrm{SOC}} = \lambda \boldsymbol{LS} = H_{\mathrm{SOC}}^{0} + H_{\mathrm{SOC}}^{1}$, where the coupling strength is represented by $\lambda$, and $\boldsymbol{L}$, $\boldsymbol{S}$ are the vectors of orbital and spin angular moments, respectively. This Hamiltonian comprises two components: the interaction between identical spin states ($H_{\mathrm{SOC}}^{0}$), and that between inverse spin states ($H_{\mathrm{SOC}}^{1}$). Depending on the crystal field, the VBM is generally derived from the $d_{z^2}$ orbitals while the CBM arises from the $d_{xy}$ and $d_{x^2-y^2}$ orbitals, or vice versa. Taking the former type as a representative, the conduction band can be represented by the superposition of the two orbitals:

$$|\varphi_C^{\tau}\rangle = \sqrt{\frac{1}{2}}(|d_{x^2-y^2}\rangle + i\tau\, |d_{xy}\rangle) \tag{29}$$

where $\tau$ is the valley index, which is 1 and -1 at K and K', respectively. The SOC Hamiltonian considering the spin direction can be further written as:

$$H_{\mathrm{SOC}} = \lambda S_{z'}(L_z \cos\theta + \frac{1}{2} L_{+} e^{-i\phi} \sin\theta + \frac{1}{2} L_{-} e^{+i\phi} \sin\theta) \tag{30}$$

where $\theta$ and $\phi$ indicate the spin orientations. The eigenvalue is $E_C^{\tau} = \langle \varphi_C^{\tau} | H_{\mathrm{SOC}} | \varphi_C^{\tau} \rangle$, and the value of valley splitting based on matrix elements for the SOC operator $\boldsymbol{L} \cdot \boldsymbol{S}$ of *s*, *p* and *d* orbitals[424] in Tables VI and VII can be described as:

$$E_C^{K'} - E_C^{K} = i\langle d_{x^2-y^2} | H_{\mathrm{SOC}} | d_{xy} \rangle - i\langle d_{xy} | H_{\mathrm{SOC}} | d_{x^2-y^2} \rangle = 2\lambda - (-2\lambda) = 4\lambda \tag{31}$$

where the spin direction is considered as pointing towards the $z$ axis (out-of-plane), $\theta$ is 0, so sin $\theta$ = 0 and cos $\theta$ = 1. However, if the spin orientation is pointing to the in-plane direction, $\theta$ is 90° ($\pi / 2$), so sin $\theta$ = 1 and cos $\theta$ = 0, the value of valley splitting is $E_C^{K'} - E_C^{K} = i\left\langle d_{x^2-y^2} \mid H_{\mathrm{SOC}} \mid d_{xy} \right\rangle - i\left\langle d_{xy} \mid H_{\mathrm{SOC}} \mid d_{x^2-y^2} \right\rangle = 0$. Therefore, the form of the spin orientation can be abbreviated as:

$$\left\langle d_{xy} \middle| H_{SOC} \middle| d_{x^2-y^2} \right\rangle = 2i\lambda S_z \quad (32)$$

in which the $S_z$ is the component of spin along the $z$ axis. Therefore, an intrinsic out-of-plane easy axis is essential for spontaneous valley polarization in 2D Janus materials in H-, T- and their derivative phases. While certain materials with IMA that may possess spontaneous valley polarization, like $W_2MnC_2O_2$ MXene,[425] and altermagnetic $V_2Se_2O$,[426] they do not belong to the 2D Janus H- and T-family and thus are beyond the scope of this discussion. Furthermore, a robust PMA is crucial for resisting external thermal perturbations and opening the magnetic oscillator excitation gap, which helps stabilize valley polarization. Weak PMA can lead to the easy axis reorienting under external disturbances, negatively impacting valley polarization.

To further investigate the valley physics and corresponding potential applications in 2D Janus materials, the Berry curvature $\Omega_z(k)$ according to the Kubo formula[427] is expressed as:

$$\Omega_z(k) = -\sum_n \sum_{n \neq m} f_n \frac{2\,\mathrm{Im}\left\langle \Psi_{nk} \mid \upsilon_x \mid \Psi_{mk} \right\rangle \left\langle \Psi_{mk} \mid \upsilon_y \mid \Psi_{nk} \right\rangle}{(E_m - E_n)^2} \quad (33)$$

in which $\upsilon_{x/y}$ is the velocity operator, $f_\mathrm{n}$ represents the Fermi-Dirac distribution function, $\Psi_{nk}$ stands for the Bloch wave function and $E_{m/n}$ is the corresponding energy eigenvalue. The opposite Berry curvatures at K and K' valleys function similarly to a magnetic field. This valley-contrasting Berry curvature enables the AVHE without external magnetic field, which is advantageous for encrypted transport and non-volatile storage.[314,428]

Experimental and theoretical investigations into the valley splitting of 2D Janus materials—both magnetic and non-magnetic—are illustrated in Figs. 18(a)-18(n) and

19. The helicity of emission peaks (~9 % at RT and 50 % at 90 K) for Janus H-MoSSe in the presence of circularly polarized PL supports its valley polarization.[28] However, non-magnetic 2D Janus materials like molybdenum and tungsten dichalcogenides exhibit only transient valley polarization. To achieve stable valley polarization, time-reversal symmetry should be broken. Incorporating exotic magnetic atoms or constructing heterojunctions with magnetic materials has been shown to stabilize valley polarization in these 2D Janus dichalcogenides.[420] As depicted in Figs. 18(a) and 18(b), doping Janus H-MoSSe with Cr/V breaks the energy merger without significant structural disruption.[47] V-doping systems achieve stable valley polarization of up to 59 meV and exhibit strain sensitivity.[47]

Heterojunctions generate interfacial magnetic proximity effects and modulate interlayer excitons to enhance valley polarization. Charge transfer, perpendicular magnetic moments from Mn, and substrate electrostatic interactions enable Janus H-WSSe on MnO (111) surface to generate substantial valley polarization of about 410 meV, equivalent to an effective Zeeman field $B_z$ of 200 meV.[429] Additionally, Janus H-WSSe/CrN heterojunctions with various interfaces produce intrinsic valley splitting.[430] By compressing the interlayer distance to 2.23 Å from 2.53 Å, interlayer interactions and electronic structures are tuned, the maximum value reaches 272 meV from 103 meV, corresponding to an effective Zeeman magnetic field of 2560 T.[430] This modulation of valley splitting stems from changes in induced spin-resolved charges around S and W atoms.[430]

2D Janus non-magnetic molybdenum and tungsten dichalcogenides can achieve decent valley polarization when time-reversal symmetry is broken,[28,431] but these external strategies have drawbacks, like strong scattering from magnetic atoms, experimental preparation difficulties of heterojunctions with magnets (possible lattice matching and unwanted chemical reaction), and energy consumption from external magnetic fields (required additional energy all the time).[420] Fortunately, some 2D Janus materials were found to be FV with spontaneous magnetism and valley polarization, breaking both time-reversal and spatial inversion symmetries with adequate SOC strength, which supplied convenience and preserve intrinsic characteristics.[46,346]

For pristine monolayers, Janus H-VSeTe possesses a valley splitting of 158 meV,[339] while that of Janus H-VSSe monolayer is about 85 meV, with the $d_{xy}$ and $d_{x^2-y^2}$ orbitals of V atoms playing a major role.[46] Its anomalous Hall conductivity (AHC) can be enhanced by strain, with a maximum AHC of 29.0 (7.9) S $cm^{-1}$ between the two valence (conduction) edges of valleys.[432] Strain distorts the structure, shifting the energy of electronic orbitals at valleys, which also modulates valley splitting. A -2 % strain improves the valley polarization of Janus H-VSeTe to 169 meV [Figs. 18(c) and 18(d)].[336]

In heterojunctions, interfacial interaction and charge transfer can change the orbital energies of valleys and modify energy degeneracy. The H-$VSe_2$/Janus H-VSeTe has a valley splitting of 114 meV.[413] The valley splitting is sensitive to twist angles, as seen in graphene/Janus H-VSeTe heterostructure, where it varies with different twist angles.[341] It is 11 meV at twist angles of 5.2° and 7.6°, while is nearly zero at twist angles of 5.8° and 13.9°.[341]

Theoretical studies have explored 2D Janus magnetic dihalides for valley polarization.[272,346,377,433] Janus H-FeClBr monolayer displays a large spontaneous valley polarization of 188 meV, which can be modulated by circularly polarized light, linear light, and hole doping.[383] Different electronic correlation strengths in Janus H-FeClF monolayer can realize FV and half-valley-metal (HVM), with a valley splitting of 109 meV under a Hubbard U of 2.5 eV.[346] Interestingly, its bilayer exhibits a valley switch effect, where the valley polarization (~108 meV) undergoes a presence-disappearance-presence transition with increasing electric field from 0 to 0.15 eV $Å^{-1}$, depending on the PMA-IMA-PMA transition.[348] Moreover, Guo et al. revealed that Janus H-FeClF monolayer possesses a quantum anomalous Hall (QAH) insulator phase between two HVM states for PMA [Figs. 18(e)-18(g)], and a semimetal phase for IMA with different electronic correlation strengths.[346] It has a unit Chern number and a chiral edge state connecting the conduction and valence bands for both left and right edges at the QAH state [Figs. 18(h) and 18(i)], indicating non-trivial topological properties.[346]

Alongside AVHE, topological states are expected to emerge when electronic structures are further modified.[434] Elements in the same main group as Fe have also been probed for valley physics and topological states as presented in Supporting Information. Insulators and semimetals with topologically non-trivial band structures have been revealed in 2D Janus materials.[346,435] These topological states are vital discoveries in condensed matter physics, as surface states protected by topology, make carriers less susceptible to scatter from impurities and enable dissipation-free transport.[436] Topological materials can also be used to construct quantum bits for stable topological quantum computation.[437] Notably, topological properties are also found in Janus H-VSSe monolayer, which shows non-trivial corner states, and is forecasted as a robust second-order topological insulator (SOTI) against magnetization directions.[438]

Berry curvatures in the 2D Brillouin zone and along the high symmetry of Janus H-RuClBr are presented in Figs. 18(j) and 18(k), and the first Brillouin zone with the high-symmetry points is displayed in Fig. 18(l).[377] Its valley peaks at K and K' points are equivalent to a magnetic field, enabling AVHE [Figs. 18(m) and 18(n)], where the "+" and "-" signals are the hole and electron, upward and downward arrows are the spin-up and down channels, respectively.[377] Under in-plane electric field, carriers can be accumulated at the sides without an external magnetic field.[377] The research on 2D Janus dihalides in lanthanides groups has also revealed significant and tunable valley polarization.[278,381,382] Some other representatives in Fig. 19[439] are included in Supporting Information.

Achieving stable valley polarization requires breaking both time-reversal and spatial inversion symmetry. The main routes to enhance valley polarization in 2D Janus materials are: (1) Strain engineering may alter valley polarization by adjusting the lattice structure and SOC strength. (2) Heterostructure building enhances valley polarization through interfacial effects and interlayer interactions. (3) Elemental doping introduces atoms to modify the electronic structure. (4) Electric field modulation affects valley polarization by regulating Fermi levels and electron distribution. These results reveal substantial and tunable valley polarization in 2D Janus materials, inspiring their cutting-edge valleytronic design and application.

## 3. Rashba, Dresselhaus and Zeeman Effects

The Rashba, Dresselhaus and Zeeman Splitting effects are pivotal in describing the splitting of energy level degeneracies based on SOC under specific physical conditions [Figs. 20(a)-20(f)]. These phenomena furnish a robust foundation for advancements in both spin transport and quantum computation.[440,441] To unveil their inner mechanisms, and compare their similarities and differences, the Hamiltonian can be written as:

$$H = \frac{\boldsymbol{p}^2}{2m^*} + V + H_{\mathrm{SOC}}$$

(34)

in which the momentum operator, electron effective mass, crystal potential, and the SOC Hamiltonian are represented by $\boldsymbol{p}$, $m^*$, $V$, and $H_{\mathrm{SOC}}$, respectively. The intrinsic electric field $\boldsymbol{E}$ is the gradient of V, whose form is $\boldsymbol{E} = -\nabla V$, $H_{\mathrm{SOC}}$ can be described as:[126]

$$H_{\mathrm{SOC}} = \frac{\hbar}{4m^2c^2}(\nabla V \times \boldsymbol{p})\boldsymbol{\sigma} \tag{35}$$

where $\hbar$, $m$, $c$, and $\boldsymbol{\sigma} = (\sigma_x, \sigma_y, \sigma_z)$ are the reduced Planck's constant, electron mass, velocity of light, and Pauli matrices, respectively. The Bloch wave function of electrons can be attained as $\psi_{\boldsymbol{k}}(\boldsymbol{r}) = \mathrm{e}^{i\boldsymbol{kr}}\phi_{\boldsymbol{k}}(\boldsymbol{r})$, in which $\boldsymbol{k}$, $\boldsymbol{r}$, and $\phi_{\boldsymbol{k}}(\boldsymbol{r})$ are the wave vector, position vector, and periodic wave function, respectively. Thus, the Schrödinger equation is written as:

$$H_0(\boldsymbol{k})\phi_{\boldsymbol{k}}^0 = \varepsilon^0(\boldsymbol{k})\phi_{\boldsymbol{k}}^0 \tag{36}$$

which is $H_0(\boldsymbol{k}) = (\hbar\boldsymbol{k} + \boldsymbol{p})^2 / 2m + V$. And $H_{\mathrm{SOC}} = \boldsymbol{\Omega}(\boldsymbol{k})\boldsymbol{\sigma}$, where $\boldsymbol{\Omega}(\boldsymbol{k})$ is the spin-orbit field depended on wave vector and can be described as:

$$\boldsymbol{\Omega}(\boldsymbol{k}) = \left\langle \phi_{\boldsymbol{k}}^0 \middle| \frac{\hbar}{4m^2c^2}[\nabla V \times (\hbar \boldsymbol{k} + \boldsymbol{p})] \middle| \phi_{\boldsymbol{k}}^0 \right\rangle \tag{37}$$

The spin operator $\boldsymbol{S} = (\hbar / 2)\langle \boldsymbol{\sigma} \rangle$, which forms spin textures stemming from SOC, can be obtained as:

$$\boldsymbol{S}_{\pm} = \frac{\hbar}{2}\langle \psi_{\pm} | \boldsymbol{\sigma} | \psi_{\pm} \rangle \tag{38}$$

Rashba effect, proposed by Emmanuel Rashba,[127] derives from SOC in systems lacking inversion symmetry. It depends on electron momentum and creates an effective magnetic field. The Rashba Hamiltonian $H_R(\boldsymbol{k})$ for 2D semiconductors can be described as:[442]

$$H_R(\boldsymbol{k}) = \alpha(\boldsymbol{\sigma} \times \boldsymbol{k})\hat{z} = \alpha(k_y\sigma_x - k_x\sigma_y) \tag{39}$$

in which the Rashba parameter is represented by $\alpha = 2E_R / k_R$, $E_R$ and $k_R$ are the energy difference and momentum offset, respectively. $\hat{z}$ is the surface normal. The eigenvalues and eigenstates can be attained as:

$$E_{R\pm}(\boldsymbol{k}) = \hbar^2 k^2 / (2m^*) \pm \alpha k \tag{40}$$

$$\psi_{R\pm}(\boldsymbol{k}) = \frac{ei\boldsymbol{kr}}{2\pi\hbar}\frac{1}{\sqrt{2}}\begin{pmatrix} \pm(ik_x + k_y)/k \\ 1 \end{pmatrix} \tag{41}$$

where the "+" and "–" symbols stand for inner and outer branches, respectively. As presented in Figs. 20(a) and 20(d), the Rashba spin texture can be written as:

$$\langle \boldsymbol{\sigma} \rangle_{R\pm} = \langle \psi_{R\pm} | \boldsymbol{\sigma} | \psi_{R\pm} \rangle = \pm \begin{pmatrix} \sin\theta \\ -\cos\theta \\ 0 \end{pmatrix} \tag{42}$$

The rotations of textures of inner and outer bands are in counterclockwise and clockwise directions, respectively.

Similarly, the Dresselhaus effect, proposed by Gilbert Dresselhaus,[126] generates analogous momentum-dependent spin splitting, but it typically originates from bulk inversion asymmetry. The Dresselhaus Hamiltonian can be described as:[443]

$$H_D(k) = \gamma[k_x(k_y^2 - k_z^2)\sigma_x + k_y(k_z^2 - k_x^2)\sigma_y + k_z(k_x^2 - k_y^2)\sigma_z] \tag{43}$$

in which $\gamma$ stands for the material constant. Accordingly, the 2D Dresselhaus Hamiltonian can be attained as:[444]

$$H_D^{2D}(\boldsymbol{k}) = \gamma[k_x(k_y^2 - \langle k_z^2 \rangle)\sigma_x + k_y(\langle k_z^2 \rangle - k_x^2)\sigma_y] = \beta(k_x\sigma_x - k_y\sigma_y) + \gamma(k_x k_y^2 \sigma_x - k_x^2 k_y \sigma_y) \tag{44}$$

where the cubic and linear Dresselhaus constants are represented by $\gamma$ and $\beta = -\gamma\langle k_z^2 \rangle$, respectively. The cubic component is ignored for 2D systems with strong confinement, $\langle k_z^2 \rangle >> k_x^2$, so the 2D Dresselhaus Hamiltonian can be simplified as:[445]

$$H_D^{2D}(\boldsymbol{k}) = \beta(k_x\sigma_x - k_y\sigma_y) \tag{45}$$

Then the eigenvalues and eigenstates can be obtained as:

$$E_{D\pm}(\boldsymbol{k}) = \hbar^2 k^2 / (2m^*) \pm \beta k = \hbar^2 (k \pm k_D)^2 / (2m^*) - E_D \tag{46}$$

$$\psi_{D\pm}(\boldsymbol{k}) = \frac{ei\boldsymbol{kr}}{2\pi\hbar} \frac{1}{\sqrt{2}} \begin{pmatrix} \pm(k_x + ik_y)/k \\ 1 \end{pmatrix} = \frac{ei\boldsymbol{kr}}{2\pi\hbar} \frac{1}{\sqrt{2}} \begin{pmatrix} \pm e^{i\theta} \\ 1 \end{pmatrix} \tag{47}$$

As presented in Figs. 20(b) and 20(e), the Dresselhaus spin texture can be written as:

$$\langle \boldsymbol{\sigma} \rangle_{D\pm} = \langle \psi_{D\pm} | \boldsymbol{\sigma} | \psi_{D\pm} \rangle = \pm \begin{pmatrix} \cos\theta \\ -\sin\theta \\ 0 \end{pmatrix} \tag{48}$$

The Zeeman effect, proposed by Pieter Zeeman,[440] involves the splitting of energy levels, when an external magnetic field is applied [Figs. 20(c) and 20(f)]. The Zeeman Hamiltonian can be described as:[446]

$$H_Z = \lambda_{x,y} P_x \sigma_y + \lambda_{x,z} P_x \sigma_z + \lambda_{y,x} P_y \sigma_x + \lambda_{z,x} P_z \sigma_x \tag{49}$$

in which $\lambda_{i,j}$ and $P_i$ are the magnetoelectric coupling and induced electronic polarization by electric field, respectively. The Zeeman-like pairs are written as:

$$E_{\boldsymbol{k},\uparrow}^{\alpha_i} \neq E_{\boldsymbol{k},\downarrow}^{\beta_i}, E_{\boldsymbol{k},\uparrow}^{S_1} \neq E_{\boldsymbol{k},\downarrow}^{S_2}$$

(50)

where $i$ = 1, 2 and $S$ = $\alpha$ , $\beta$ . On account of $P_i \propto E_{ext}$ , $H_Z \propto \gamma_{z,x} E_z \sigma_x$ . $E_{ext}$ is the external electric field and $\gamma_{z,x}$ is the splitting strength. The Zeeman splitting can be regulated effectively by electric field.[446,447]

Rashba, Dresselhaus, and Zeeman effects provide approaches of spin manipulation and contribute to spintronics.[441,447,448] The combined Rashba-Zeeman model has predicted 2D systems with insulator-to-conductor transition driven by exchange fields, which is useful for switching and sensing.[449] The Dresselhaus-Zeeman model can also control spin textures in topological insulators.[450] 2D Janus materials, with their natural lack of spatial inversion symmetry, could exhibit these splitting effects, especially the Rashba effect, if there is sufficient SOC strength.[45,451] This makes them promising for spin utilization and worthy of further investigation.

The Rashba, Dresselhaus, and Zeeman effects offer valuable approaches for spin manipulation and integral to spintronics. Some representative effects and Rashba parameters in 2D Janus materials are shown in Figs. 20(g)-20(n) and 21. Experimental 2D Janus materials are uncovered with these splitting species.[45,216] Janus H-MoSSe, with broken spatial symmetry, demonstrates Rashba spin splitting contributed by $d_{z^2}$ orbitals around the Γ point.[28] The Rashba parameters for Janus H-MoSSe and WSSe monolayers are 0.54 (0.42) and 0.75 (0.86) eV Å along Γ-K (Γ-M) directions, respectively.[216] As displayed in Figs. 20(g) and 20(h), the Rashba SOC and the spin Hall conductivity (SHC) in Janus H-MoSSe is enhanced, and the SHC of MoSSe in the valence band is an offspring of competition between Zeeman and Rashba types.[452] It can be modulated by adjusting strain and Fermi level, the built-in electric field is reduced and the Rashba splitting energy is correspondingly decreased with the increased strain [Figs. 20(i) and 20(j)].[452] Similarly, Janus H-MoSTe, MoSeTe, WSTe and WSeTe also exhibit Rashba effect near the Γ point, and Janus molybdenum and tungsten dichalcogenide monolayers all possess spin splitting of Zeeman type at K and K' points.[45]

External stimuli such as electric field and strain effectively enhance the Rashba splitting. For instance, the Rashba parameter of Janus H-MoSeTe increases from pristine 0.48 eV Å to 1.12 eV Å at -3 % strain.[45] The $d_{xz}$ and $d_{yz}$ orbitals dominate the Rashba SOC in stabilized Janus distorted-T-WSSe (not T' phase), which is weakened by macroscopic charge transfer, lowering potential gradient, and increasing macroscopic polarization, for strain modulation.[453]

Rashba splitting of Janus H-WSSe bilayer and multilayer depends on interlayer electrostatic interactions, which can be regulated by interlayer spacing and stacking patterns.[454] The Pb adsorption on Janus H-WSeTe reduces symmetry and introduces both Rashba and Zeeman species at Γ point, with a Rashba parameter of 0.75 eV Å, arising from strong hybridization between the W-$d_{z^2}$ and Pb-$p_y$ and $p_z$ orbitals.[455] It can be further boosted by decreasing adsorption distance, increasing adsorption concentration, and compression, and reaches a maximum of 1.55 eV Å under -8 % strain.[455] Janus H-MoSSe/WSSe VHT augments out-of-plane electric polarity, and consequently strengthen the Rashba effect, with corresponding parameter of 1.22 eV Å.[194] Positive electric field and in-plane compression can also strengthen the Rashba spin splitting, and the largest Rashba parameter can reach 0.54 eV Å in -4 % strained Janus H-MoSTe/WSTe.[456] The Rashba effects in Janus H-WSSe/MoSSe, WSTe/MoSTe, and WSeTe/MoSeTe heterojunctions are weaker than those of the corresponding Janus monolayers, attributed to the negative effect of interlayer stacking interactions and the decrease in intralayer/interlayer potentials, although WSeTe/MoSeTe reaches 0.54 eV Å of Rashba parameter.[457] Under the combined Rashba-Zeeman splitting in Janus H-MoSSe/GaN heterojunction, splitting sub-bands provide different free-carrier densities and enable net charge currents, which facilitates efficient charge transport, and manipulates photoelectrochemical water-splitting.[458] The W atom exhibits a stronger SOC effect than the Mo atom.[216] In GeC/Janus H-MoSSe heterostructure, the Rashba parameters are lower than those of H-MoSSe monolayer, while in GeC/Janus H-WSSe, they can be increased to 0.81 (0.62) eV Å along Γ-K (Γ-M) direction.[216] Heterostructure redistributes charges, creating a built-in electric field that favors the Rashba effect [Fig. 20(n)].[459]

In addition to the earliest studies on molybdenum and tungsten dichalcogenides mentioned above, other 2D Janus materials in the chromium, vanadium, and platinum groups that have been experimentally prepared, were also predicted to exhibit Rashba splitting.[261,451,460] Janus H-CrSSe, CrSTe, and CrSeTe possess appreciable Rashba parameters of 0.26, 0.31, and 1.23 eV Å (0.66, 0.50, and 2.11 eV Å under -2 % strain), respectively, owing to broken symmetry and strong SOC.[261] The band structure with Rashba splitting in Janus H-CrSeTe is displayed in Figs. 20(k) and 20(l), and further confirmed by the spin textures in Fig. 20(m).[261] Janus H-VSeTe, with ample SOC and a built-in electric field, has a Rashba spin-splitting parameter of 0.66 eV Å, while Janus H-VSSe and VSTe with weaker SOC, do not exhibit Rashba splitting, both presenting parameters of 0.[460]

Janus platinum dichalcogenides show Rashba splitting near the M point, with the Dresselhaus effect playing a secondary role.[451] Among them, Janus T-PtSSe has the Rashba parameters of up to 1.65 and 1.33 eV Å at the M-Γ and M-K regions, respectively.[451] However, the contribution of heavy elements in the sulfur group is not favorable for SOC-induced Rashba splitting, as seen in Janus T-PtSeTe, which has Rashba parameters of 0.44 and 0.75 eV Å at the M-Γ and M-K regions, respectively.[451] Additionally, theoretical investigations have also delved into some 2D Janus dichalcogenides for Rashba and Zeeman effects.[459,461] Some other SOC-induced splitting of 2D Janus family in Fig. 21[462-465] are discussed in Supporting Information to avoid repetition of similar content.

2D Janus materials hold significant potential for their Rashba, Dresselhaus, and Zeeman splitting effects, which describe energy-level splitting based on SOC under various physical conditions. The Rashba effect occurs in systems lacking inversion symmetry, the Dresselhaus effect stems from bulk inversion asymmetry, and the Zeeman effect involves energy-level splitting in external magnetic fields. These phenomena are crucial for spin manipulation in spintronics. 2D Janus materials, with their inherent lack of spatial inversion symmetry, naturally exhibit these splitting effects, particularly the Rashba effect, provided there is sufficient SOC strength. The primary methods to enhance these splitting effects in 2D Janus materials include: (1) Strain engineering can enhance Rashba and Dresselhaus splitting by altering the lattice structure and SOC strength. (2) Heterostructure building enhances Rashba and Dresselhaus splitting through interfacial effects and interlayer interactions. (3) Elemental doping enhances Rashba splitting by introducing heavy atoms or highly electronegative elements to boost SOC. (4) External electric and magnetic fields affect Rashba, Dresselhaus, and Zeeman splitting by modulating electron distribution and energy levels. These findings demonstrate that 2D Janus materials possess a rich variety of SOC-induced spin splitting, offering great promise for carrier regulation and transport in advanced spintronic nano-devices.

## V. CONCLUSIONS AND PERSPECTIVES

### A. Conclusions

In conclusion, this review has comprehensively delved into the optical, catalytic, electrochemical, thermoelectric, piezoelectric, magnetic, electronic, and valleytronic properties and applications of 2D Janus materials with experimental H- and T-structures [Fig. 2(a)]. These materials, distinguished by their broken spatial symmetry and varied chemical compositions, possess a broad spectrum of tunable electronic configurations that make them highly suitable for a multitude of applications. They demonstrate unique optical responses, enhanced catalytic efficiency, favorable thermoelectric and piezoelectric properties, and significant electrochemical performance. Furthermore, their magnetic properties, including magnetic anisotropy, DMI, and spin and valley splitting, present excellent opportunities for spintronics and quantum computing.

This review also synthesizes both experimental and theoretical advancements, along with their underlying mechanisms. This provides a robust framework and a solid foundation for designing, regulating, and applying 2D Janus family. Furthermore, the ability to manipulate their versatile properties through strategies like strain, electric field, and heterojunction construction significantly broadens their potential application landscape. Overall, the unique asymmetric structure, and highly tunable properties of 2D Janus materials position them as pivotal candidates for addressing the pressing demands in next-generation energy and information technologies.

## B. Perspectives

Research on 2D Janus materials presents several challenges and opportunities. (1) Experimental investigations on these materials are still in the early stages, with only a limited number of experimentally prepared species. Thus, more 2D Janus materials need to be synthesized based on theoretically predictions, utilizing existing growth methods. In parallel, more experimental fabrication strategies should be developed. (2) Their properties are not yet fully characterized experimentally. Many features currently remain theoretical, and require experimental validation. A combination of experimental and theoretical approaches is encouraged to better understand the principles governing these materials and guide future experiment and application. (3) There are structural limitations, as H- and T-phases and their derivatives are the main experimentally explored structures for 2D Janus family, although diverse 2D Janus phases have been unveiled in theory. The experimental scope needs to be expanded to encompass a wider variety of structural phases and chemical components. In addition to chalcogens, elements from the halogen, carbon, and nitrogen groups, should also be incorporated into the material design and device applications. (4) Despite their vast properties in academic research, 2D Janus materials have received limited attention in industrial and commercial sectors. Given their potential in addressing energy, environment, and information issues in the post-Moore era, more practical applications should be developed.

Overall, 2D Janus materials hold immense promise due to their wide-ranging physical, chemical, and biological properties. They offer opportunities in advanced materials science and manipulation technologies. However, research in this field is still in its infancy. Unlocking the full potential of 2D Janus materials will require sustained and interdisciplinary research efforts in the years to come.

## SUPPLEMENTARY MATERIAL

See the supplementary material for supporting reviews on 2D Janus in optics, catalysis, thermoelectricity, piezoelectricity, electrochemistry, magnetic anisotropy, magnetic state, Dzyaloshinskii-Moriya interaction, and spin and valley.

## ACKNOWLEDGMENTS

We express our sincere gratitude to Prof. Zheng-Wen Fu from Shanghai Key Laboratory of Molecular Catalysts and Innovative Materials, Department of Chemistry, Fudan University, for his insightful discussions on 2D CrSSe.

Guoying Gao acknowledges support from the National Natural Science Foundation of China (Grant No. 12174127).

## AUTHOR DECLARATIONS

### Conflict of Interest

The authors have no conflicts to disclose.

### Author Contributions

**Long Zhang:** Conceptualization (equal); Data curation (equal); Investigation (equal); Supervision (equal); Writing - original draft (equal). **Ziqi Ren:** Formal analysis (equal); Visualization (equal). **Li Sun:** Methodology (equal); Visualization (equal). **Yihua Gao:** Supervision (equal); Validation (equal). **Deli Wang:** Visualization (equal); Writing - review & editing (equal). **Junjie He:** Formal analysis (equal); Validation (equal). **Guoying Gao:** Conceptualization (equal); Funding acquisition (equal); Project administration (equal); Resources (equal); Supervision (equal); Writing - review & editing (equal).

## DATA AVAILABILITY

The data that support the findings of this study are available from the corresponding author upon reasonable request.

# Figures and Tables

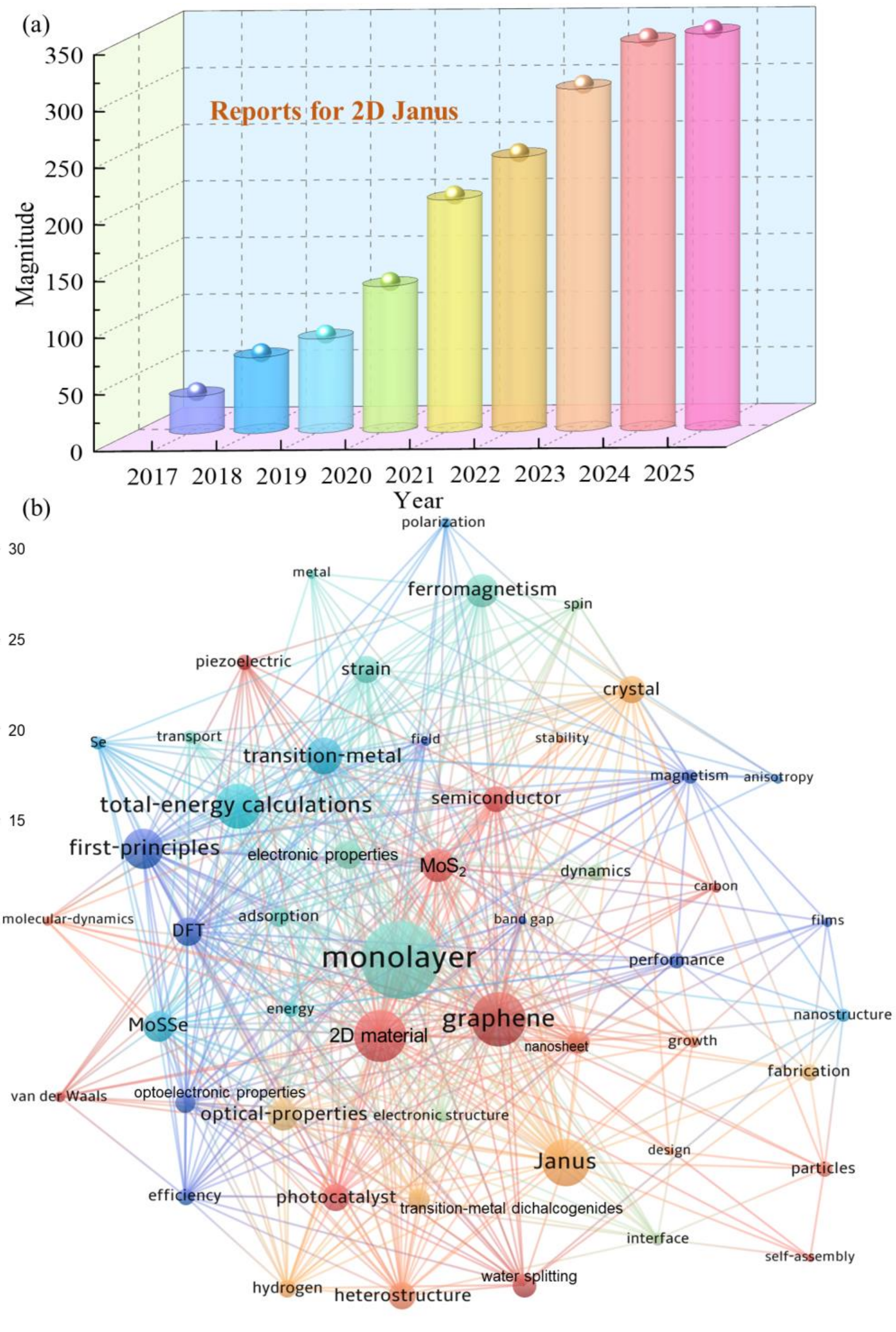

**FIG. 1.** The number of annual reports for 2D Janus (a). The overlay visualization of the co-occurrence network of keywords in view of both “Janus” and “2D” (b).

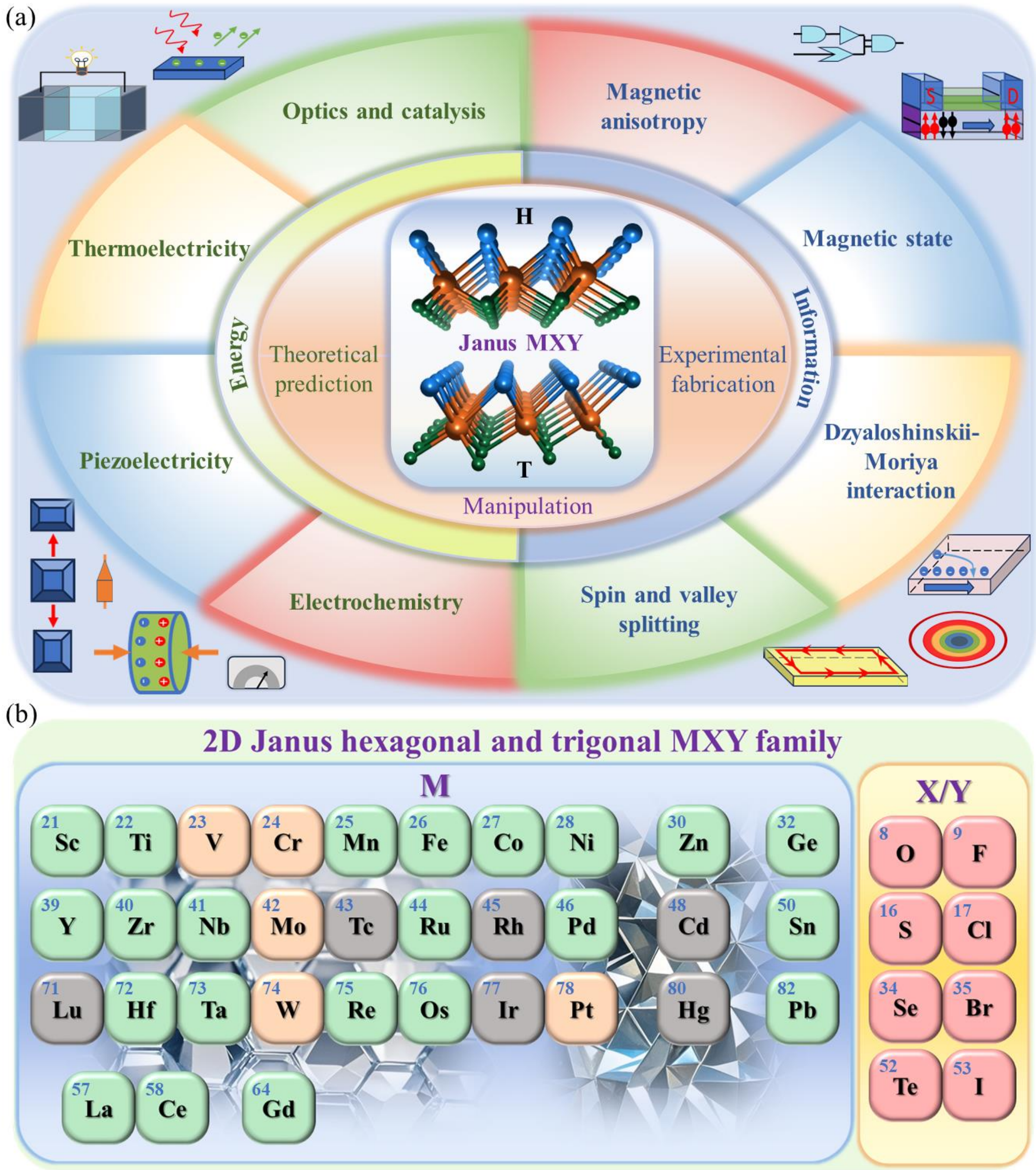

**FIG. 2.** The overview of theoretical and experimental investigation and manipulation (a), and the diagrammatic keyboard (b) for 2D Janus H- and T-MXY structures.

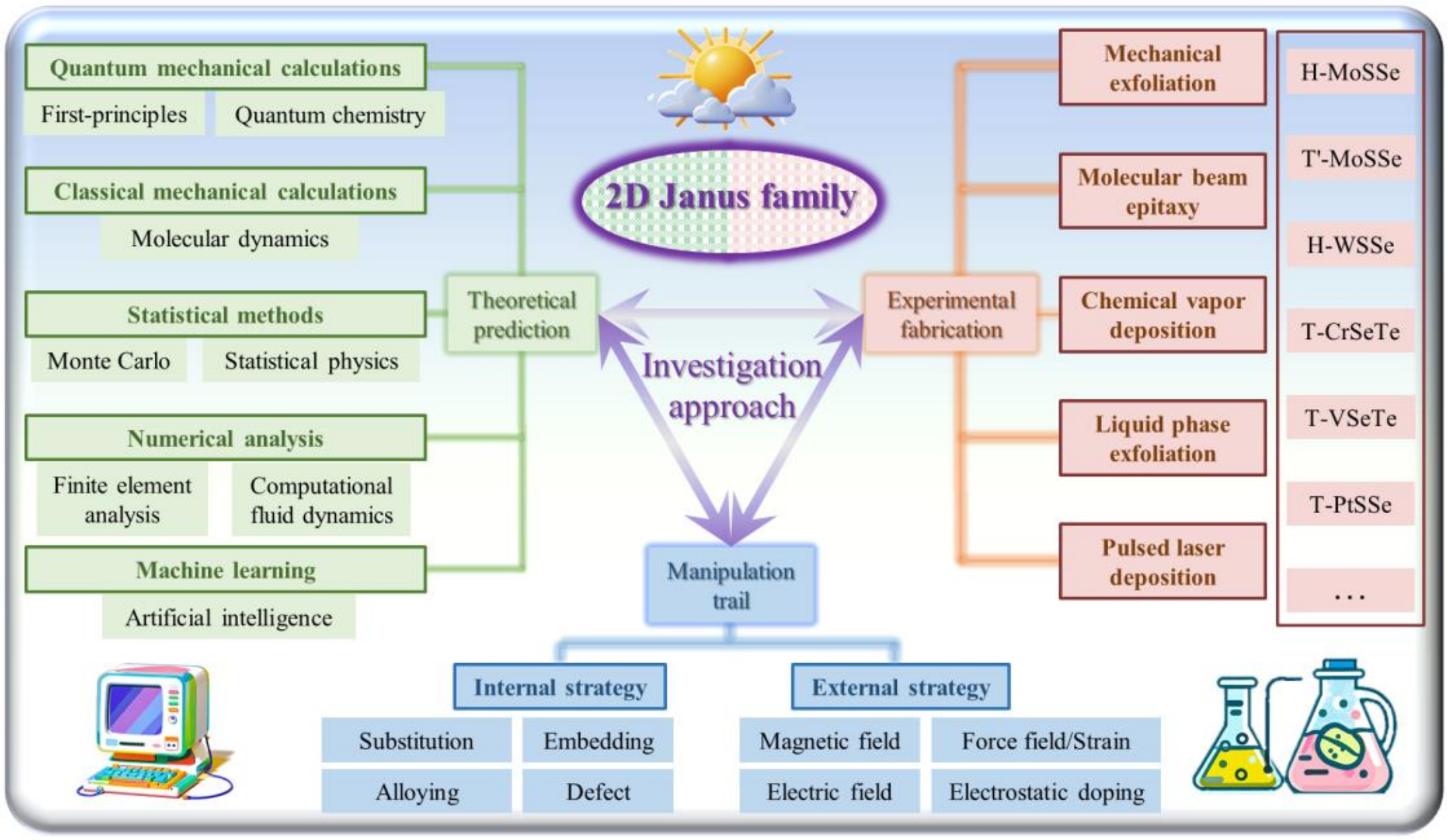

**FIG. 3.** The theoretical prediction, experimental fabrication, and manipulation trail for 2D Janus family.

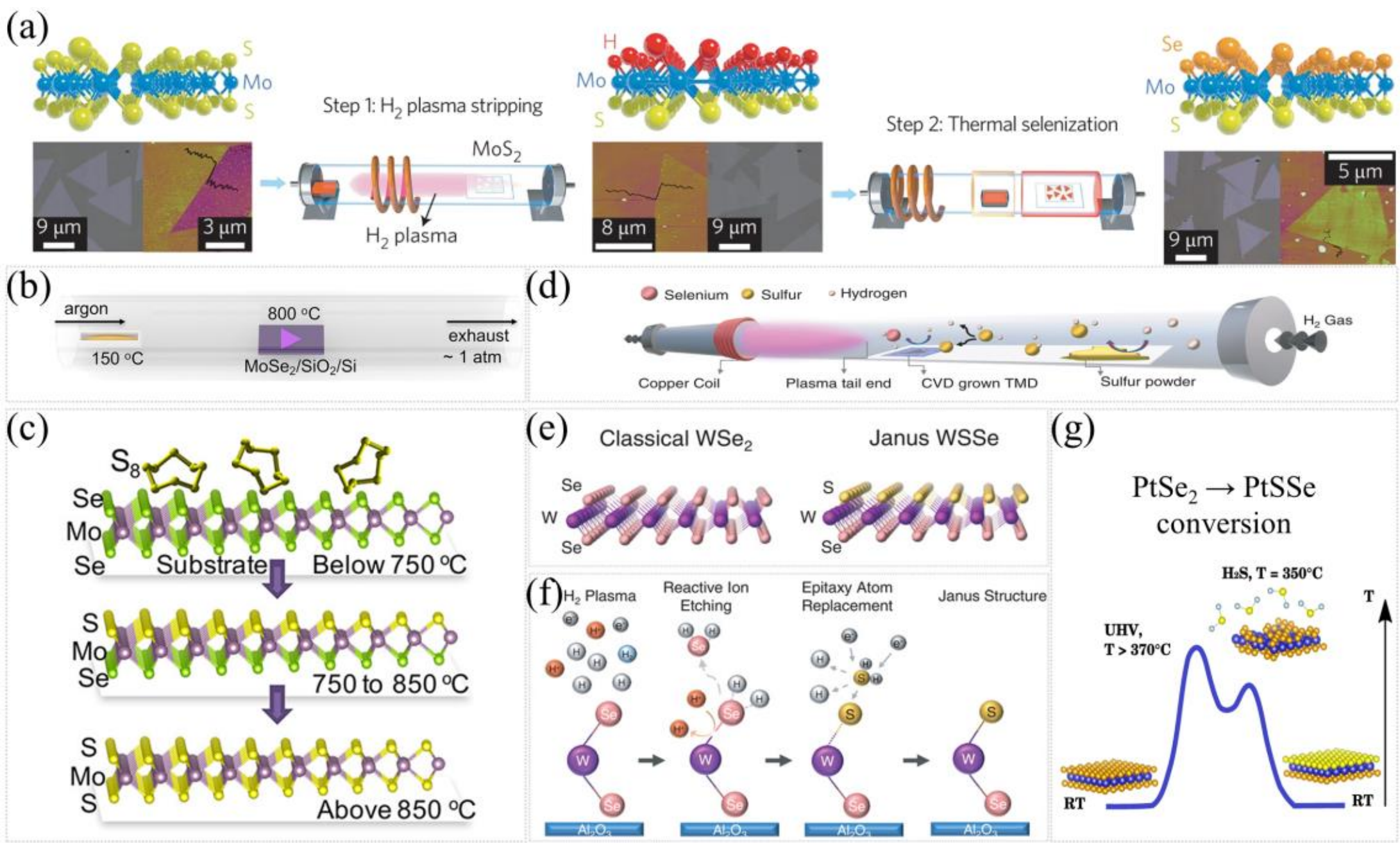

**FIG. 4.** Synthesis of the Janus H-MoSSe monolayer (a).[28] Schematic illustration of the reaction setup (b) and proposed reaction mechanism for the sulfurization of monolayer $MoSe_2$ on $SiO_2$/Si substrate at different temperatures (c).[29] Schematic demonstration of the selective epitaxy atomic replacement (SEAR) process through inductively coupled plasma (d), the crystal structure of $WSe_2$ monolayer and Janus H-WSSe monolayer (e), and working scheme of room temperature (RT) SEAR process (f).[32] The conversion of the pristine $PtSe_2$ into a Janus T-PtSSe material (g).[34] (a) Readapted with permission from Lu *et al.*, Nature Nanotech. **12**, 744 (2017). Copyright 2017 Spring Nature. (b) and (c) Readapted with permission from Zhang *et al.*, ACS Nano **11**(8), 8192 (2017). Copyright 2017 American Chemical Society. (d)-(f) Readapted with permission from Trivedi *et al.*, Adv. Mater. **32**(50), 2006320 (2020). Copyright 2020 John Wiley and Sons. (g) Readapted with permission from Sant *et al.*, npj 2D Mater. Appl. **4**, 41 (2020). Copyright 2020 Author(s); licensed under a Creative Commons Attribution (CC BY) license.

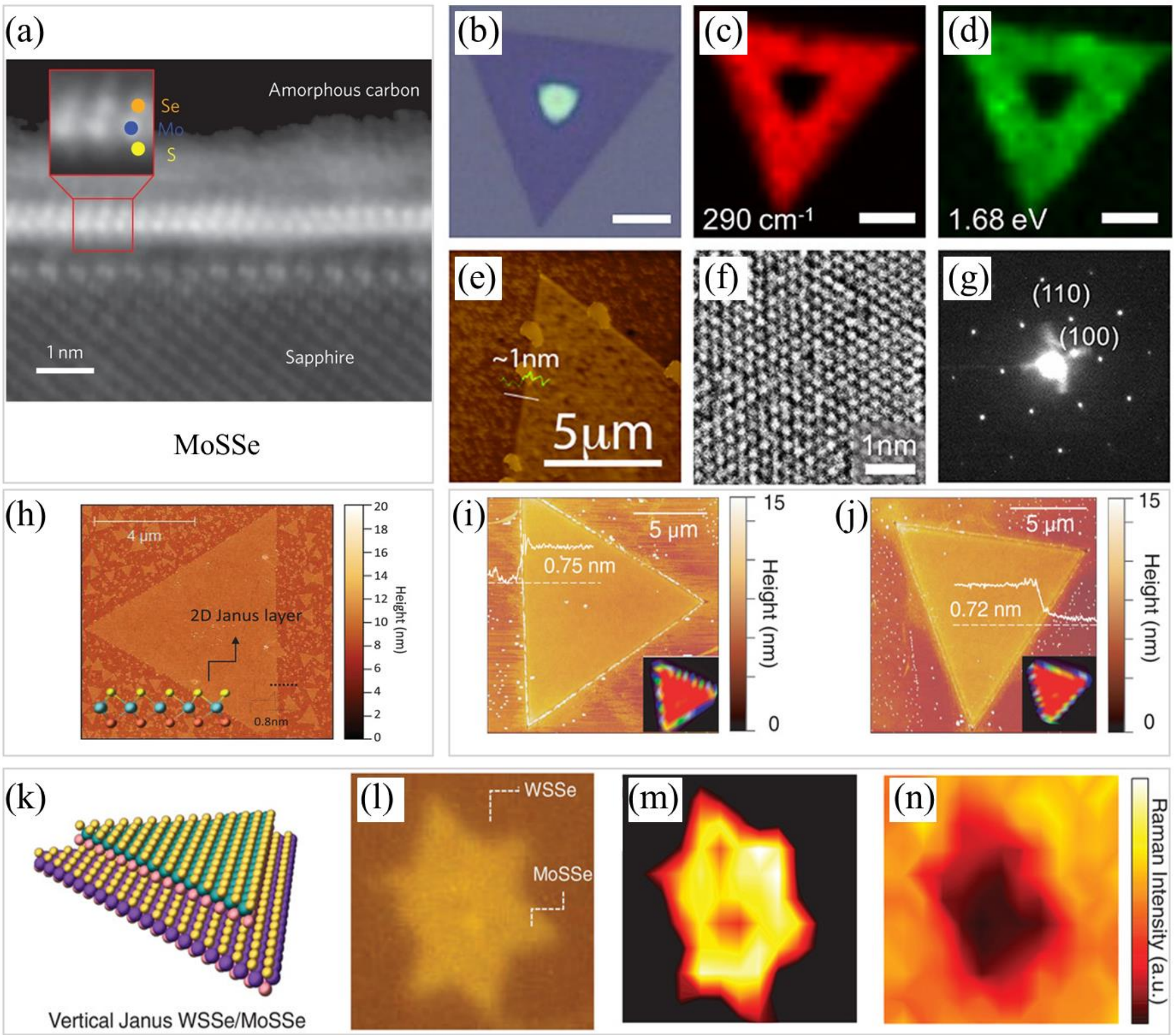

**FIG. 5.** Annular dark-field scanning transmission electron microscopy image of the sample cross-section, showing the asymmetric Janus H-MoSSe monolayer structure with Se (orange) on top and S (yellow) at the bottom of the Mo atoms (blue) (a).[28] Optical image of a Janus H-MoSSe triangle, where the purple and the central island with high contrast is the monolayer and bulk crystal region, respectively (b); Raman (c) and photoluminescence (PL) (d) peak intensity mappings of the Janus H-MoSSe triangle, the mapping shows uniform distribution of the identical Raman peak at 287 $cm^{-1}$ and PL peak at 1.68 eV; atomic force microscopy topography image of the Janus H-MoSSe triangle (e) and the profile shows that the thickness of the flake is < 1 nm; high resolution Transmission electron microscope (HRTEM) image of the Janus H-MoSSe lattice (f); the atom arrangement indicates the 2H structure of the monolayer, and the corresponding selected area electron diffraction pattern of the monolayer (g).[29] Atomic force microscopy image of 2D Janus H-WSSe monolayer (h).[31] Atomic force microscopy profile of $WSe_2$ before (i) and after (j) the SEAR process, and the atomic representation (k) and the optical image (l) of vertical Janus H-MoSSe/WSSe heterostructure, and Raman mapping of MoSSe at 290 $cm^{-1}$ (m) and WSSe peak at 284 $cm^{-1}$ (n).[32] (a) Readapted with permission from Lu *et al.*, Nature Nanotech. **12**, 744 (2017). Copyright 2017 Spring Nature. (b)-(g) Readapted with permission from Zhang *et al.*, ACS Nano **11**(8), 8192 (2017). Copyright 2017 American Chemical Society. (h) Readapted with permission from Li *et al.*, Adv. Mater. **32**(33), 2002401 (2020). Copyright 2020 John Wiley and Sons. (i)-(n) Readapted with permission from Trivedi *et al.*, Adv. Mater. **32**(50), 2006320 (2020). Copyright 2020 John Wiley and Sons.

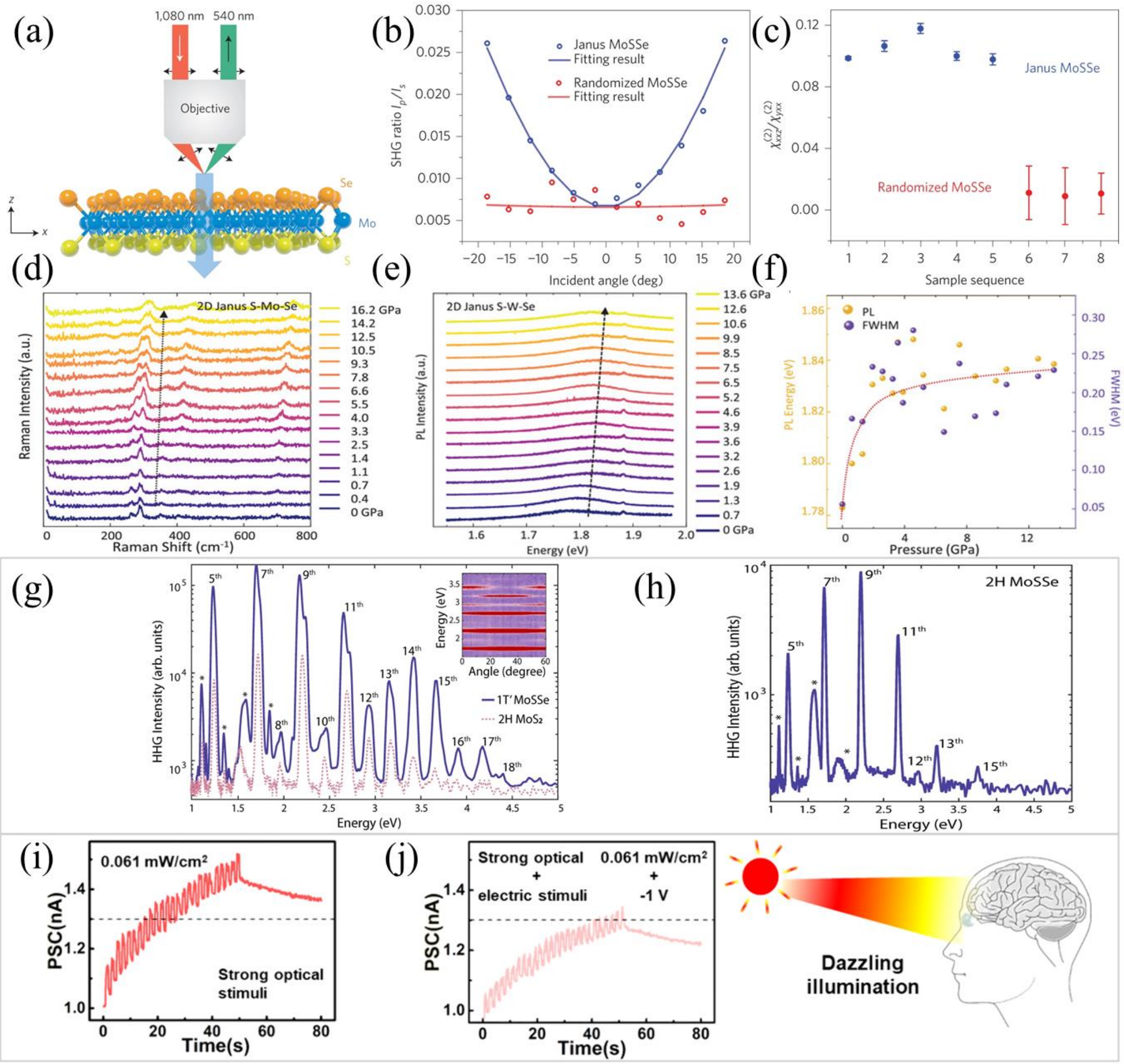

**FIG. 6.** Schematics of out-of-plane induced second harmonic generation (SHG) (a), angle-dependent SHG intensity ratio between p and s polarization in Janus H-MoSSe and randomized alloy samples (b) and second-order susceptibility ratio statistics (c).[28] Collected Raman spectra for 2D Janus MoSSe monolayers from low to high pressures during compression (d), optical properties and band renormalization under high pressure (e), and PL spectra collected at different pressures for Janus H-WSSe (f).[31] High harmonic generation (HHG) spectrum of Janus T'-MoSSe is over an order of magnitude stronger than the HHG from macroscopic monolayer H-$MoS_2$ (g), where the cancellation of a few orders at some polarization angles indicates the signal is generated from a single flake, and the HHG spectrum of Janus H-MoSSe taken under the same conditions (h).[30] Simulation of light adaptation of human visual system under excessively powerful optical intensity (0.061 mW $cm^{-2}$), the current could increase higher than the threshold (>1.3 nA), which mimics the injury of biological eyes exposed to harsh light (i), and the artificial retina could adapt to the strong light intensity through the modulation of negative voltage pulses (−1 V) and the current fell below the threshold (<1.3 nA) (j) based on Janus H-MoSSe.[202] (a)-(c) Readapted with permission from Lu *et al.*, Nature Nanotech. **12**, 744 (2017). Copyright 2017 Spring Nature. (d)-(f) Readapted with permission from Li *et al.*, Adv. Mater. **32**(33), 2002401 (2020). Copyright 2020 John Wiley and Sons. (g) and (h) Readapted with permission from Shi *et al.*, Nat. Commun. **14**, 4953 (2023). Copyright 2023 Author(s); licensed under a Creative Commons Attribution (CC BY) license. (i) and (j) Readapted with permission from Meng *et al.*, Nano Lett. **22**(1), 81-89 (2022). Copyright 2022 American Chemical Society.

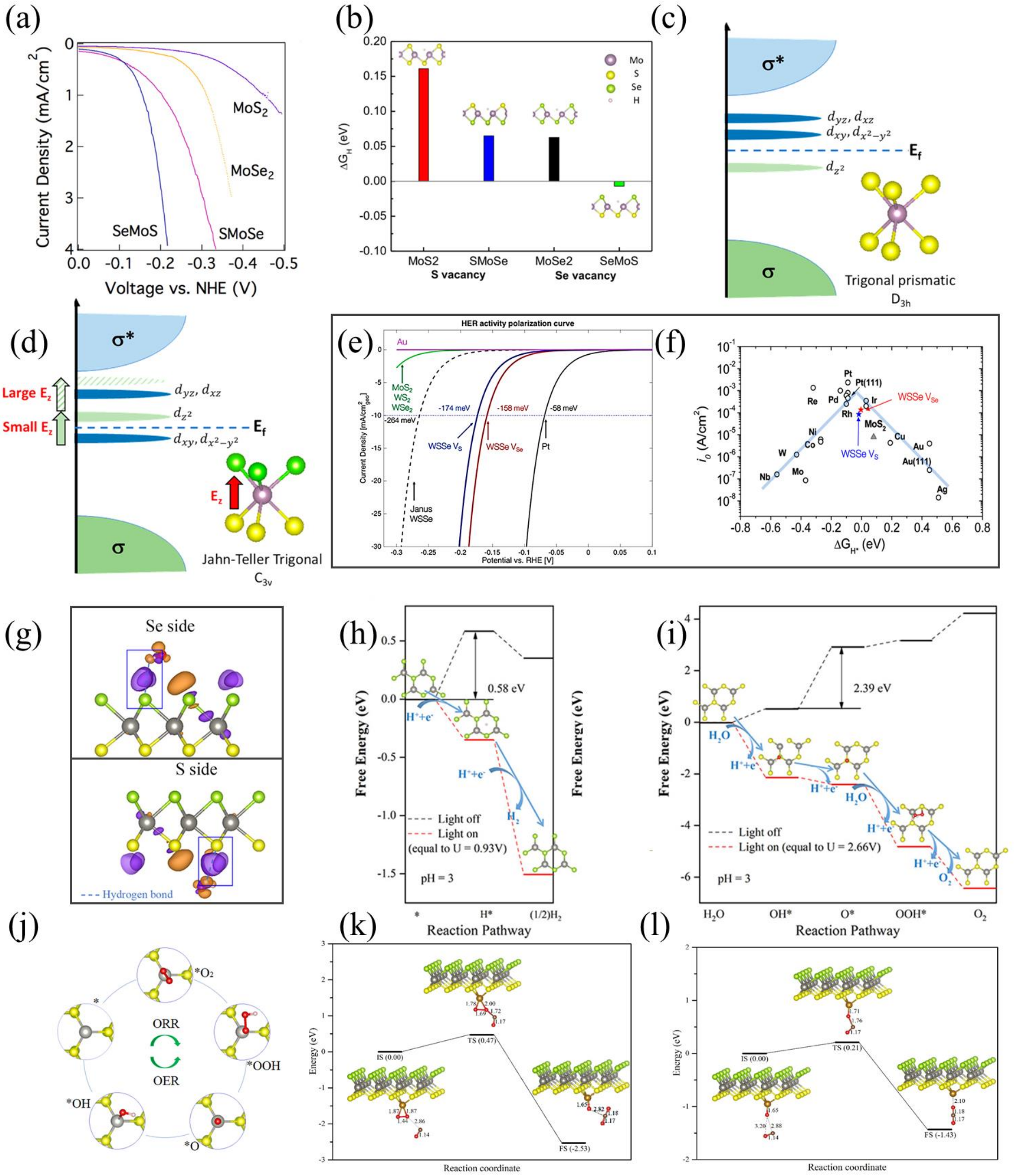

**FIG. 7.** Hydrogen evolution reaction (HER) polarization curves of $MoS_2$, $MoSe_2$, Janus H-SMoSe and SeMoS (a), and hydrogen adsorption free energy for $MoS_2$ and SMoSe with S vacancy and for $MoSe_2$ and SeMoS with Se vacancy (b).[29] Conventional H with prismatic (c) and Janus (d) structures. Current density (e) and HER volcano curve (f) of Janus H-WSSe with S and Se vacancies versus other HER catalysts.[159] Charge density difference for a $H_2O$ molecule adsorbed on the Se and S sides (g), and free energy steps of HER (h) and oxidation evolution reaction (OER) (i) on Janus H-WSSe.[200] Schemes of oxygen reduction reaction (ORR) and OER on anchored Pd and Pt Janus H-MoSSe (j).[211] The potential energy diagrams and configurations for Fe-WSSe monolayer with CO + $O_2$ → OOCO → $O_{ads}$ + $CO_2$ (k) and CO + $O_{ads}$ → $CO_2$ (l), where all bond lengths are in Å.[213] (a) and (b) Readapted with permission from Zhang *et al.*, ACS Nano **11**(8), 8192 (2017). Copyright 2017 American Chemical Society. (c)-(f) Readapted with permission from Er *et al.*, Nano Lett. **18**(6), 3943 (2018). Copyright 2018 American Chemical Society. (g)-(i) Readapted with permission from Ju *et al.*, ACS Appl. Mater. Inter. **12**(26), 29335 (2020). Copyright 2020 American Chemical Society. (j) Readapted with permission from Guo *et al.*, J. Colloid Interface Sci. **616**, 177 (2022). Copyright 2022 Elsevier. (k) and (l) Readapted with permission from Guo *et al.*, Appl. Surf. Sci. **565**, 150558 (2021). Copyright 2021 Elsevier.

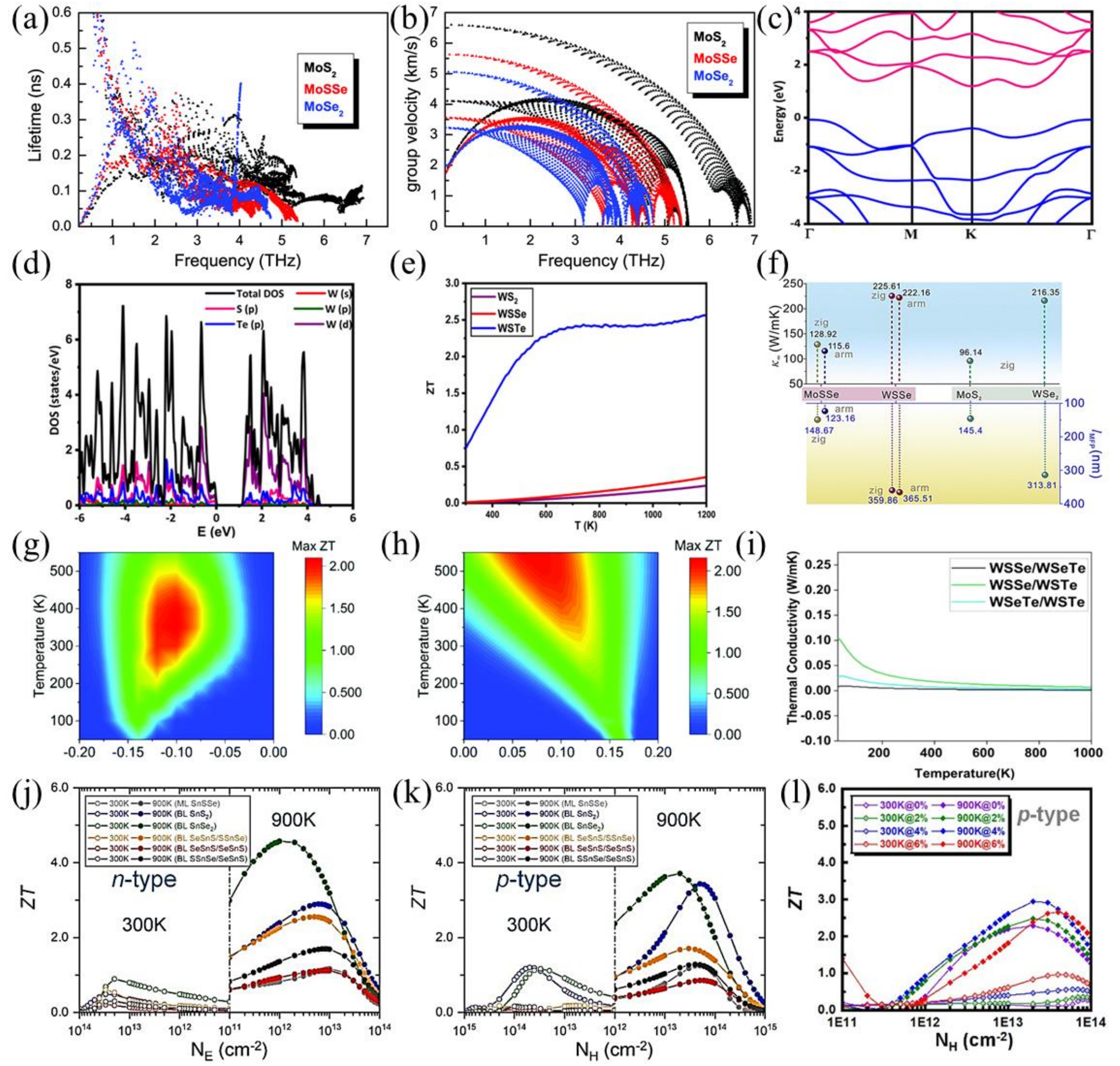

**FIG. 8.** Phonon mode lifetimes (a) and group velocities (b) of H-$MoS_2$ (black), Janus MoSSe (red) and $MoSe_2$ (blue) monolayers for the out-of-plane acoustic (ZA) (square symbol), transversal acoustic (TA) (circle symbol) and longitudinal acoustic (LA) (triangle symbol) branches.[240] Band structures (c) and density of states (DOS) (d) of Janus H-WSTe, and figure of merit (ZT) for H-$WS_2$, Janus WSSe and WSTe (e).[242] The infinitely long sample $\kappa_\infty$ and phonon mean free path $l_{MFP}$ of Janus H-MoSSe/WSSe and $MoS_2$/$WSe_2$ (f).[241] Max ZT of symmetric armchair Janus H-MoSSe nanoribbon (g) and graphene/MoSSe heterostructure nanoribbon (h).[43] Thermal conductivity of Janus H-WSeTe/WSTe, WSSe/WSeTe and WSSe/WSTe heterostructures (i).[245] The concentration dependence of the ZT for n-type (j) and p-type (k) pristine and Janus T-SnXY (X, Y = S, Se) monolayers and bilayers.[238] The ZT for strain-tunable p-type Janus T-PtSSe monolayer as a function of carrier concentration at 300 and 900 K (l).[251] (a) and (b) Readapted with permission from Guo, Phys. Chem. Chem. Phys. **20**(10), 7236 (2018). Copyright 2018 Royal Society of Chemistry. (c)-(e) Readapted with permission from Patel *et al.*, ACS Appl. Mater. Inter. **12**(41), 46212 (2020). Copyright 2020 Author(s); licensed under a Creative Commons Attribution (CC BY) license. (f) Readapted with permission from Qin *et al.*, Phys. Chem. Chem. Phys. **24**(34), 20437 (2022). Copyright 2022 Royal Society of Chemistry. (g) and (h) Readapted with permission from Deng *et al.*, Phys. Chem. Chem. Phys. **21**(33), 18161 (2019). Copyright 2019 Royal Society of Chemistry. (i) Readapted with permission from Sharma *et al.*, Nanoscale **16**(6), 3091-3100 (2024). Copyright 2024 Royal Society of Chemistry. (j) and (k) Readapted with permission from Bai et al., Chem. Eng. J. **455**, 140832 (2023). Copyright 2023 Elsevier. (l) Readapted with permission from Bai et al., Appl. Surf. Sci. **599**, 153962 (2022). Copyright 2022 Elsevier.

**Table I.** The reported values for figure of merit (ZT) in 2D Janus materials, which are monolayers unless specifically noted.

| 2D Janus | Condition | ZT | Refs. |
|---|---|---|---|
| H-MoSSe nanoribbon | 300 K | 1.64 | [43] |
| H-MoSSe/graphene nanoribbon | 300 K | 2.01 | [43] |
| H-WSSe | 600 K | 0.32 | [157] |
| H-WSSe | 900 K | 0.50 | [157] |
| H-WSTe | 300 K | 0.74 | [242] |
| H-WSTe | 600 K | ~2.25 | [242] |
| H-WSeTe | 800 K | 1.53 | [243] |
| H-WSeTe/MoSSe heterostructure | 300 K, -3 % strain | 1.62 | [243] |
| T-TiSSe | 300 K, 4 % strain | 1.04 | [466] |
| T-TiSTe | 300 K, 6% strain | 0.95 | [466] |
| T-TiSeTe | 300 K, 10% strain | 0.88 | [466] |
| T-HfSSe | 300 K | 1.18 | [467] |
| T-HfSSe | 600 K | 3.24 | [467] |
| T-HfSSe bilayer | 300 K | 2.33 | [467] |
| T-HfSSe bilayer | 600 K | 5.54 | [467] |
| H-HfSTe | 300 K | 0.50 | [468] |
| H-HfSTe | 900 K | 1.98 | [468] |
| H-HfSeTe | 300 K | 0.51 | [468] |
| H-HfSeTe | 900 K | 2.57 | [468] |
| H-HfBrCl | 300 K | 0.88 | [469] |
| H-HfBrI | 300 K | 1.80 | [469] |
| H-HfClI | 300 K | 2.15 | [469] |
| T-ZrOS | 900 K | 0.82 | [470] |
| T-ZrSSe | 900 K, 6% strain | 4.88 | [250] |
| H-ZrSTe | 300 K | 0.64 | [468] |
| H-ZrSTe | 900 K | 2.72 | [468] |
| H-ZrSeTe | 300 K | 0.94 | [468] |
| H-ZrSeTe | 900 K | 3.63 | [468] |
| T-NiOS | 300 K | 0.90 | [471] |
| T-NiSSe | 300 K | 0.88 | [471] |
| T-PdSSe | 900 K | 0.66 | [246] |
| T-PdSTe | 900 K | 0.58 | [246] |
| T-PdSeTe | 900 K | 0.65 | [246] |
| T-PtSeTe | 300 K | 0.91 | [247] |
| T-PtSeTe | 900 K | 2.54 | [247] |
| T-SnSSe | 900 K | 1.20 | [251] |
| T-SnSSe | 900 K, 6% strain | 1.75 | [472] |
| T-SnSSe bilayer | 900 K | 2.55 | [238] |
| T-PbSSe | 900 K | 2.99 | [251] |
| T-PbSSe | 900 K, 4% strain | 3.77 | [251] |

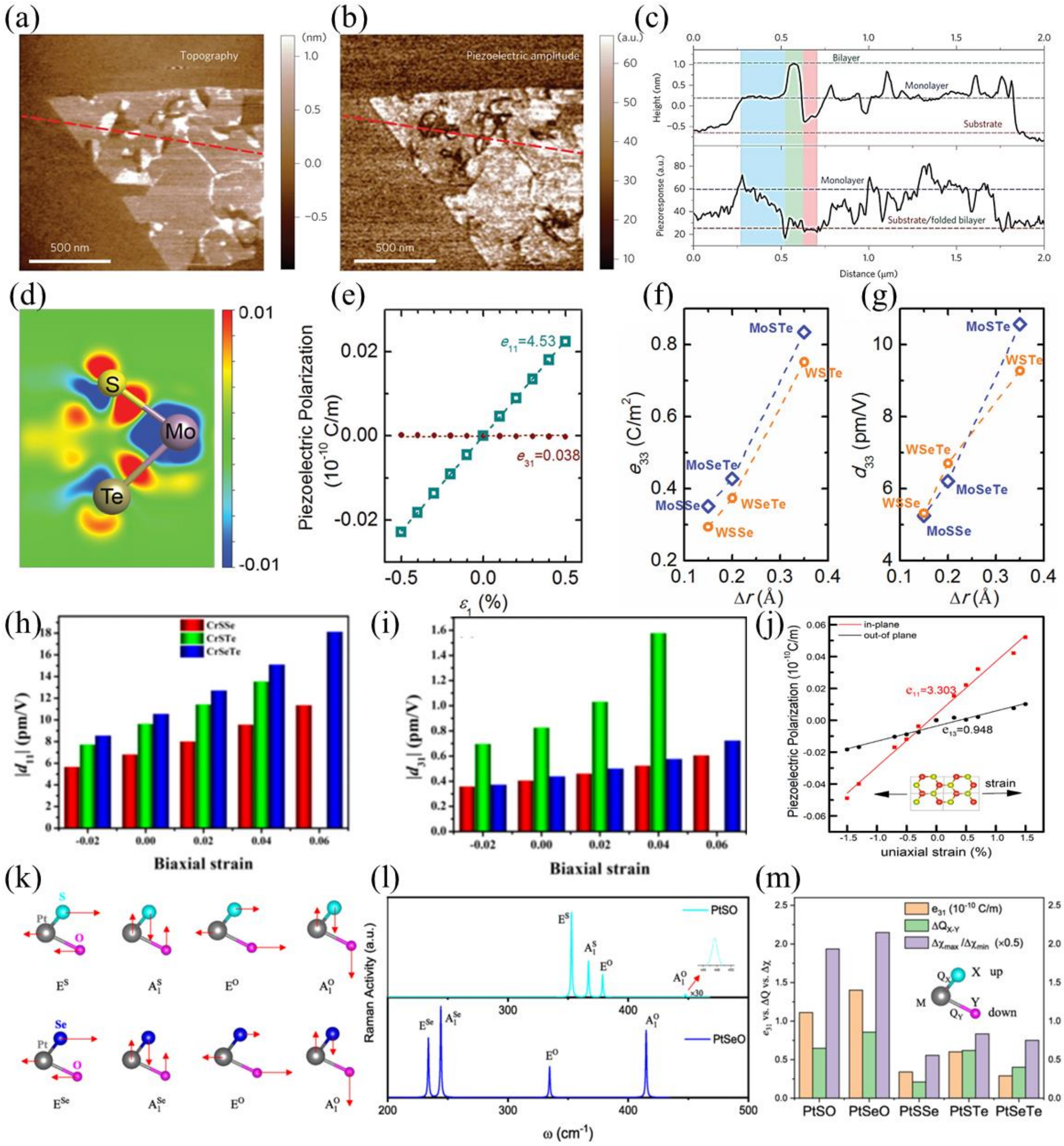

**FIG. 9.** Topography (a) and piezoelectric amplitude (b) of an isolated Janus H-MoSSe monolayer directly grown on highly oriented pyrolytic graphite (HOPG), measured by resonance-enhanced piezo-response force microscopy, and height and piezo-response of Janus monolayer and backfolded bilayer (c).[28] Bonding charge density of Janus H-MoSTe monolayer with unit of e $Bohr^{-3}$ (d), and piezoelectric polarization of Janus H-MoSTe monolayer (e) and MXY (M = Mo, W; X ≠ Y = S, Se and Te) multilayers (f,g).[259] In-plane (h) and out-of-plane (i) piezoelectric stress coefficient of the Janus H-CrXY monolayer.[261] Linear changes in the piezoelectric polarization of Janus H-VSSe monolayer (j).[46] Atoms vibrations (k) and Raman spectrum (l) for Janus T-PtOS and PtOSe, the Bader charge difference, electronegativity difference ratio and out-of-plane piezoelectricity of Janus T-platinum dichalcogenides (m).[269] (a)-(c) Readapted with permission from Lu *et al.*, Nature Nanotech. **12**, 744 (2017). Copyright 2017 Spring Nature. (d)-(g) Readapted with permission from Dong *et al.*, ACS Nano **11**(8), 8242 (2017). Copyright 2017 American Chemical Society. (h) and (i) Readapted with permission from Chen *et al.*, Phys. Rev. B **106**(11), 115307 (2022). Copyright 2022 American Physical Society. (j) Readapted with permission from Zhang *et al.*, Nano Lett. **19**(2), 1366 (2019). Copyright 2019 American Chemical Society. (k)-(m) Readapted with permission from Zhang *et al.*, Phys. Rev. B **108**(3), 035411 (2023). Copyright 2023 American Physical Society.

**Table II.** The absolute values of the piezoelectric strain coefficients with parallel and perpendicular directions of strain (stress) and electric polarization (electric field) in 2D Janus materials, which are monolayers unless specifically noted.

| 2D Janus | Parallel (pm $V^{-1}$) | Perpendicular (pm $V^{-1}$) | Refs. |
|---|---|---|---|
| H-MoSSe | $d_{11}$ = 3.76 | $d_{31}$ = 0.02 | [259] |
| Janus H-MoSSe/BP heterostructure | $d_{33}$ = 14.91 | - | [197] |
| Janus H-MoSSe/BAs heterostructure | $d_{33}$ = 7.63 | - | [197] |
| H-MoSTe | $d_{11}$ = 5.04 (5.10) | $d_{31}$ = 0.03, $d_{13}$ = 0.40 | [259,262] |
| T'-MoSTe | - | $d_{14}$ = 17.80 | [262] |
| H-MoSeTe | $d_{11}$ = 5.30 | $d_{31}$ = 0.03 | [259] |
| H-WSSe | $d_{11}$ = 2.26 | $d_{31}$ = 0.01 | [259] |
| H-WSTe | $d_{11}$ = 3.33 | $d_{31}$ = 0.01 | [259] |
| H-MoSeTe/WSTe heterostructure | $d_{33}$ = 13.91 | $d_{31}$ = 0.26 | [266] |
| H-WSeTe | $d_{11}$ = 3.52 | $d_{31}$ = 0.01 | [259] |
| H-CrSSe | $d_{11}$ = 6.79 (11.34, 6 % strain) | $d_{31}$ = 0.40 (0.61, 6 % strain) | [261] |
| H-CrSTe | $d_{11}$ = 9.62 (13.53, 4 % strain) | $d_{31}$ = 0.83 (1.58, 4 % strain) | [261] |
| H-CrSeTe | $d_{11}$ = 10.53 (18.11, 6 % strain) | $d_{31}$ = 0.44 (0.72, 6 % strain) | [261] |
| H-VSSe | $d_{11}$ = 2.30 | - | [267] |
| H-VClBr | $d_{11}$ = 7.98 | $d_{31}$ = 0.34 | [272] |
| T-PtOS | $d_{11}$ = 4.30 | $d_{31}$ = 0.92 | [269] |
| T-PtOSe | $d_{11}$ = 8.80 | $d_{31}$ = 1.54 | [269] |
| T-PtSSe | $d_{11}$ = 1.17 | $d_{31}$ = 0.35 | [269] |
| H-ScClI | $d_{11}$ = 7.39 | $d_{31}$ = 1.14 | [273] |
| H-YBrI | $d_{11}$ = 5.61 (11.14, 6 % strain) | $d_{31}$ = ~0.10 | [274] |
| H-TiClI | $d_{11}$ = 4.41 | $d_{31}$ = 1.63 | [275] |
| H-TiBrI | $d_{11}$ = 4.58 | $d_{31}$ = 1.05 | [275] |
| T-ZrSTe | $d_{22}$ = 14.58 | $d_{31}$ < 0.01 (0.004) | [270] |
| T-HfSTe | $d_{22}$ = 11.64 | $d_{31}$ = 0.41 | [270] |
| T-HfSeTe | $d_{22}$ = 12.86 | $d_{31}$ = 0.18 | [270] |
| T-NiClI | $d_{11}$ = 5.21 | $d_{31}$ = 1.89 | [276] |
| T-ZnBrI | - | $d_{31}$ = 0.40 | [277] |
| T-GeSSe | $d_{33}$ = 0.59 (5.16/4.70, 7 % armchair/biaxial strain) | $d_{15}$ = 7.90 | [260] |
| T-GeSSe bilayer | $d_{33}$ = 0.26 - 0.37 | - | [260] |
| T-SnSSe | $d_{11}$ = 2.20 | $d_{31}$ = 0.11 | [260,271] |
| T-SnOSe | $d_{11}$ = 27.30 | $d_{31}$ = 0.50 | [271] |
| H-CeClBr | $d_{11}$ = 2.95 (4.03, 6 % strain) | - | [278] |

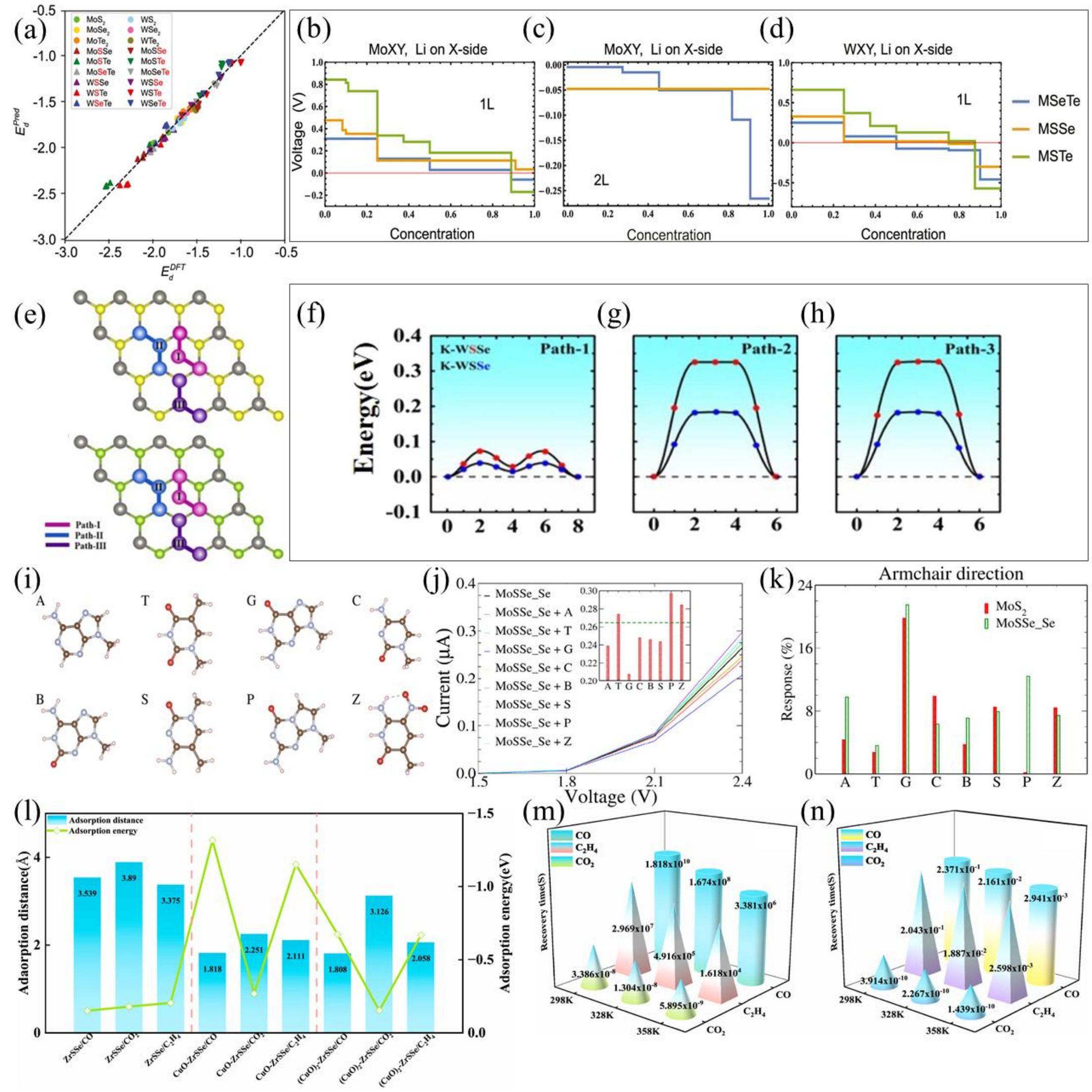

**FIG. 10.** The Li adsorption energies from clusterwise linear regression (CLR) vs DFT values for pristine and Janus H-molybdenum and tungsten dichalcogenides (a), and voltage profiles as a function of Li concentration on the Janus ones (b,c,d).[158] The migration paths (e) and diffusion barriers (f,g,h) for K on S and Se sides of Janus H-WSSe.[295] Molecules of Hachimoji natural and modified deoxyribonucleic acid (DNA) bases (i), and current-voltage curve (j) and sensitivity (k) of the Se side of Janus H-MoSSe in armchair direction.[307] Gas adsorption on various systems (l), and the recovery time for CuO-doped (m) and $(CuO)_2$-doped (n) Janus T-ZrSSe.[311] (a)-(d) Readapted with permission from Chaney *et al.*, ACS Appl. Mater. Inter. **13**(30), 36388 (2021). Copyright 2021 American Chemical Society. (e)-(h) Readapted with permission from Ahmad et al., Appl. Surf. Sci. **632**, 157545 (2023). Copyright 2023 Elsevier. (i)-(k) Readapted with permission from Badar *et al.*, ACS Appl. Mater. Inter. **16**(17), 21427 (2024). Copyright 2024 Author(s); licensed under a Creative Commons Attribution (CC BY) license. (l)-(n) Readapted with permission from Wu et al., Mater. Today Chem. **38**, 102038 (2024). Copyright 2024 Elsevier.

**Table III.** The differences of spin-orbital angular momentum matrix elements ($\left|\langle o^{+}|L_x|u^{+}\rangle\right|^2-\left|\langle o^{+}|L_z|u^{+}\rangle\right|^2$) and ($\left|\langle o^{-}|L_x|u^{+}\rangle\right|^2-\left|\langle o^{-}|L_z|u^{+}\rangle\right|^2$) for *p* orbitals between magnetization along *x/y* (in-plane) and *z* (out-of-plane) axes.

| $u^{+}$ | $o^{+}(o^{-})$ | | |
|---|---|---|---|
| | $p_y$ | $p_z$ | $p_x$ |
| $p_y$ | 0 | -1 | 1 |
| $p_z$ | -1 | 0 | 0 |
| $p_x$ | 1 | 0 | 0 |

**Table IV.** The differences of spin-orbital angular momentum matrix elements ($\left|\langle o^{+}|L_x|u^{-}\rangle\right|^2-\left|\langle o^{+}|L_z|u^{-}\rangle\right|^2$) and ($\left|\langle o^{-}|L_x|u^{-}\rangle\right|^2-\left|\langle o^{-}|L_z|u^{-}\rangle\right|^2$) for *d* orbitals between magnetization along *x*/y (in-plane) and *z* (out-of-plane) axes.

| $u^{-}$ | $o^{+}(o^{-})$ | | | | |
|---|---|---|---|---|---|
| | $d_{xy}$ | $d_{yz}$ | $d_{z^2}$ | $d_{xz}$ | $d_{x^2-y^2}$ |
| $d_{xy}$ | 0 | 0 | 0 | -1 | 4 |
| $d_{yz}$ | 0 | 0 | -3 | 1 | -1 |
| $d_{z^2}$ | 0 | -3 | 0 | 0 | 0 |
| $d_{xz}$ | -1 | 1 | 0 | 0 | 0 |
| $d_{x^2-y^2}$ | 4 | -1 | 0 | 0 | 0 |

**Table V.** The reported values for magnetic anisotropy energy (MAE = $E_{x/y}$ - $E_z$) in 2D Janus materials, which are monolayers unless specifically noted.

| 2D Janus | MAE (meV f.u.$^{-1}$) | Refs. |
|---|---|---|
| T-CrSeTe | -0.176 (1.110, 6 % strain) | [334] |
| T-CrSSe | ~1.400 | [335] |
| H-VSSe | ~-0.330 | [338] |
| H-VSTe | -1.890 | [375] |
| H-VSeTe | ~-1.160 (~1.950, 3 % strain) | [336,338] |
| T-VSTe/$Cr_2I_3Br_3$ heterostructure | ~0.013 - ~-0.133 | [340] |
| T-NbSeTe | -1.140 | [337] |
| T-MnSeTe | 0.389 (3.100, 5 % strain) | [343] |
| T-MnSTe | 0.374 (2.000, 5 % strain; 2.750, $E_f$ = -0.3 V/Å) | [343,473] |
| T-MnSSe | ~0.220 | [415] |
| T-MnSeTe/$In_2Se_3$ heterostructure | 1.970 | [395] |
| T-$In_2Se_3$/T-MnSeTe/$In_2Se_3$ heterostructure | 2.240 | [395] |
| T-ReSeTe | -3.721 (3.240, 2 % strain) | [337] |
| T-FeSSe | -0.590 | [474] |
| H-ScBrI | -0.228 (~0.350, 0.4 *h* doping) | [273] |
| H-YBrI | ~-0.310 (U=0) - ~-0.240 (U=3) | [274] |
| T-FeClI | 0.200 | [344] |
| T-FeBrI | -0.146 (~2.269, -6 % strain) | [344] |
| H-FeClF | -0.760 (0.172, -10 % strain) | [347] |
| H-RuClF | 0.187 | [377] |
| T-CoClBr | -0.542 (0.392, -2 % strain) | [379] |
| T-NiClI | -1.439 | [276] |
| H-LaBrI | -0.100 | [381] |
| H-CeClBr | 0.052 (~0.130, -4 % strain) | [278] |
| H-GdClF | 0.139 (~0.160, -2 % strain) | [475] |
| H-GdBrI | -0.420 (-0.210, bilayer) | [476,477] |

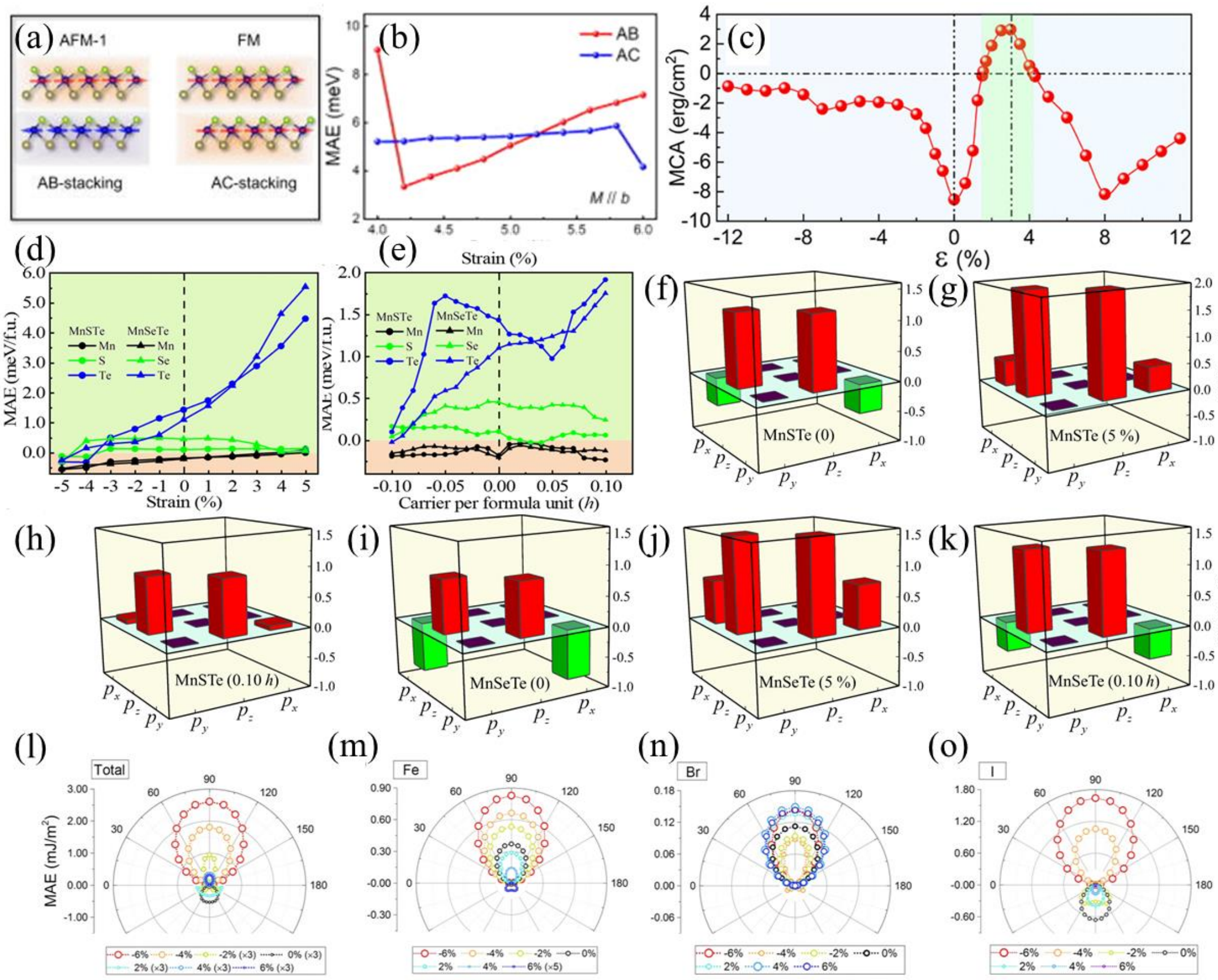

**FIG. 11.** The AB and AC stacking Janus T-CrSTe bilayer (a) and corresponding magnetic anisotropy energy (MAE) (b).[50] Change of magnetocrystalline anisotropy (MCA) energy of Janus H-VSeTe with strain (c).[336] The atom-resolved MAE with strain (d) and carrier doping (e) and Te-p intraorbital hybridization (f,g,h,i,j,k) of Janus T-MnSTe and MnSeTe.[343] Angular dependence of the total (l) and atom-resolved (m,n,o) MAE of Janus T-FeBrI.[344] (a) and (b) Readapted from Li *et al.*, Appl. Phys. Lett. **122**(12), 121902 (2023), with the permission of AIP Publishing. (c) Readapted with permission from Guan *et al.*, ACS Appl. Mater. Inter. **12**(47), 53067 (2020). Copyright 2020 American Chemical Society. (d)-(k) Readapted with permission from Zhang *et al.*, Nanoscale **15**(46), 18910 (2023). Copyright 2023 Royal Society of Chemistry. (l)-(o) Readapted with permission from Li *et al.*, ACS Appl. Mater. Inter. **13**(32), 38897 (2021). Copyright 2021 American Chemical Society.

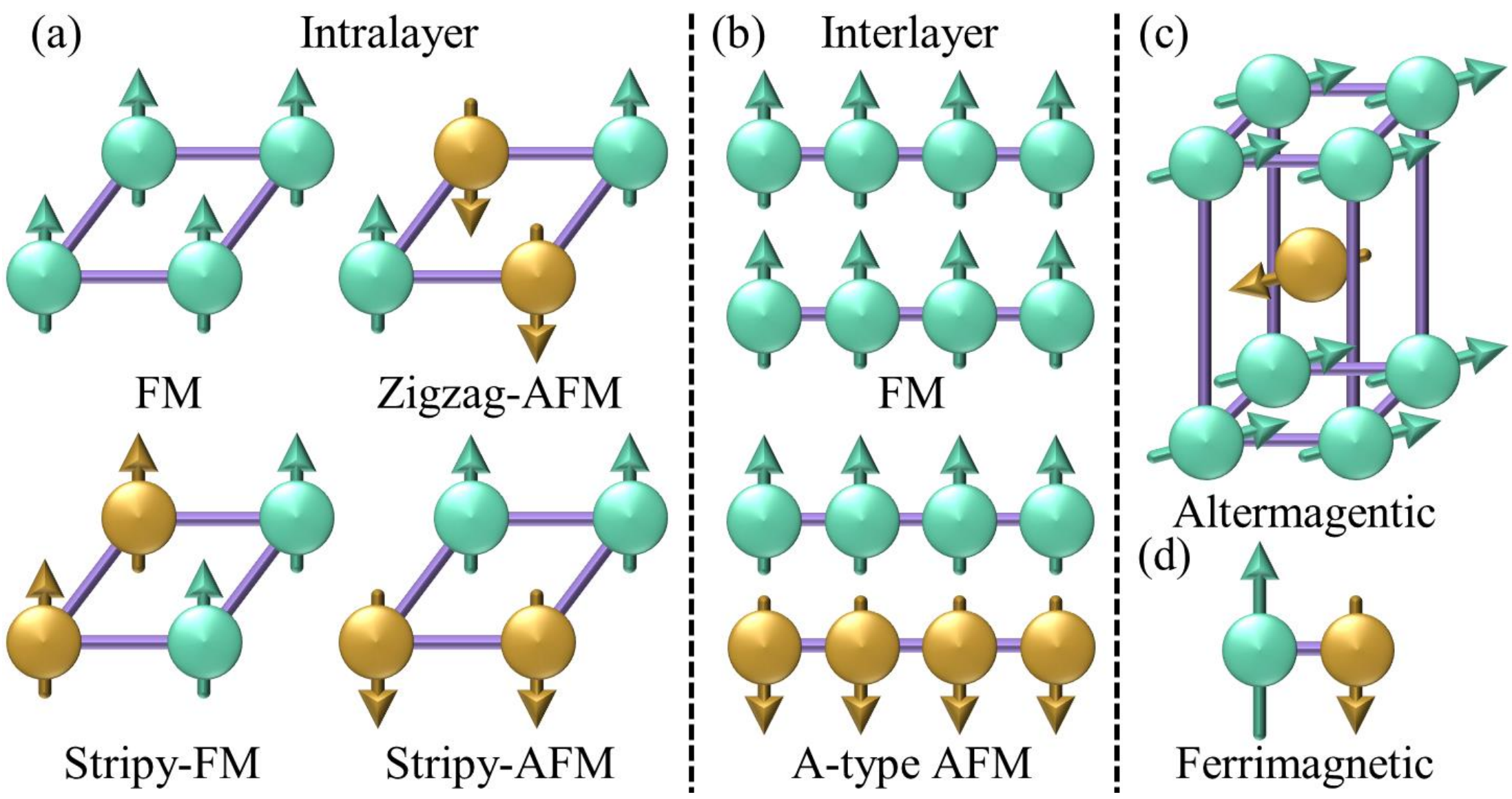

**FIG. 12.** The illustration of the intralayer FM, zigzag-AFM and stripy-FM (a) and the interlayer FM and A-type AFM configurations (b) for 2D Janus materials. The illustration of altermagnetic (c) and ferrimagnetic (d) configurations.

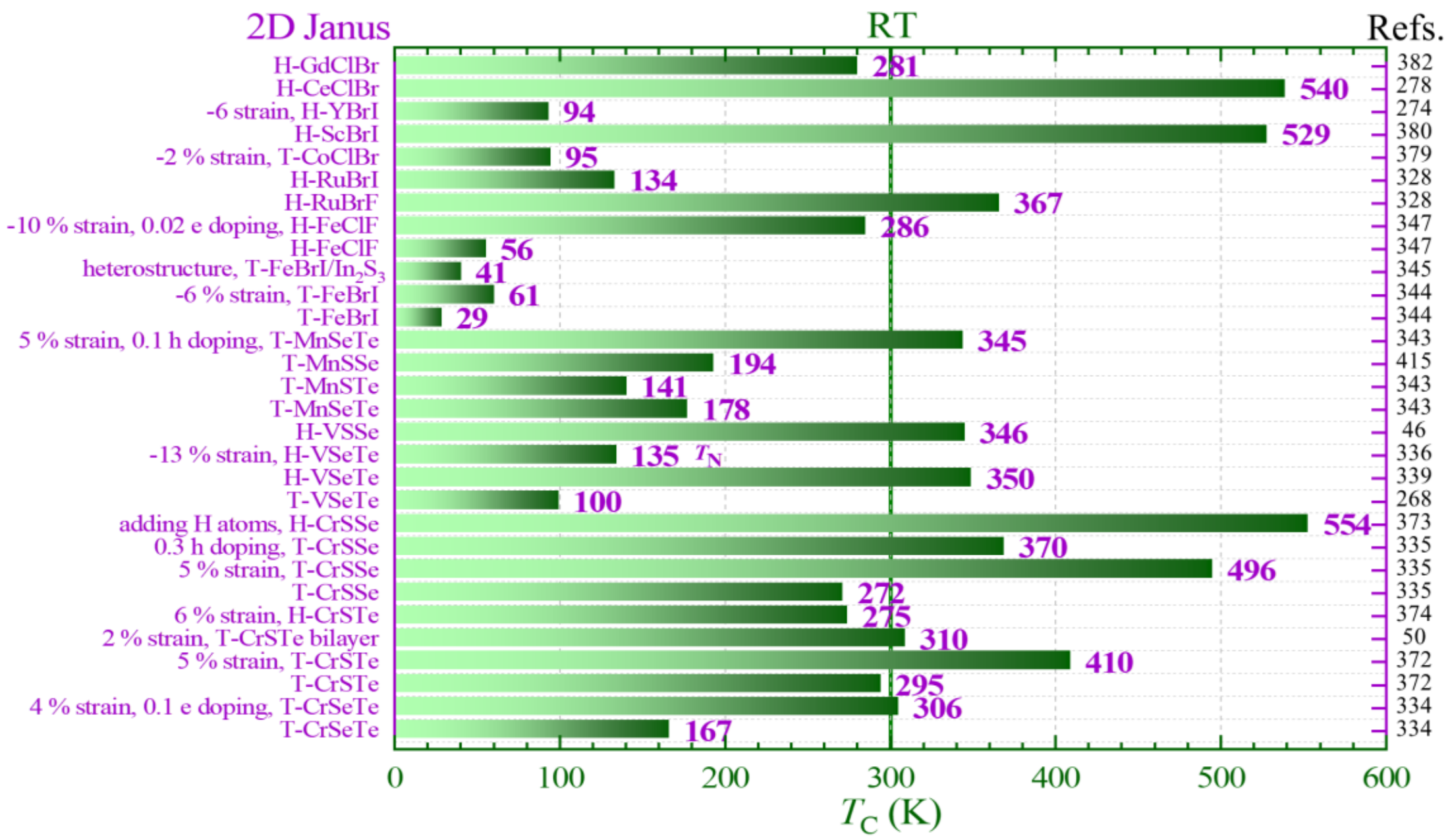

**FIG. 13.** The reported Curie temperatures ($T_C$s) including labeled Néel temperatures ($T_N$s) in 2D Janus materials, which are for monolayers unless specifically noted.

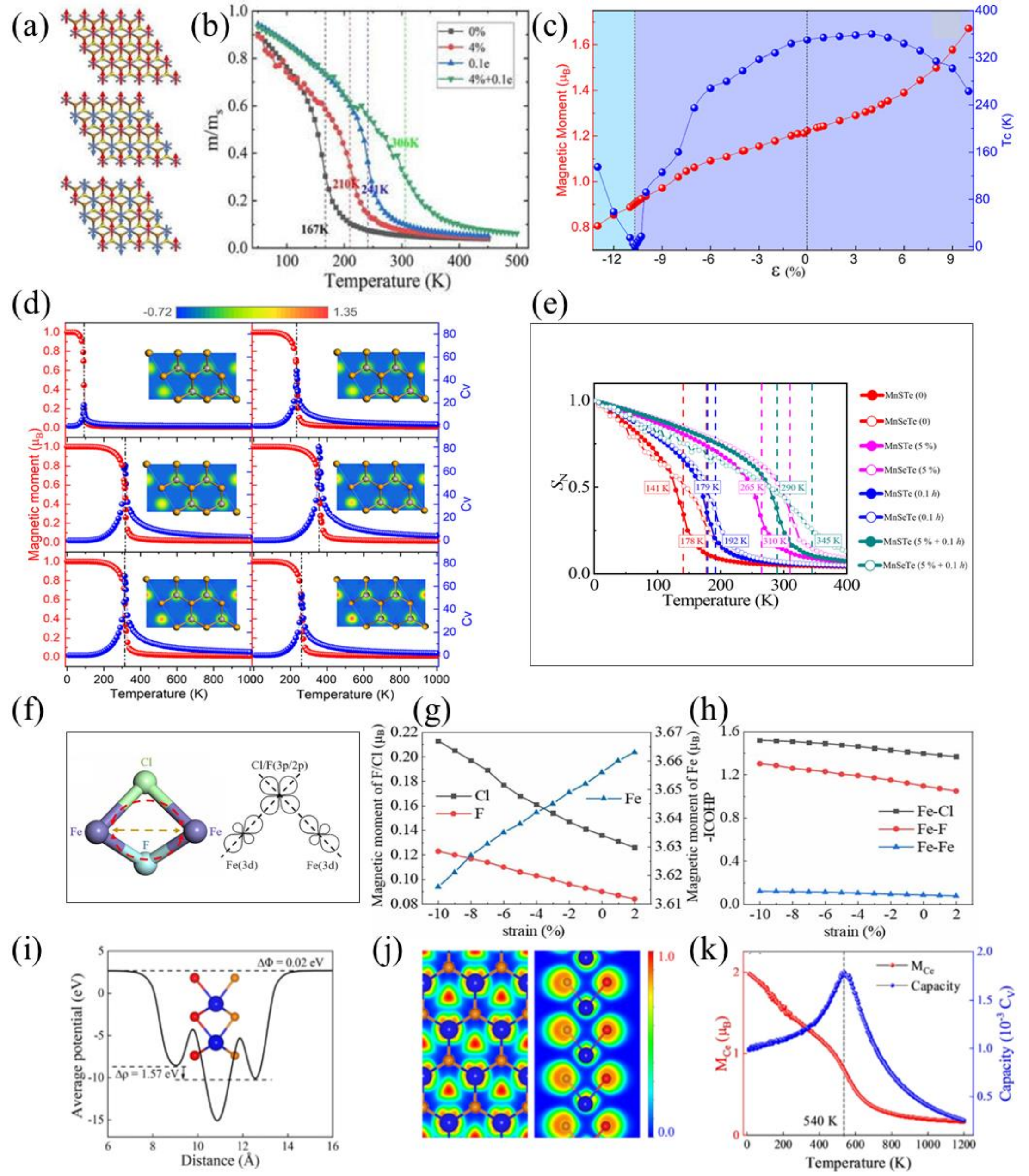

**FIG. 14.** The FM, stripy-AFM and zigzag-AFM states (a) and the changes of normalized magnetic moment with temperature for Janus T-CrSeTe (b).[334] Evolution of magnetic moment and Curie temperature ($T_C$) of Janus H-VSeTe (c,d).[336] Normalized spin operator varying with temperature for Janus T-MnSTe and MnSeTe (e).[343] Super- (tan) and direct- (red) exchange interactions and super-exchange interaction for the nearly-90° bond angle (f), the changes of atomic magnetic moments (g) and bond lengths (h) with strain in Janus H-FeClF.[347] The average electrostatic potential (i), the top and side views of electron localization function (ELF) (j) and the changes of magnetic moment and capacity with temperature (k) in Janus H-CeBrCl.[278] (a) and (b) Readapted from Liu *et al.*, Appl. Phys. Lett. **123**(19), 192407 (2023), with the permission of AIP Publishing. (c) and (d) Readapted with permission from Guan *et al.*, ACS Appl. Mater. Inter. **12**(47), 53067 (2020). Copyright 2020 American Chemical Society. (e) Readapted with permission from Zhang *et al.*, Nanoscale **15**(46), 18910 (2023). Copyright 2023 Royal Society of Chemistry. (f)-(h) Readapted with permission from Zhang *et al.*, 2D Mater. **10**(4), 045005 (2023). Copyright 2023 IOP Publishing. (i)-(k) Readapted with permission from Chen *et al.*, Phys. Rev. B **109**(12), 125421 (2024). Copyright 2024 American Physical Society.

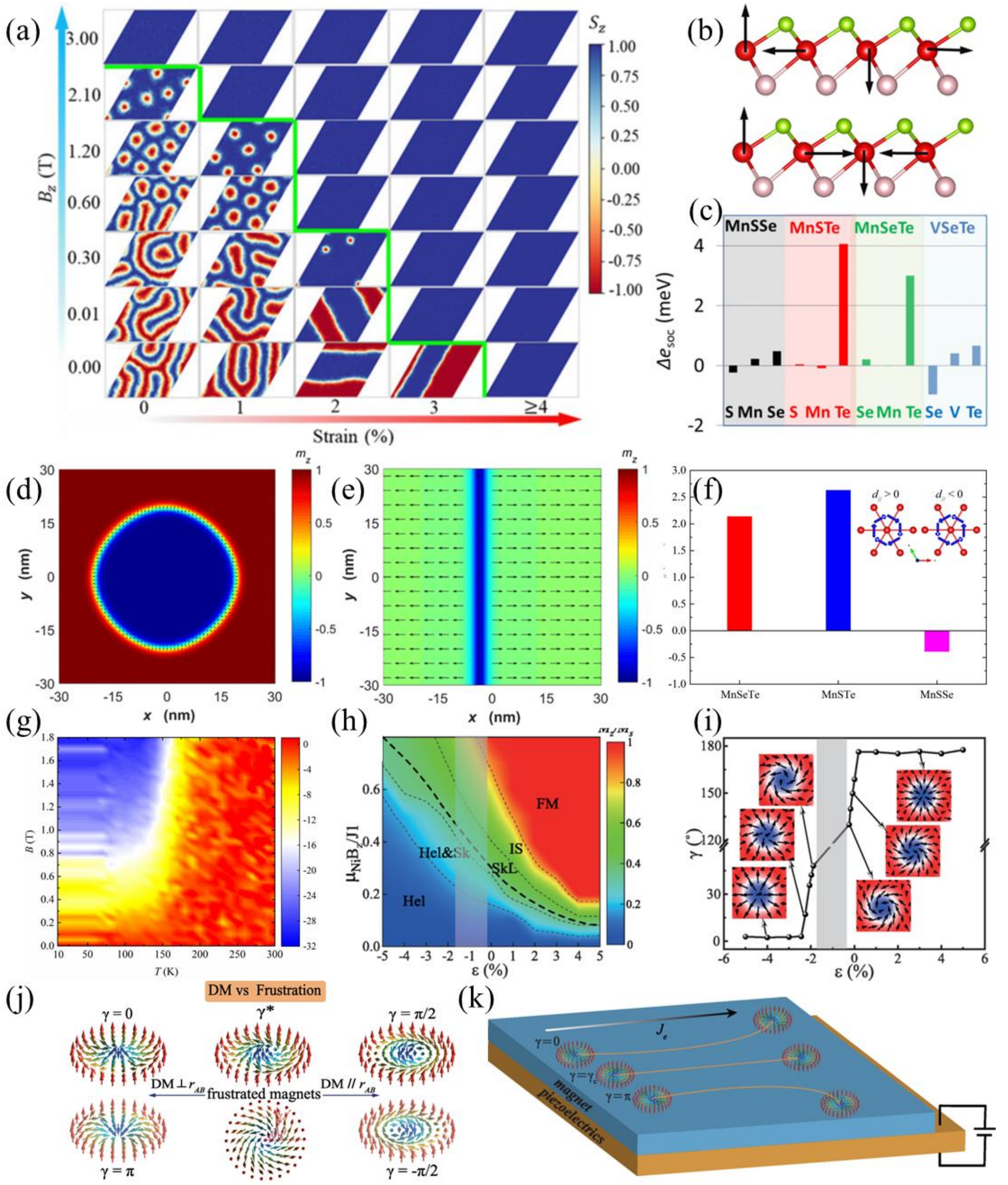

**FIG. 15.** Spin textures for Janus T-CrSeTe with strain and magnetic field at 10 K (a).[372] Left-hand and right-hand spin-spiral configurations (b) and atom-resolved localization of the associated SOC energy (c) in Janus T-MnXY (X ≠ Y = S, Se and Te) and VSeTe, and Magnetization distribution for the relaxed states at zero magnetic field in Janus T-MnSTe (d) and VSeTe (e).[326] The DMI parameter $d_{//}$ of Janus T-MnXY (f) and the topological charge $Q$ of MnSTe with temperature and magnetic field (g).[327] Phase diagram with strain and magnetic field (h) and skyrmion helicity with strain (i) in Janus T-NiClBr, the competition between DMI and exchange frustration for skyrmion helicity (j) and schematic diagram of skyrmion motion under electric current (k).[396] (a) Readapted with permission from Cui *et al.*, Phys. Rev. B **102**(9), 094425 (2020). Copyright 2020 American Physical Society. (b)-(e) Readapted with permission from Yuan *et al.*, Phys. Rev. B **101**(9), 094420 (2020). Copyright 2020 American Physical Society. (f) and (g) Readapted with permission from Liang *et al.*, Phys. Rev. B **101**(18), 184401 (2020). Copyright 2020 American Physical Society. (h)-(k) Readapted with permission from Liu *et al.*, Phys. Rev. B **109**(9), 094431 (2024). Copyright 2024 American Physical Society.

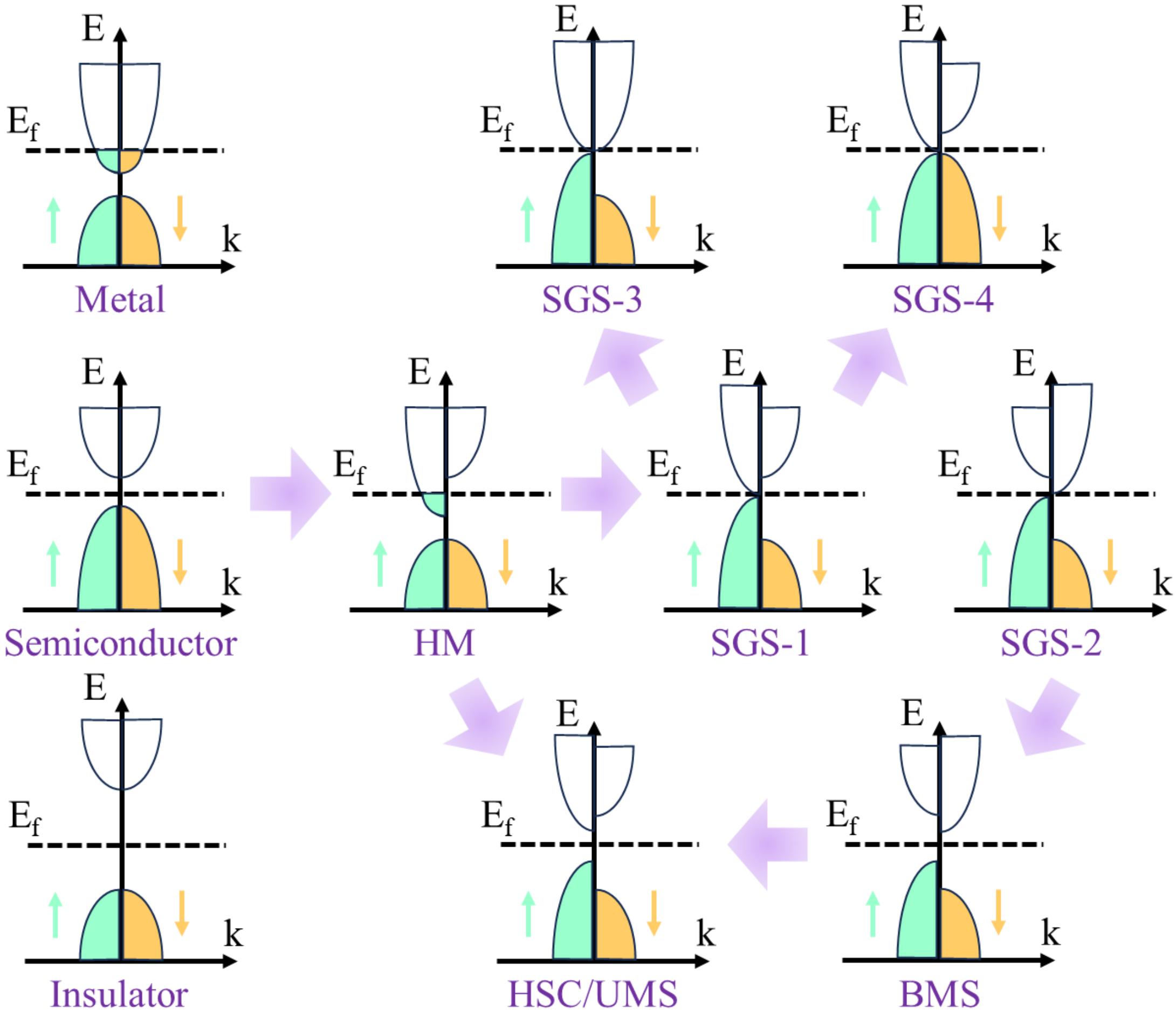

**FIG. 16.** Schematics for several special types of spin-resolved band structures.

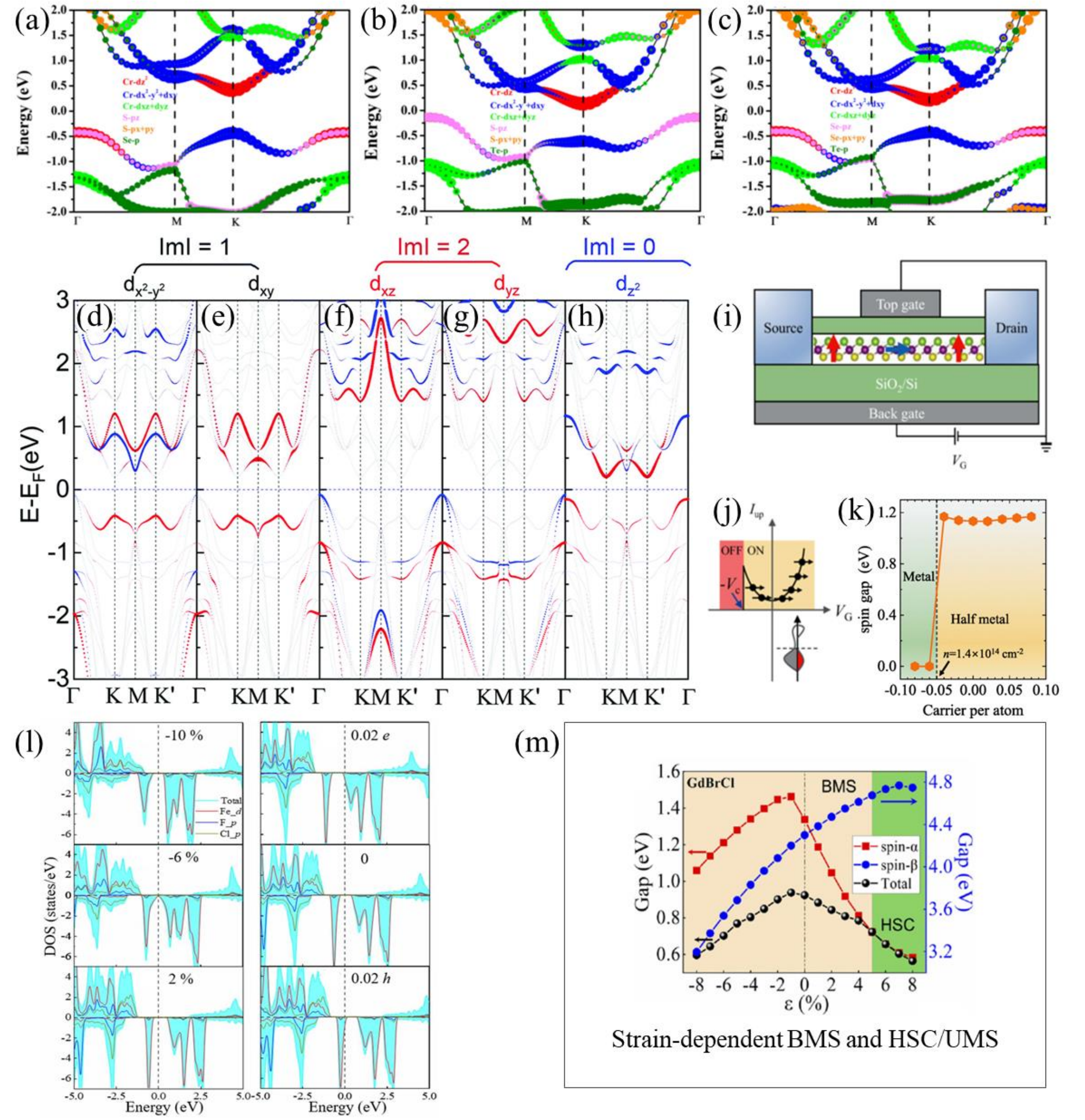

**FIG. 17.** The projected band structures of Janus H-CrSSe (a), CrSTe (b) and CrSeTe (c).[261] The band structures for V-*d* orbitals in Janus H-VSeTe (d,e,f,g,h).[339] Schematic illustration of proposed spin field effect transistor (FET) (i), spin-down current with voltage (j) and variation of spin-down band gaps with carrier concentration (k) in Janus T-MnSSe.[415] The total and main projected density of states (DOS) for Janus H-FeClF with strain and carrier doping (l).[347] Strain dependencies on band gaps for Janus H-GdBrCl (m).[382] (a)-(c) Readapted with permission from Chen *et al.*, Phys. Rev. B **106**(11), 115307 (2022). Copyright 2022 American Physical Society. (d)-(h) Readapted with permission from Guan *et al.*, Nanoscale **12**(44), 22735 (2020). Copyright 2020 Royal Society of Chemistry. (i)-(k) Readapted with permission from Chen *et al.*, Phys. Rev. B **105**(19), 195410 (2022). Copyright 2022 American Physical Society. (l) Readapted with permission from Zhang *et al.*, 2D Mater. **10**(4), 045005 (2023). Copyright 2023 IOP Publishing. (m) Readapted with permission from Li *et al.*, Phys. Rev. B **107**(11), 115428 (2023). Copyright 2023 American Physical Society.

**Table VI.** The matrix elements for the SOC operator $\boldsymbol{L} \cdot \boldsymbol{S}$ of *s* and *p* orbitals.

| Orbital | $s$ | $p_x$ | $p_y$ | $p_z$ |
|---|---|---|---|---|
| $s$ | 0 | 0 | 0 | 0 |
| $p_x$ | 0 | 0 | $-iS_z$ | $iS_y$ |
| $p_y$ | 0 | $iS_z$ | 0 | $-iS_x$ |
| $p_z$ | 0 | $-iS_y$ | $iS_x$ | 0 |

**Table VII.** The matrix elements for the SOC operator $\boldsymbol{L} \cdot \boldsymbol{S}$ of *d* orbitals.

| Orbital | $d_{xy}$ | $d_{x^2-y^2}$ | $d_{xz}$ | $d_{yz}$ | $d_{z^2}$ |
|---|---|---|---|---|---|
| $d_{xy}$ | 0 | $2iS_z$ | $-iS_x$ | $iS_y$ | 0 |
| $d_{x^2-y^2}$ | $-2iS_z$ | 0 | $iS_y$ | $iS_x$ | 0 |
| $d_{xz}$ | $iS_x$ | $-iS_y$ | 0 | $-iS_z$ | $i\sqrt{3}\,S_y$ |
| $d_{yz}$ | $-iS_y$ | $-iS_x$ | $iS_z$ | 0 | $-i\sqrt{3}\,S_x$ |
| $d_{z^2}$ | 0 | 0 | $-i\sqrt{3}\,S_y$ | $i\sqrt{3}\,S_x$ | 0 |

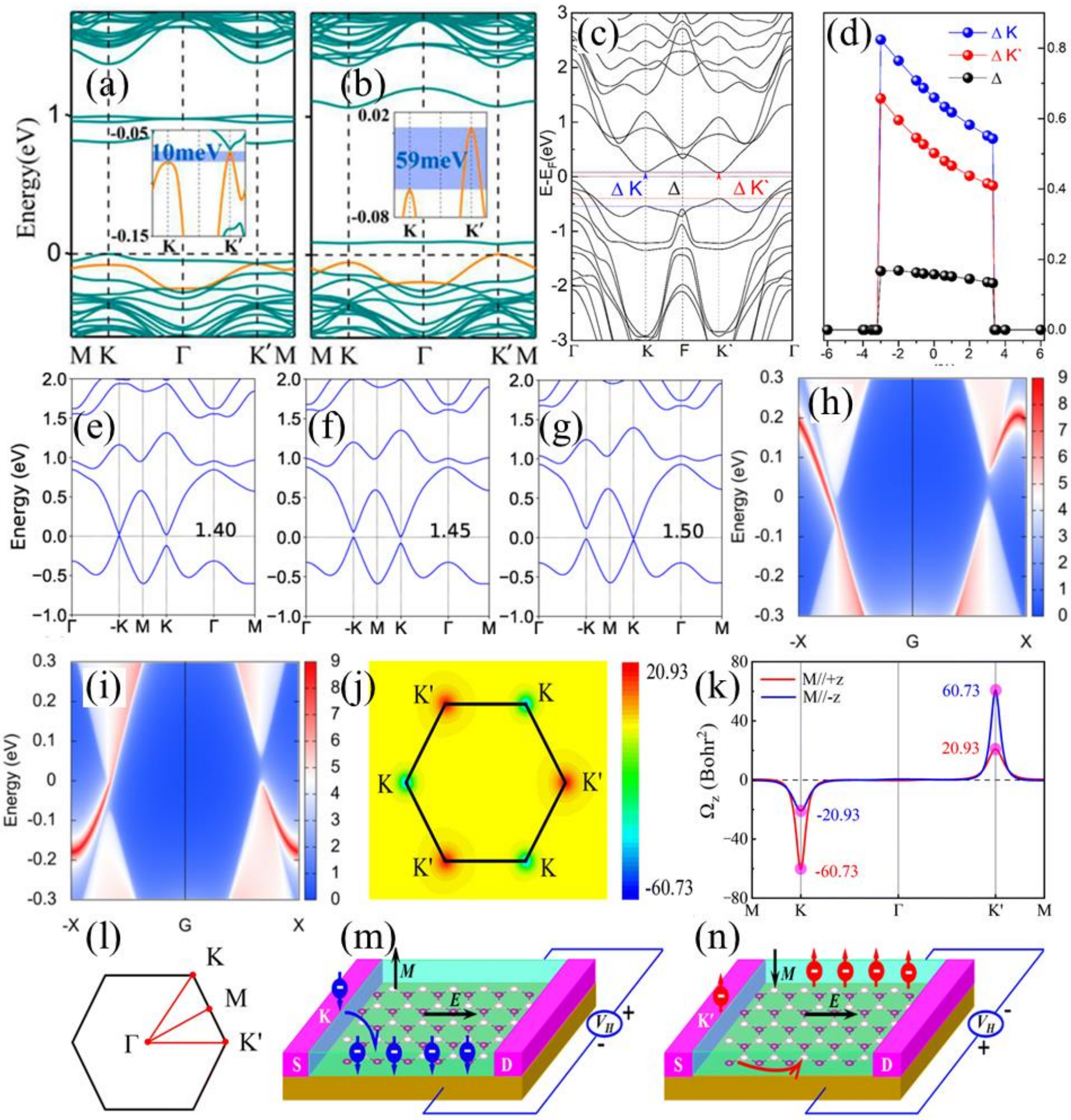

**FIG. 18.** Band structures of Cr-doped (a) and V-doped (b) Janus H-MoSSe.[47] Band structures (c) and valley splitting (d) with strain in Janus H-VSeTe.[336] For out of plane easy axis, band structures with U of 1.40 (e), 1.45 (f) and 1.50 (g) eV and topological states of left (h) and right (i) edges in Janus H-FeClF.[346] Berry curvatures in the 2D Brillouin zone (j) and along the high symmetry (k), the first Brillouin zone with the high-symmetry points (l), and schematic diagrams of anomalous valley Hall effect (AVHE) devices (m,n) in Janus H-RuClBr.[377] (a) and (b) Readapted with permission from Peng *et al.*, J. Phys. Chem. Lett. **9**(13), 3612-3617 (2018). Copyright 2018 American Chemical Society. (c) and (d) Readapted with permission from Guan *et al.*, ACS Appl. Mater. Inter. **12**(47), 53067 (2020). Copyright 2020 American Chemical Society. (e)-(i) Readapted with permission from Guo *et al.*, Phys. Rev. B **105**(10), 104416 (2022). Copyright 2022 American Physical Society. (j)-(n) Readapted with permission from Ma *et al.*, Nanoscale **15**(18), 8278 (2023). Copyright 2023 Royal Society of Chemistry.

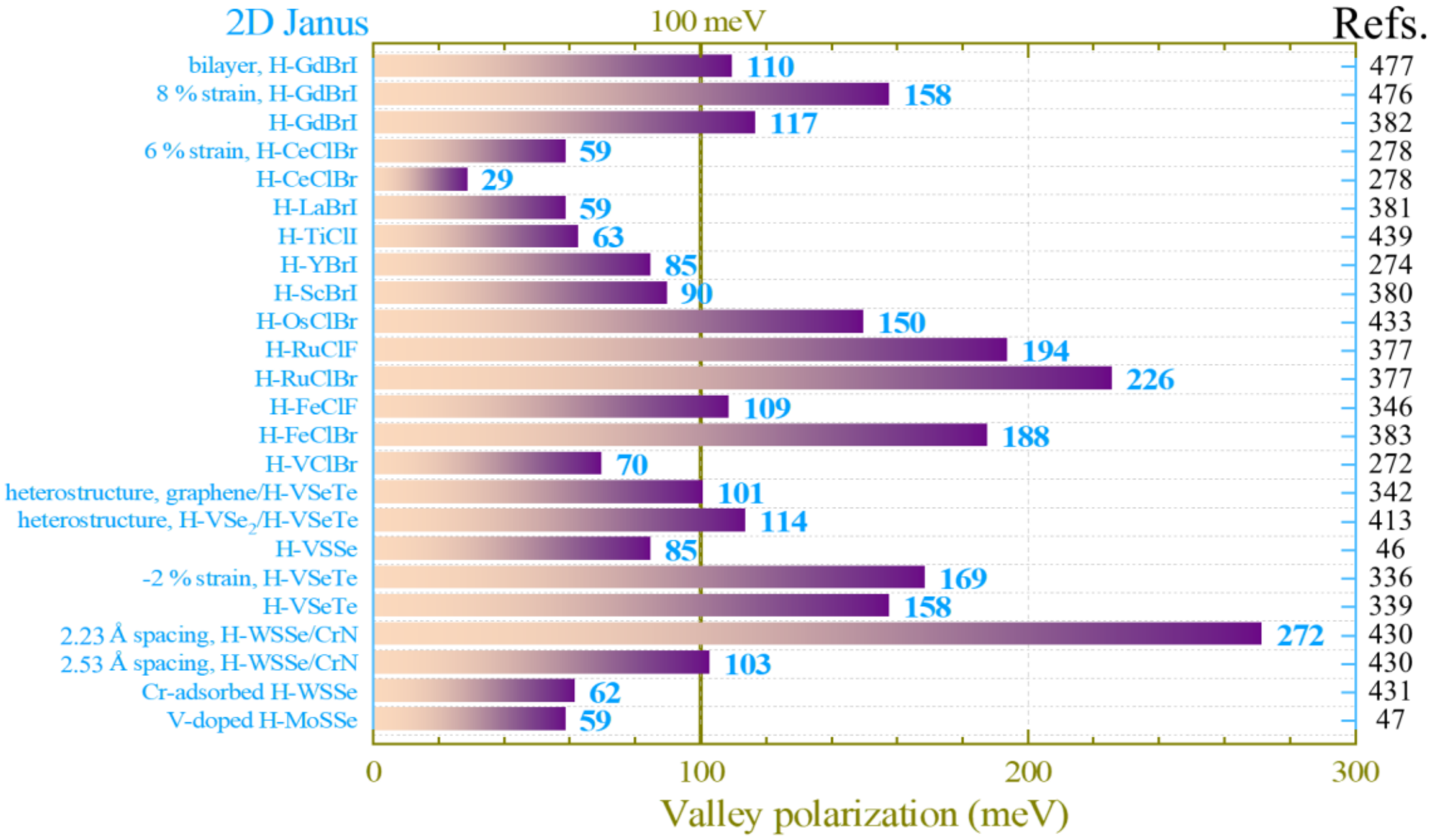

**FIG. 19.** The reported valley polarization in 2D Janus materials, which are for monolayers unless specifically noted.

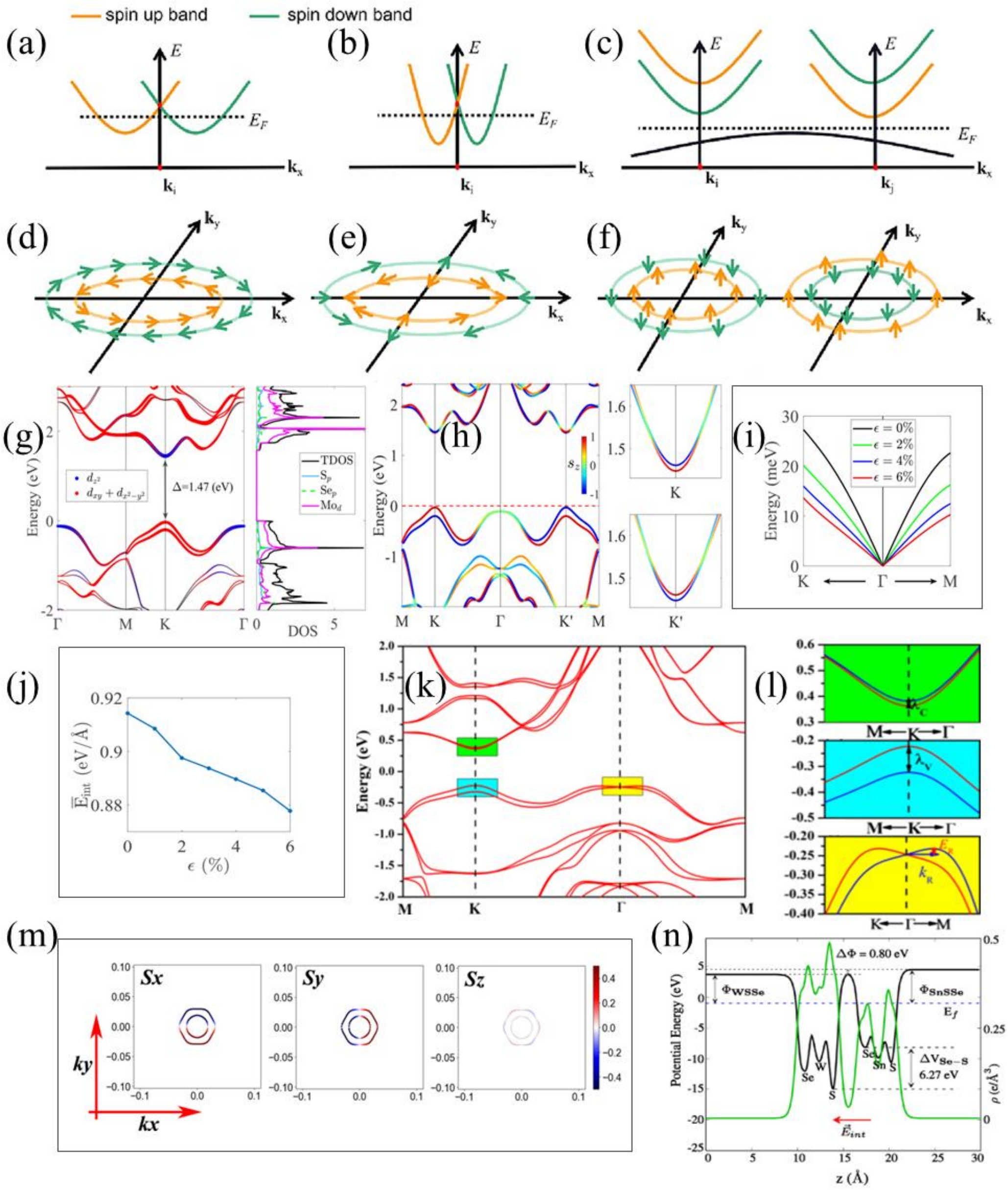

**FIG. 20.** Schematic bands (a,b,c) and spin textures (d,e,f) spin splitting of Rashba, Dresselhaus and Zeeman types.[448] The main projected band structures and DOS (g), band structure projection of z axis (h), Rashba splitting energy (i), and perpendicular built-in electric field (j) in Janus H-MoSSe.[452] Total (k) and local (l) band structures, and spin textures at energy level of -0.227 eV (m) of Janus H-CrSeTe.[261] The planar average for electrostatic potential energy (black) and charge density (green) in Janus T-SnSSe/H-WSSe heterostructure (n).[459] (a)-(f) Readapted with permission from He *et al.*, Results Phys. **49**, 106490 (2023). Copyright 2023 Elsevier. (g)-(j) Readapted with permission from Yu *et al.*, Phys. Rev. B **104**(7), 075435 (2021). Copyright 2021 American Physical Society. (k)-(m) Readapted with permission from Chen *et al.*, Phys. Rev. B **106**(11), 115307 (2022). Copyright 2022 American Physical Society. (n) Readapted with permission from Bhat, J. Phys.: Condens. Matter **35**(43), 435301 (2023). Copyright 2023 IOP Publishing.

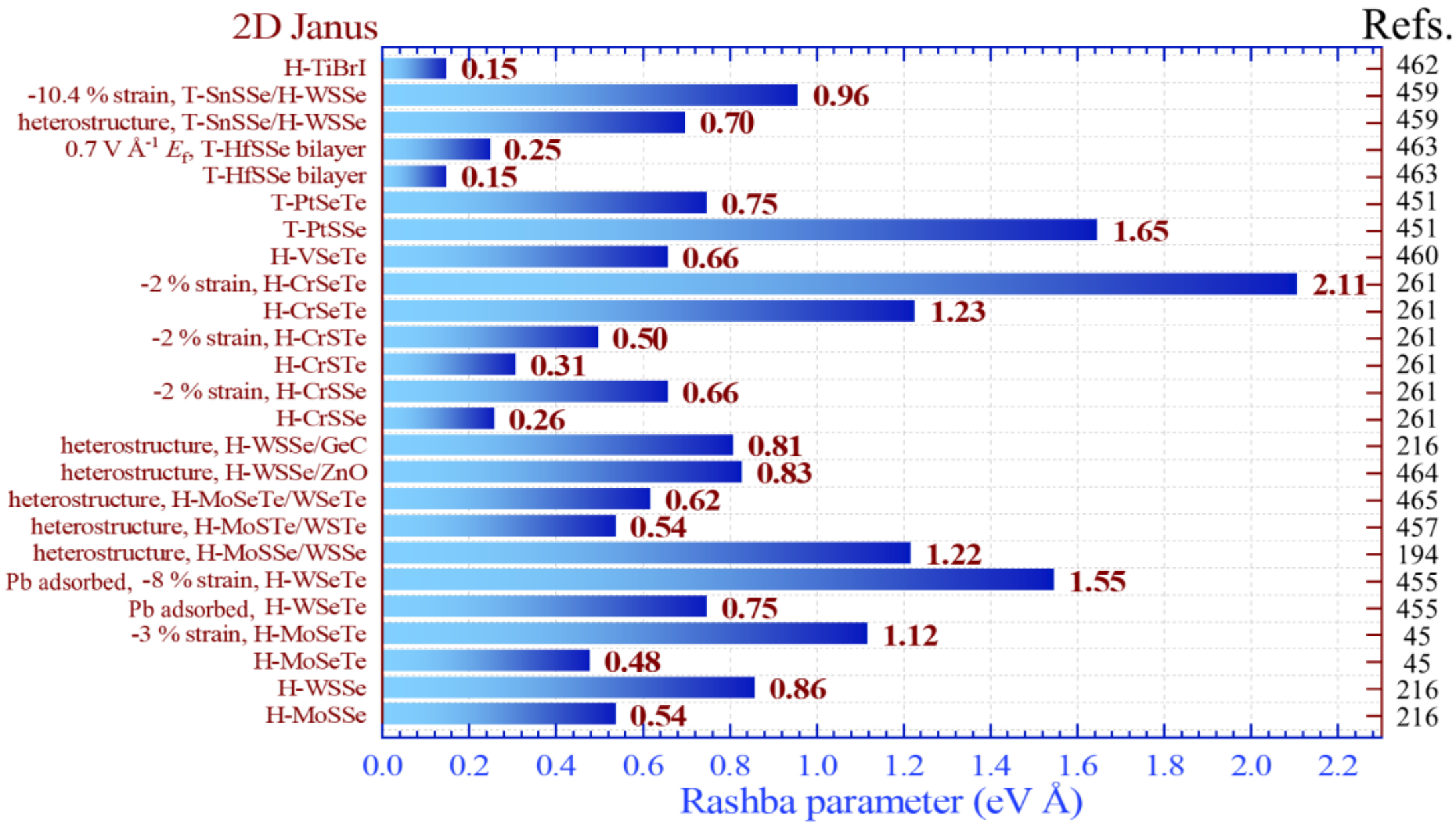

**FIG. 21.** The reported Rashba parameters in 2D Janus materials, which are for monolayers unless specifically noted.